\documentclass{aa}  

\usepackage{graphicx}
\usepackage{txfonts}
\usepackage{epstopdf}
\usepackage[utf8]{inputenc}
\usepackage[english]{babel}
\usepackage{booktabs}
\bibpunct{(}{)}{;}{a}{}{,}

\usepackage{hyperref}
\usepackage{array,multirow}
\usepackage{amsmath}	
\usepackage{amssymb}	
\usepackage[flushleft]{threeparttable}
\usepackage{comment}

\newcommand{\ihor}{$\iota$\,Hor}
\newcommand{\vect}[1]{\mathit{\mathbf{#1}}}	

\usepackage{xcolor, fontawesome}
\definecolor{twitterblue}{RGB}{64,153,255}

\hypersetup{colorlinks,linkcolor={red!60!black},citecolor={blue!60!black},urlcolor={blue!60!black}}
\usepackage{breakurl}

\usepackage{natbib,twoopt}
\bibpunct{(}{)}{;}{a}{}{,} 
\makeatletter
\newcommandtwoopt{\citeads}[3][][]{\href{http://ui.adsabs.harvard.edu/abs/#3}%
{\def\hyper@linkstart##1##2{}%
\let\hyper@linkend\@empty\citealp[#1][#2]{#3}}}
\newcommandtwoopt{\citepads}[3][][]{\href{http://ui.adsabs.harvard.edu/abs/#3}%
{\def\hyper@linkstart##1##2{}%
\let\hyper@linkend\@empty\citep[#1][#2]{#3}}}
\newcommandtwoopt{\citetads}[3][][]{\href{http://ui.adsabs.harvard.edu/abs/#3}%
{\def\hyper@linkstart##1##2{}%
\let\hyper@linkend\@empty\citet[#1][#2]{#3}}}
\newcommandtwoopt{\citeyearads}[3][][]%
{\href{http://ui.adsabs.harvard.edu/\#abs/#3}
{\def\hyper@linkstart##1##2{}%
\let\hyper@linkend\@empty\citeyear[#1][#2]{#3}}}
\makeatother

\begin{document} 

   \title{Far beyond the Sun: III. The magnetic cycle of $\boldsymbol\iota$ Horologii}

   \author{J. D. Alvarado-G\'omez\inst{1}\fnmsep\thanks{julian.alvarado-gomez@aip.de} \and G. A. J. Hussain\inst{2} \and E. M. Amazo-G\'omez\inst{1} \and Y. Xu\inst{1,3} \and K.~Poppenh\"{a}ger\inst{1,4} \and J.~Chebly\inst{1,5} \and J.-F.~Donati\inst{6} \and 
   B.~Stelzer\inst{7} \and J. Sanz-Forcada\inst{8}
          }
   \institute{Leibniz Institute for Astrophysics Potsdam, An der Sternwarte 16, 14482 Potsdam, Germany
    \and
    European Space Agency, ESTEC, Keplerlan 1, 2201 AZ Noordwijk, The Netherlands
    \and
    School of Earth and Space Sciences, Peking University, Beijing 100781, China
    \and
    Universit\"{a}t Potsdam, Institut f\"{u}r Physik und Astronomie, Karl-Liebknecht-Straße 24/25, 14476 Potsdam-Golm, Germany
    \and
    Universit\'e Paris-Saclay, Universit\'e Paris Cit\'e, CEA, CNRS, AIM, 91191,Gif-sur-Yvette, France
    \and
    CNRS-IRAP, 14, avenue Edouard Belin, F-31400 Toulouse, France
    \and
    Eberhard Karls Universit\"{a}t, Institut f\"{u}r Astronomie und Astrophysik, Sand 1, 72076 T\"{u}bingen, Germany
    \and
    Centro de Astrobiolog\'ia (CSIC-INTA), ESAC Campus, Camino Bajo del Castillo, E-28692 Villanueva de la Ca{\~n}ada, Madrid, Spain
             }

   \date{Received ---; accepted ---}

 \abstract{We present a comprehensive investigation of the magnetic cycle of the young, active solar analogue $\iota$ Horologii (\ihor) based on intensive spectropolarimetric monitoring using HARPSpol. Over a nearly three-year campaign, the technique of Zeeman-Doppler Imaging (ZDI) was used to reconstruct 18 maps of the large-scale surface magnetic field of the star. These maps trace the evolution of the magnetic field morphology over approximately 139 stellar rotations. Our analysis uncovers pronounced temporal evolution, including multiple polarity reversals and changes in field strength and geometry. We examine the evolution of the poloidal and toroidal field components, with the toroidal component showing strong modulation in concert with the chromospheric activity. Furthermore, for the first time, we reconstruct stellar magnetic butterfly diagrams which are used to trace the migration of large-scale magnetic features across the stellar surface, determining a magnetic polarity reversal timescale of roughly 100 rotations ($\sim$\,$773$~d). In addition, by tracking the field-weighted latitudinal positions, we obtain the first estimates of the large-scale flow properties on a star other than the Sun, identifying possible pole-ward and equator-ward drift speeds for different field polarities. These results provide critical insights into the dynamo processes operating in young solar-type stars and offer a direct comparison with the solar magnetic~cycle.}    

   \keywords{stars: individual ($\iota$\,Hor, HD~17051) -- stars: activity -- stars: magnetic field -- stars: solar-type -- techniques: polarimetric}

\maketitle

\section{Introduction}\label{sec:intro}
\nolinenumbers
Over the past 25 years, the study of magnetic fields in late-type stars has benefited from improvements in both instrumentation and observational methods. Data from high-resolution spectropolarimeters such as NARVAL (\citeads{2003EAS.....9..105A}), ESPaDOnS (\citeads{2003ASPC..307...41D}), PEPSI (\citeads{2002AN....323..510H,2015AN....336..324S}), and HARPS\-pol (\citeads{2011Msngr.143....7P}), combined with procedures like Least-Squares Deconvolution (LSD, \citeads{1997MNRAS.291..658D}; \citeads{2010A&A...524A...5K}) and Single-Value Decomposition (SVD, \citeads{2012A&A...548A..95C}), have shown remarkable success detecting the very weak polarization signatures induced by magnetic fields via the Zeeman effect (\citeads{2009ARA&A..47..333D}; \citeads{2012LRSP....9....1R}). Furthermore, the technique of Zeeman-Doppler Imaging (ZDI, \citeads{1989A&A...225..456S}; \citeads{1997A&A...326.1135D}; \citeads{2002A&A...381..736P}; \citeads{2002A&A...388..868K}) has provided a way to reconstruct the spatial distribution of large-scale surface magnetic fields in stars of various spectral types (see~\citeads{2011IAUS..271...23D}; \citeads{2017AN....338..428K}). 

The aforementioned spectropolarimetric observations require a significant amount of telescope time, limiting current survey-scale efforts to magnetic field detections alone (\citeads{2014MNRAS.444.3517M}; \citeads{2017MNRAS.465.2734M}). Still, ZDI snapshots exist for a growing number of stars (e.g.,~\citeads{2008MNRAS.388...80P}; \citeads{2013MNRAS.435.1451F}; \citeads{2016ApJ...820L..15D}), with some systems having their large-scale fields mapped over multiple epochs, typically on yearly time-scales (e.g.,~Morin~et~al.~\citeyearads{2008MNRAS.390..567M}, \citeyearads{2010MNRAS.407.2269M}; Fares~et~al.~\citeyearads{2010MNRAS.406..409F}, \citeyearads{2012MNRAS.423.1006F}). These ZDI monitoring studies have revealed a variety of temporal behaviors of the large-scale field across spectral types. In the case of Sun-like stars, some objects display field strength variability without changes of polarity in the observed time span ($T_{\rm obs}$). Examples of this are HD~141943, (G0IV, $T_{\rm obs} \sim 4$~yr, \citeads{2011MNRAS.413.1922M}), HN~Peg (G0V, $T_{\rm obs} \sim 6$~yr, \citeads{2015A&A...573A..17B}), EK~Dra (G5V, $T_{\rm obs} \sim 5.5$ yr, \citeads{2017MNRAS.465.2076W}). In other cases, single polarity reversals in one or more components of the large-scale magnetic field have been observed. Stars in this group include HD~78366 (G0V, $T_{\rm obs} \sim 3.0$ yr, \citeads{2011AN....332..866M}), $\xi$~Boo~A (G7V, $T_{\rm obs} \sim 3.5$ yr, \citeads{2012A&A...540A.138M}), HD~29615 (G3V, $T_{\rm obs} \sim 3.5$ yr, \citeads{2016A&A...587A..28H}), HD~75332 (F7V, $T_{\rm obs} \sim 12$ yr, \citeads{2021MNRAS.501.3981B}) and LQ Hya (K1V, $T_{\rm obs} \sim 6.0$ yr, \citeads{2022A&A...660A.141L}), among others (see \citeads{2025A&A...693A.269B}). 

So far stellar magnetic cycles, understood as a double (or more) polarity reversals in the temporal large-scale magnetic evolution, have been found using ZDI in the following objects: $\tau$~Boo (F7V, $T_{\rm obs} \sim 8.5$~yr, \citeads{2008MNRAS.385.1179D}; Fares~et.~al. \citeyearads{2009MNRAS.398.1383F}, \citeyearads{2013MNRAS.435.1451F}; \citeads{2016MNRAS.459.4325M}), $\epsilon$~Eri (K2V, $T_{\rm obs} \sim 7.5$ yr, Jeffers~et~al. \citeyearads{2014A&A...569A..79J}, \citeyearads{2017MNRAS.471L..96J}), 61~Cyg~A (K5V, $T_{\rm obs} \sim 11$~yr, Boro Saikia~et~al. \citeyearads{2016A&A...594A..29B}, \citeyearads{2018A&A...620L..11B}), $\chi^1$ Ori (G0V, $T_{\rm obs} \sim 10$~yr \citeads{2016A&A...593A..35R}, \citeads{2022A&A...659A..71W}), $\kappa$~Ceti (G5V, $T_{\rm obs} \sim 10$~yr, \citeads{2022A&A...658A..16B}), HD~9986 (G5V, $T_{\rm obs} \sim 15$~yr, \citeads{2025A&A...693A.269B}) and HD~56124 (G0V, $T_{\rm obs} \sim 13$~yr, \citeads{2025A&A...693A.269B}). The long baseline of spectropolarimetric observations has allowed to start connecting the properties of these ZDI magnetic cycles with previously published stellar activity cycles (\citeads{2016MNRAS.462.4442S}). The latter are typically detected in magnetic proxies such as the chromospheric Ca\,II H\&K S-index from Mount Wilson (e.g., \citeads{1995ApJ...438..269B}; Hall~et~al. \citeyearads{2007AJ....133..862H}, \citeyearads{2009AJ....138..312H}), the coronal behavior in X-ray (see \citeads{2017MNRAS.464.3281W} and references therein) and, in a very few cases, via simultaneous activity diagnostics from the photosphere to the corona \citepads{2023MNRAS.524.5725A}.

Despite the progress, it is clear that ZDI monitoring of additional systems is required to improve our understanding of magnetic and activity cycles in cool stars. This is one of the objectives of the \textit{Far Beyond the Sun} campaign, through which we conducted an unprecedented intensive spectropolarimetric monitoring of the young Sun-like star $\iota$\,Horologii (G0V, Age: $\sim$$600$\,Myr, \citeads{2008A&A...482L...5V}; \citeads{2010MNRAS.405.1907B}). As discussed in the first papers of this series, $\iota$\,Hor is the youngest and most active planet-hosting star with reported cyclic chromospheric and coronal activity behavior (\citeads{2010ApJ...723L.213M}; \citeads{2018MNRAS.473.4326A}). The star also displays one of the shortest an best sampled X-ray cycles to date ($P_{\rm X}^{\rm cyc} \sim1.6$~yr; Sanz-Forcada~et~al. \citeyearads{2013A&A...553L...6S}, \citeyearads{2019A&A...631A..45S}), making it an exceptional target to follow-up over the time-scale of its short cycle. 

In this paper we present $18$~separate ZDI large-scale magnetic field maps of $\iota$\,Hor spanning nearly three years, sufficient to cover two complete coronal cycles and resolve the magnetic cycle. Details on the observations and ZDI reconstructions are presented in Sect.~\ref{sec:observations}. Results and analysis are included in Sect.~\ref{sec:results}. We discuss and summarize our findings in Sects.~\ref{sec:discussion} and \ref{sec:summary}.

\section{Observations}\label{sec:observations}

The observations in this project were acquired using the spectro-polarimetric mode of the HARPS spectrograph (HARPSpol), mounted on the ESO $3.6$~m telescope at the La Silla Observatory in Chile (\citeads{2003Msngr.114...20M}, \citeads{2011Msngr.143....7P}). We obtained circularly polarized spectra (Stokes\,V) of $\iota$~Hor on $199$ separate nights over the course of six consecutive ESO observing periods ($96$ to $101$). As described in \citetads{2018MNRAS.473.4326A}, two-week\footnote[1]{In practice, each epoch had an average of 11 quasi-consecutive nights (minimum of 8, maximum of 14) owing to bad weather conditions.} epochs were scheduled three times per semester, yielding a total of 18 separate visits with sufficient rotational phase coverage for individual ZDI reconstructions. 

\subsection{Data reduction}\label{ss:data}

The data processing was described in detail in \citetads{2018MNRAS.473.4326A}, and will be summarized here briefly. The circularly polarized (Stokes\,V) and diagnostic null (N) spectra were derived by combining consecutive exposures (each one consisting of four sub-exposures matching orthogonal polarization angles) using the ratio method (see \citeads{1997MNRAS.291..658D}; \citeads{2009PASP..121..993B} for details). Between two and three consecutive Stokes\,V exposures were acquired each night, with a combined integration time of approximately $1$~h. 

The observations were reduced employing the HARPSpol-compatible version of the {\sc libre-esprit} pipeline \citepads{1997MNRAS.291..658D}, which applies standard spectroscopic corrections (e.g., bias subtraction, flat-fielding, masking bad pixels, cosmic ray removal), wavelength calibration (using two reference ThAr arc spectra per night), and the polarization calculation and subtraction of continuum polarization. As was shown in \citetads{2023MNRAS.524.5725A}, this procedure retains the nominal RV precision of the HARPS spectrograph (see also \citeads{2016MNRAS.461.1465H}, \citeads{2016A&A...585A..77H}, \citeads{2018MNRAS.473.4326A}). A two-step continuum normalization was carried out, using first HARPS settings ($R \sim 115000$) for the automatic procedure included in the {\sc ispec} package \citepads{2014A&A...569A.111B}, and then performing a manual fitting of cubic-splines over $15$~nm windows throughout the entire spectral range ($378-691$~nm). The reduced data products consisted of unpolarized intensity profiles (Stokes\,I), circularly polarized spectra (Stokes\,V), and diagnostic null profiles (N) which are used to check for instrumental noise and artifacts. 
 
In moderately active solar-type stars such as $\iota$~Hor, the detection of surface magnetic features via Zeeman splitting using individual spectral lines is currently unfeasible (see \citeads{2009ARA&A..47..333D}; \citeads{2012LRSP....9....1R}). In order to maximize the information contained in thousands of photospheric individual line profiles, we employ the multi-line technique of Least-Squares Deconvolution (\citeads{1997MNRAS.291..658D}; \citeads{2010A&A...524A...5K}). Commonly used in ZDI studies, this cross-correlation technique adds coherently the polarimetric signal from multiple lines, yielding mean profiles per observation (LSD Stokes\,I, V, and N) with very high signal-to-noise ratios (S/N). 

A photospheric line list, with information on rest wavelengths, line depths, and Land\'e factors, is required to perform the LSD procedure. The Vienna Atomic Line Database (VALD3\footnote[2]{\url{http://vald.astro.uu.se}}, \citeads{2000BaltA...9..590K}; \citeads{2015PhyS...90e4005R}), was used for this purpose using the stellar properties reported by \citetads{2010MNRAS.405.1907B}, and imposing a line-depth detection threshold of $0.05$ (in normalized units). The original photospheric model ($\sim$\,$9500$ spectral lines in the HARPS wavelength range) was improved following the procedure discussed in \citetads{2015A&A...582A..38A}, tailoring all the line depths to an observed spectrum of the star, and removing spectral lines that largely deviate from the self-similarity condition required by LSD. The final line list for $\iota$~Hor has $8834$ entries and served to retrieve LSD profiles over the velocity range between $-5.0$~km~s$^{-1}$ and $45.4$~km~s$^{-1}$, with a velocity step of $\Delta v = 0.8$ km~s$^{-1}$, closely matching the HARPS detectors' pixel width. 

\subsection{Mapping the large-scale magnetic field: Zeeman-Doppler Imaging}\label{ss:zdi}

As mentioned earlier, we use ZDI to map the large-scale surface magnetic field of $\iota$~Hor for each observing run during the $\sim$\,3-year campaign. ZDI is an indirect imaging technique that inverts time series of spectropolarimetric observations to reconstruct the photospheric magnetic field in a 2-dimensional projection. Our analysis uses circularly polarized profiles, which are only sensitive to the line-of-sight component of the stellar magnetic field and therefore subject to cancellation effects in the presence of complex multipolar small-scale fields. The resulting maps can therefore only reconstruct the large-scale component of the residual surface magnetic field (see e.g.~\citeads{2012LRSP....9....1R}; \citeads{2019MNRAS.483.5246L}). As in other image inversion methods, a regularization procedure (such as maximum smoothness or minimum information), is imposed on the reconstruction to ensure the uniqueness of the solution (see \citeads{1989A&A...225..456S}; \citeads{1991A&A...250..463B}, \citeads{2000MNRAS.318..961H}; \citeads{2002A&A...381..736P}; \citeads{2012A&A...548A..95C}). The maps are regularized using maximum entropy  \citepads{1984MNRAS.211..111S}, fitting the spectropolarimetric datasets up to an optimal reduced chi-square ($\chi^2_{\rm R}$) value given by the entropy-growth stopping criterion introduced by \citetads{2015A&A...582A..38A}. 

The technique of ZDI can be used to further refine several of the key stellar parameters and spectral line properties of $\iota$~Hor reported by \citeads{2018MNRAS.473.4326A} (and references therein). Parameters such as rotation period, $v\sin(i)$, and inclination angle are fine-tuned by allowing them to vary over the ranges reported in the literature and minimizing the $\chi^2_{\rm R}$-fit to the data (e.g.,~\citeads{2008MNRAS.390..635J}, \citeads{2015MNRAS.449....8W}). Using this approach we find $v\sin(i) = 6.3$~km~s$^{-1}$ and a rotation period $P_{\rm rot} = 7.73$~d; both consistent with previous estimates and with values derived from our photometric analysis in \citetads{2023MNRAS.524.5725A}. Given the published radius of 1.16~$R_\odot$, this corresponds to an inclination angle of $\sim 56^{\circ}$. The following results were obtained assuming an inclination angle of $60^{\circ}$. The magnetic field maps presented here employ the ZDI code first described in \citetads{2002ApJ...575.1078H} which has since been updated to implement a spherical harmonic description of the large-scale magnetic field (see \citeads{2021MNRAS.500.1243L}) and briefly summarised again here. Essentially, the surface field vector is decomposed as the sum of poloidal (potential, $\alpha$ and $\beta$) and toroidal (non-potential,  $\gamma$) components, which are expressed in terms of spherical-harmonics: Equations 1-4 from \citetads{2021MNRAS.500.1243L} are reproduced here for completeness. 

The vector magnetic field $\vect{B}$ is described in terms of its constituent radial, azimuthal and meridional field components $(B_{r}, B_{\phi}, B_{\theta})$, following \citetads{1946PhRv...70..202E} and Appendix III of \citetads{1961hhs..book.....C}, but applying a left-handed coordinate system, 

\begin{equation}
B_{r}(\phi, \theta) = \sum_{\ell m} \alpha_{\ell m} P_{\ell m} e^{im\phi}, \label{Eq:B_rad}
\end{equation}
\begin{equation}
B_{\phi}(\phi, \theta) = - \sum_{\ell m} \beta_{\ell m} \frac{im P_{\ell m} e^{im\phi}}{(\ell + 1) \sin \theta} + \sum_{\ell m} \gamma_{\ell m} \frac{1}{\ell+1} \frac{\mathrm{d}P_{\ell m}}{\mathrm{d}\theta} e^{im\phi}, \label{Eq:B_azi}
\end{equation}
\begin{equation}
B_{\theta}(\phi, \theta) = \sum_{\ell m} \beta_{\ell m} \frac{1}{\ell+1} \frac{\mathrm{d}P_{\ell m}}{\mathrm{d}\theta} e^{im\phi} + \sum_{\ell m} \gamma_{\ell m} \frac{im P_{\ell m} e^{im\phi}}{(\ell + 1) \sin \theta}, \label{Eq:B_mer}
\end{equation}

\noindent so that $(B_{r}, B_{\phi}, B_{\theta}) = \vect{B}$, where $P_{\ell m} \equiv c_{\ell m}P_{\ell m}(\cos \theta)$ are the associated Legendre polynomial of mode $\ell$ and order $m$ and  $c_{\ell m}$ is a normalization constant:
\begin{equation}
c_{\ell m} = \sqrt{\frac{2\ell+1}{4\pi}\frac{(\ell - m)!}{(\ell + m)!}}.
\end{equation}

\noindent In the left-handed coordinate system, the radial $B_r$ component points radially outwards, the azimuthal  component $B_\phi$  runs in the clockwise direction (as viewed from North pole), and the meridional component $B_\theta$ runs with colatitude from North to South.
In the ($r,\phi,\theta$) spherical harmonic system used here, the $\alpha_{\ell m}$ and $\beta_{\ell m}$ components describe the poloidal field whereas the $\gamma_{\ell m}$ components describe the toroidal field (also see \citeads{2021MNRAS.500.1243L}).

By using spherical harmonics to describe the mapped magnetic field, it is relatively simple to reconstruct and examine the large-scale magnetic field on different length scales. An important first step in reconstructing the magnetic field morphology of stellar surfaces is to identify the smallest length scale to which the spectral time series are sensitive -- this is largely a function of $v\,\sin\,i$ of the star, the instrumental resolution, the intrinsic line profile width, phase coverage, and the exposure time of the observations \citepads{2016LNP...914..177K}. For $\iota$\,Hor, we find that allowing modes beyond $\ell=7$ does not yield additional information, hence all reconstructions presented here are restricted to $\ell_{\rm max}=7$, corresponding well to $\sim 25^\circ$ resolution at the equator and closely matching the optimal spatial resolution allowed due to \ihor's $v\,\sin\,i$.

During the reconstruction it is also possible to apply higher weights or penalties to even or odd $\ell$ modes, thus pushing for antisymmetric and symmetric large-scale fields, respectively. In this study, we produce three sets of solutions for each epoch: unconstrained (i.e. in which all modes are allowed and no weighting is applied), symmetric, and antisymmetric. This enables an evaluation of how robust the large-scale field reconstructions are. In addition, we learn whether an antisymmetric or symmetric description better applies to $\iota$\,Hor based on a comparison of the information content of each set of~maps.

The large-scale field can be characterised in terms of several key properties, e.g. polarity reversal, axisymmetry, the evolution of the magnetic field strength, size of mean $B^2$ (proportional to the magnetic energy), etc. We examine the evolution of these characteristics over time and their relation to the chromospheric activity indicators analyzed in Papers I and II. In addition, we evaluate the cycle variability of the dipole, quadrupole, and octupole components (i.e. $\ell=1-3$) of the toroidal and poloidal large-scale magnetic field. 

\section{Results and Analysis}\label{sec:results}

\begin{figure*}
\includegraphics[trim=0.0cm 1.0cm 1.0cm 0.0cm, clip=true, width=\linewidth]{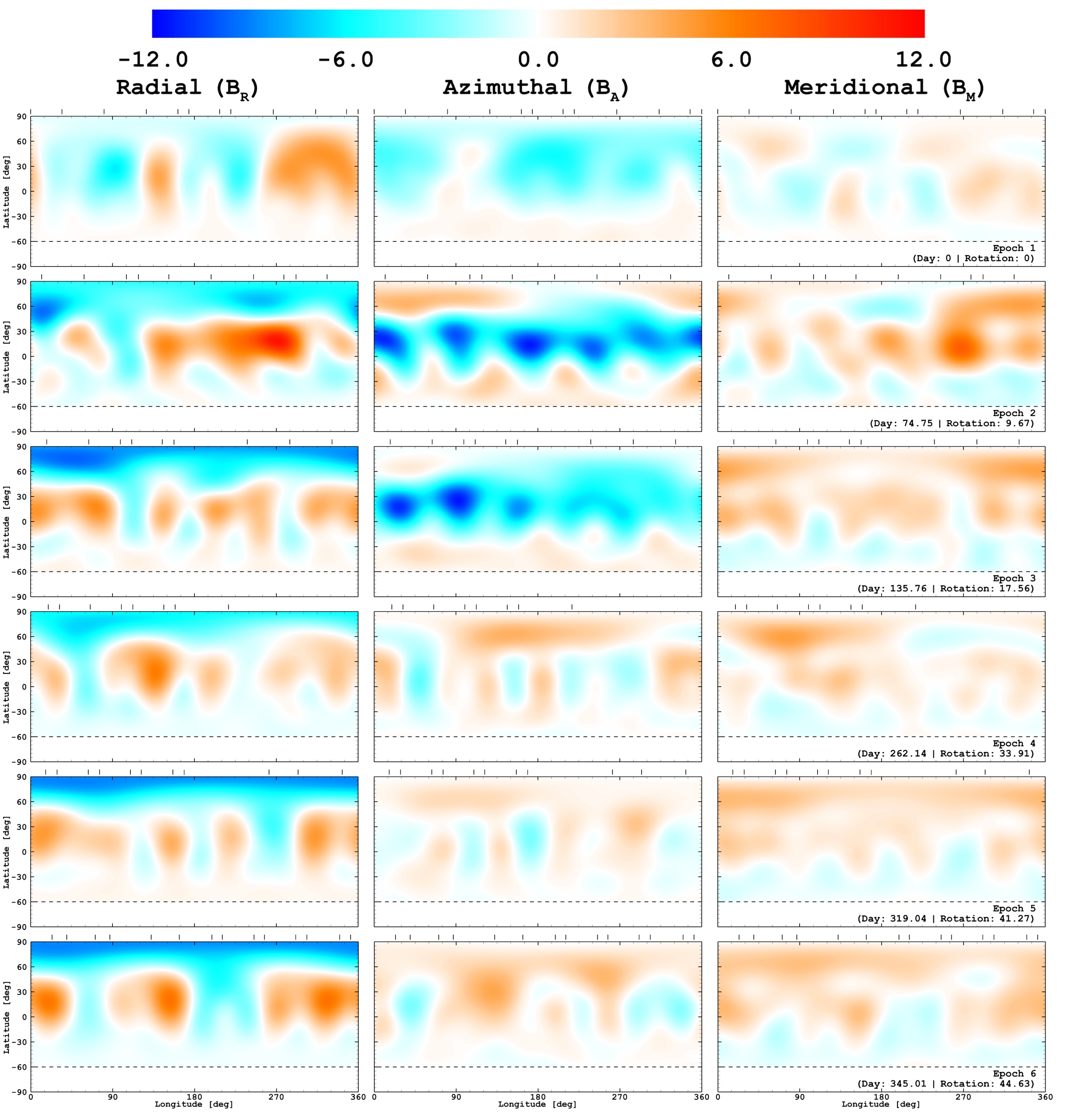}\vspace{-0.1cm}
\caption{Unconstrained ZDI reconstruction of the large-scale magnetic field maps of $\iota$~Hor in latitude-longitude equirectangular projection. Columns contain the radial ($B_{\rm R}$, left), azimuthal ($B_{\rm A}$, middle), and meridional ($B_{\rm M}$, right) field components, with the color scale indicating the magnitude (in Gauss) and field polarity in each case. Each row corresponds to a different observing epoch as indicated (Epochs 1 to 6), with the observed phases denoted by black tick marks in longitude (upper x-axes). The campaign day and number of stellar rotations (measured from ${\rm BJD} = 2457300.78580$ to the first night of observations of a given epoch) are listed in each map. The segmented line shows the visibility limit imposed by the inclination of the star ($i = 60$~deg).}\label{fig_1}
\end{figure*}

\begin{figure*}
\includegraphics[trim=0.0cm 1.0cm 1.0cm 0.0cm, clip=true, width=\linewidth]{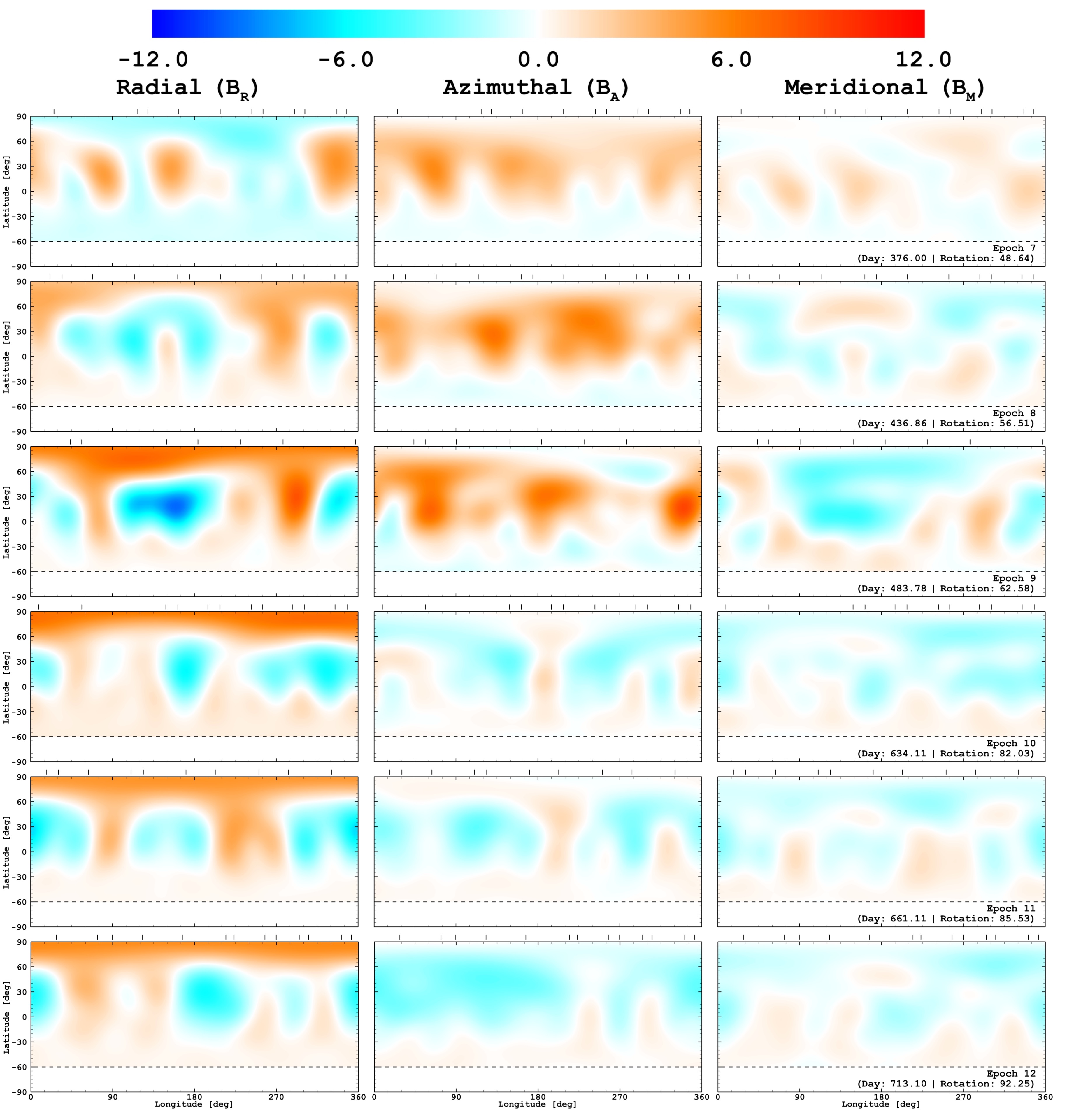}\vspace{-0.1cm}
\caption{Reconstructed large-scale magnetic field maps of $\iota$~Hor using ZDI (Epochs 7 to 12). See caption of Fig.~\ref{fig_1}.}\label{fig_2}
\end{figure*}

\begin{figure*}
\includegraphics[trim=0.0cm 1.0cm 1.0cm 0.0cm, clip=true, width=\linewidth]{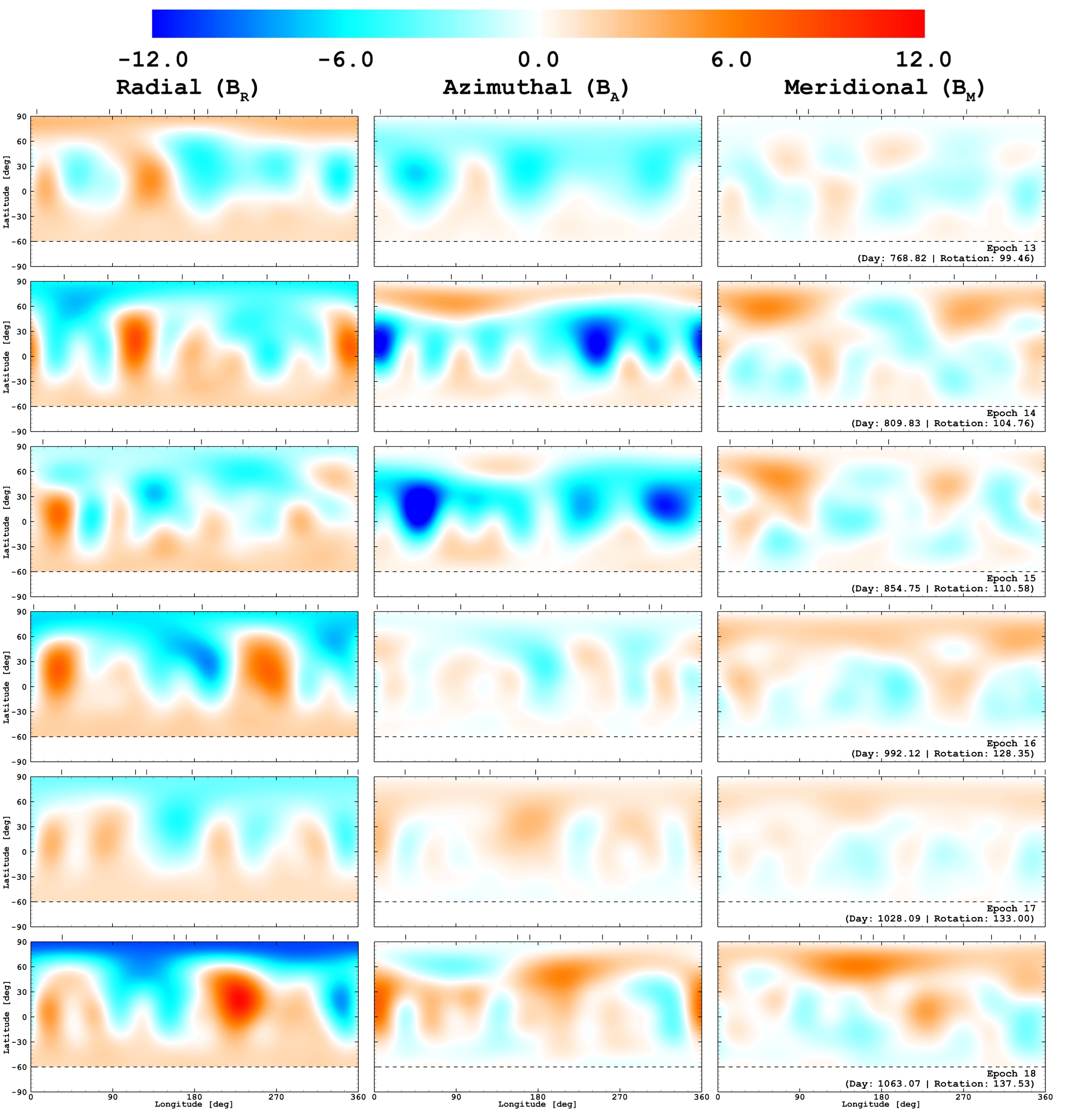}\vspace{-0.1cm}
\caption{Reconstructed large-scale magnetic field maps of $\iota$~Hor using ZDI (Epochs 13 to 18). See caption of Fig.~\ref{fig_1}.}\label{fig_3}
\end{figure*}

\begin{figure*}[!h]
\includegraphics[trim=0.5cm 0.8cm 1.0cm 1.2cm, clip=true, width=0.496\linewidth]{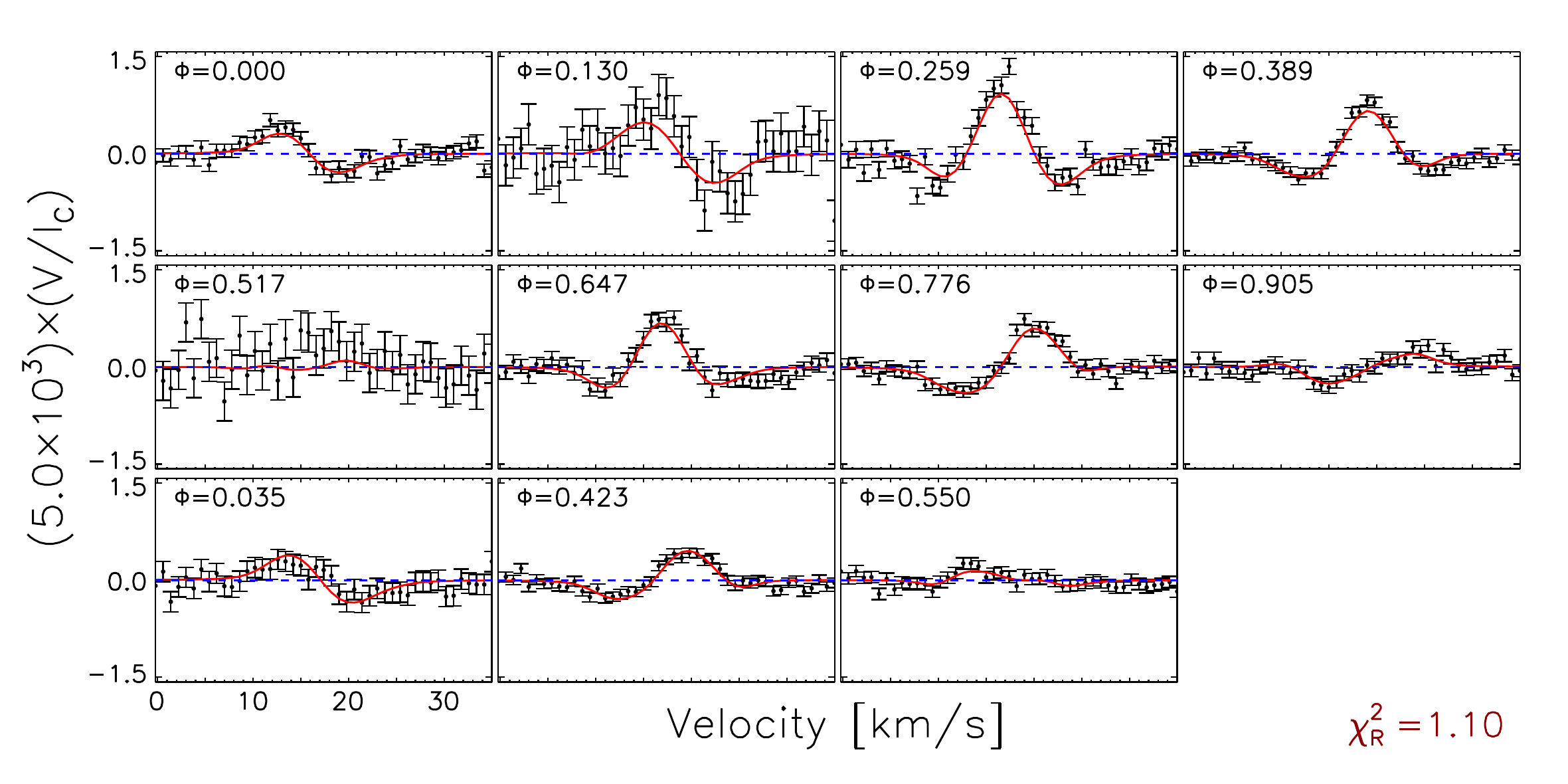}\hspace{2.5pt}\includegraphics[trim=0.5cm 0.8cm 1.0cm 1.2cm, clip=true, width=0.496\linewidth]{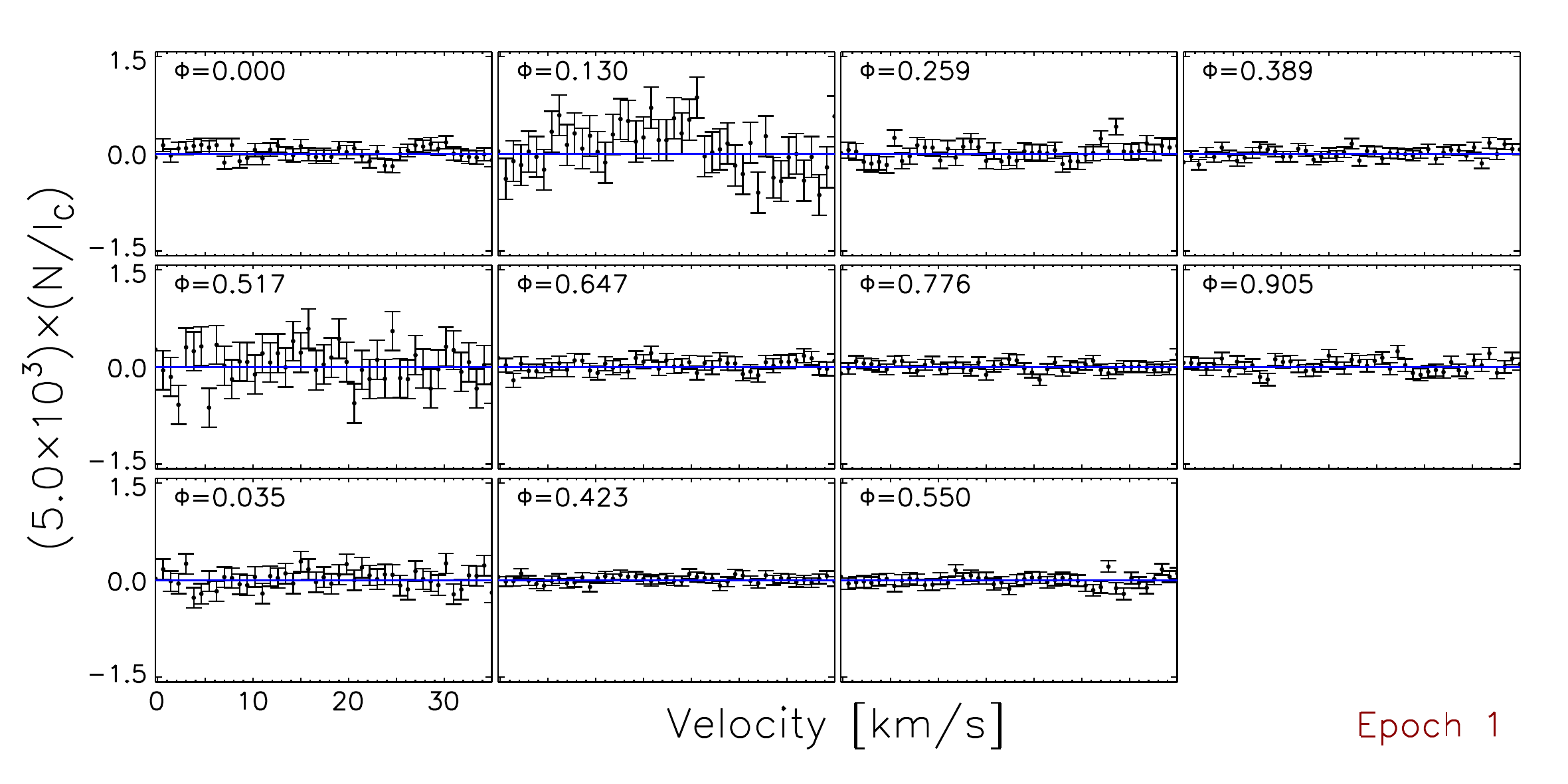}
\includegraphics[trim=0.5cm 0.8cm 1.0cm 0.2cm, clip=true, width=0.496\linewidth]{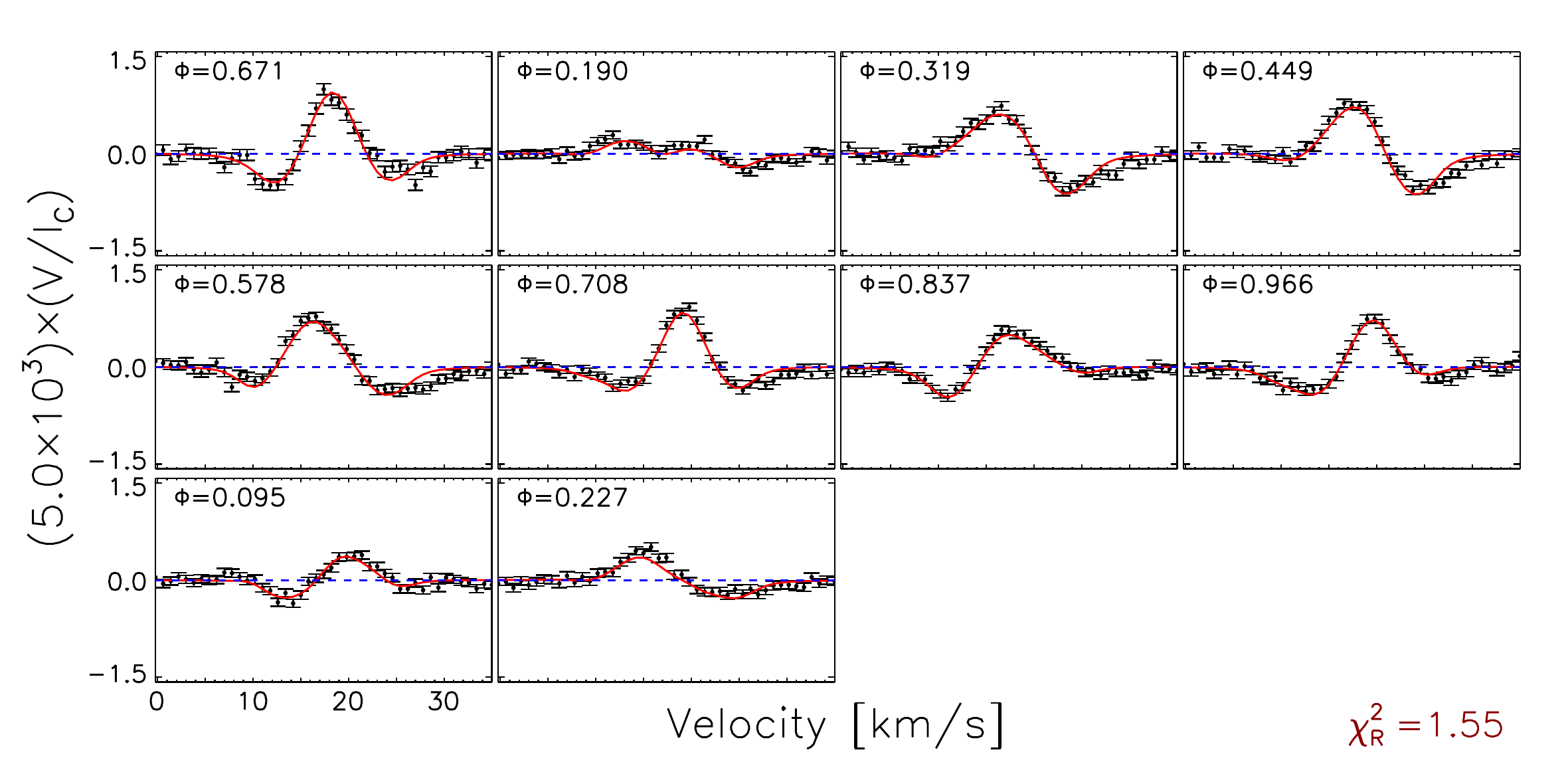}\hspace{2.5pt}\includegraphics[trim=0.5cm 0.8cm 1.0cm 0.2cm, clip=true, width=0.496\linewidth]{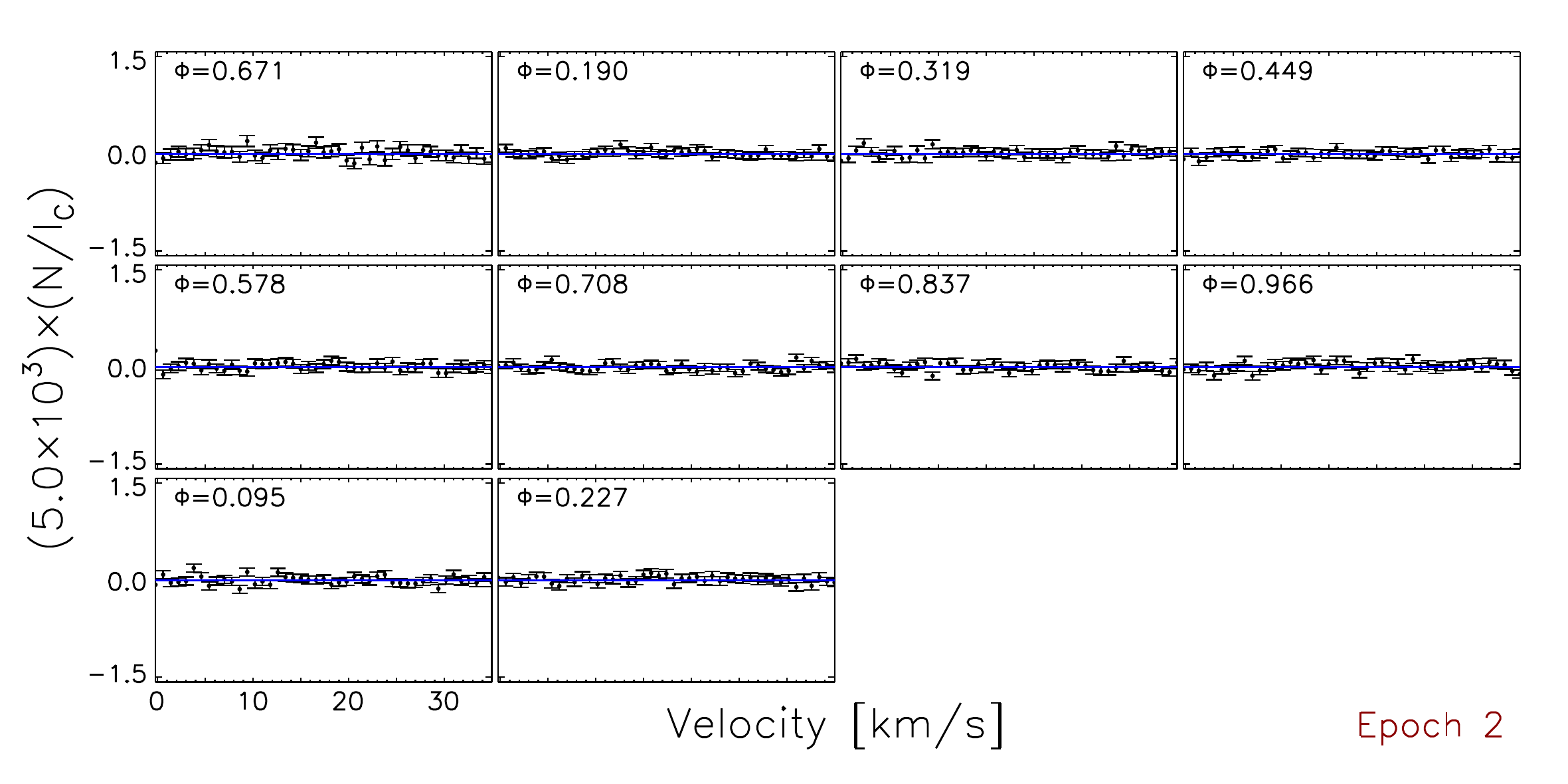}
\includegraphics[trim=0.5cm 0.8cm 1.0cm 0.2cm, clip=true, width=0.496\linewidth]{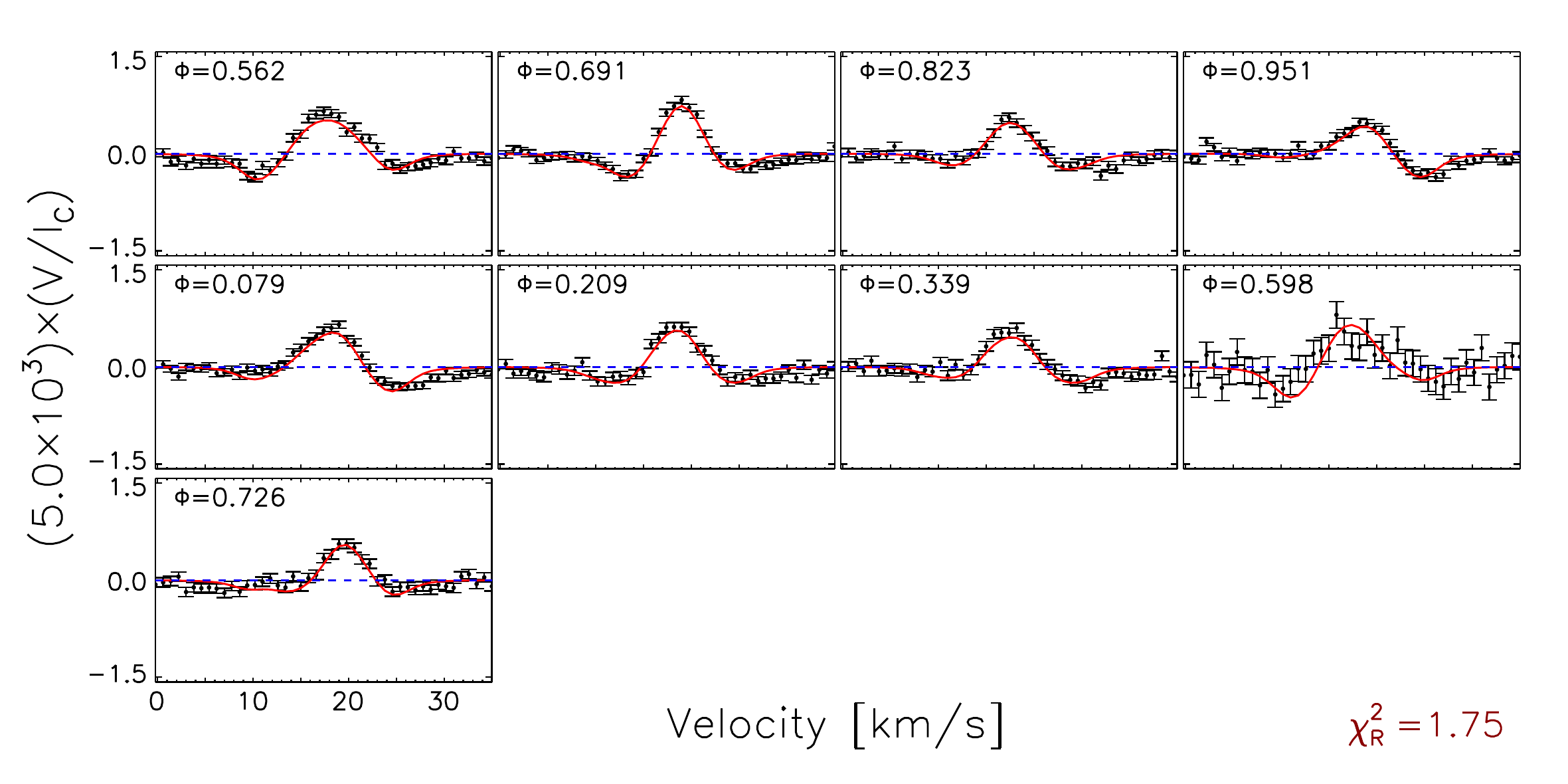}\hspace{2.5pt}\includegraphics[trim=0.5cm 0.8cm 1.0cm 0.2cm, clip=true, width=0.496\linewidth]{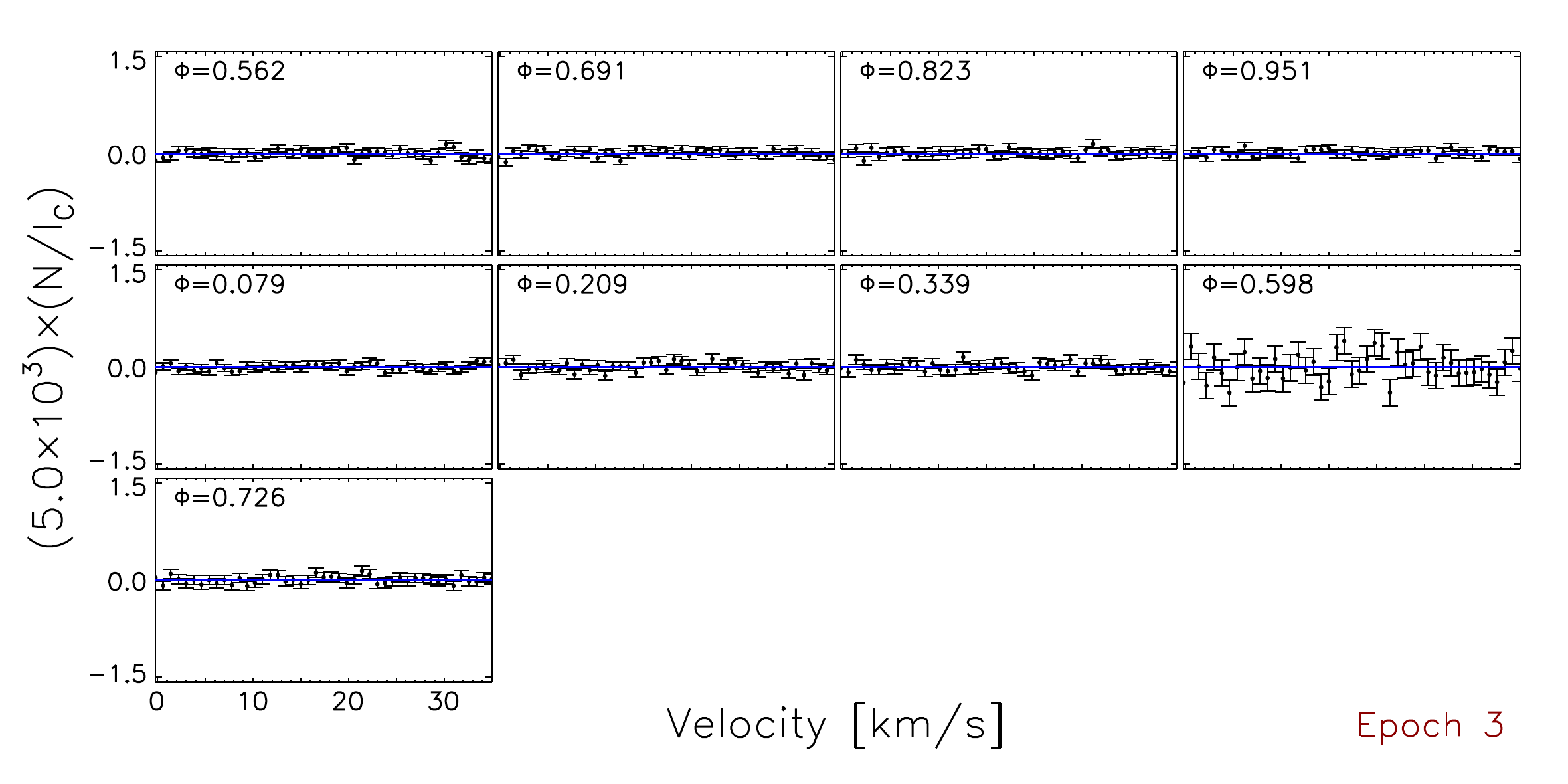}
\includegraphics[trim=0.5cm 6.1cm 1.0cm 0.2cm, clip=true, width=0.496\linewidth]{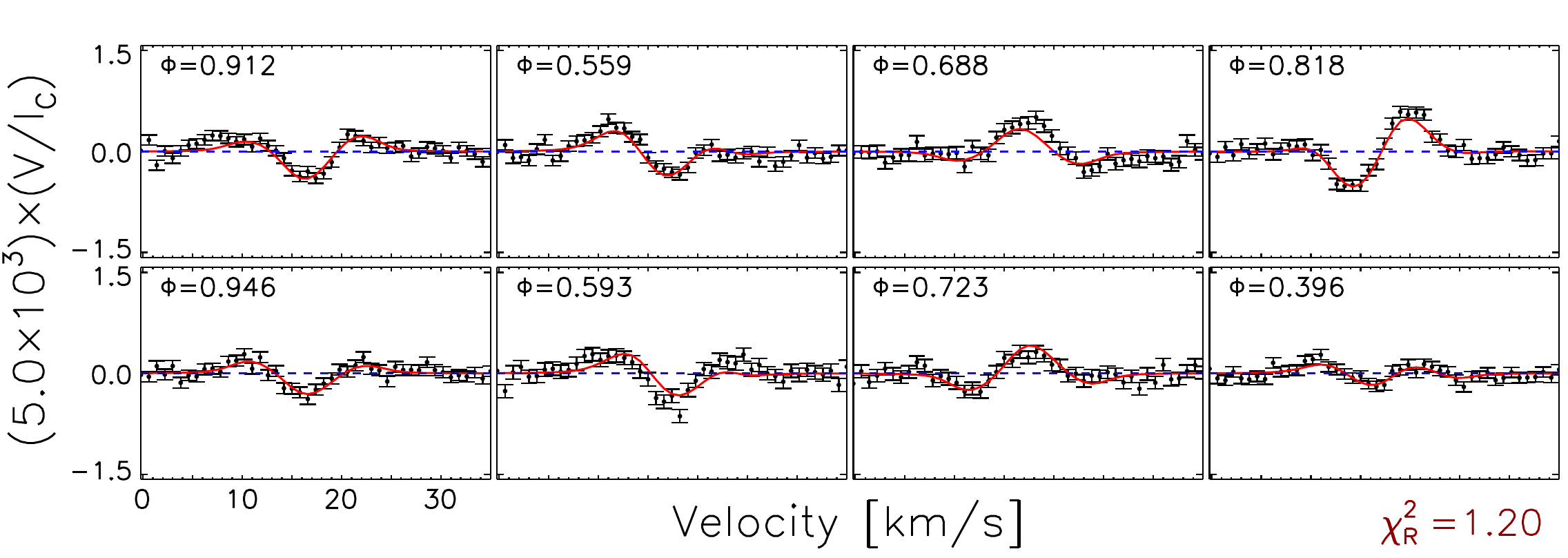}\hspace{2.5pt}\includegraphics[trim=0.5cm 6.1cm 1.0cm 0.2cm, clip=true, width=0.496\linewidth]{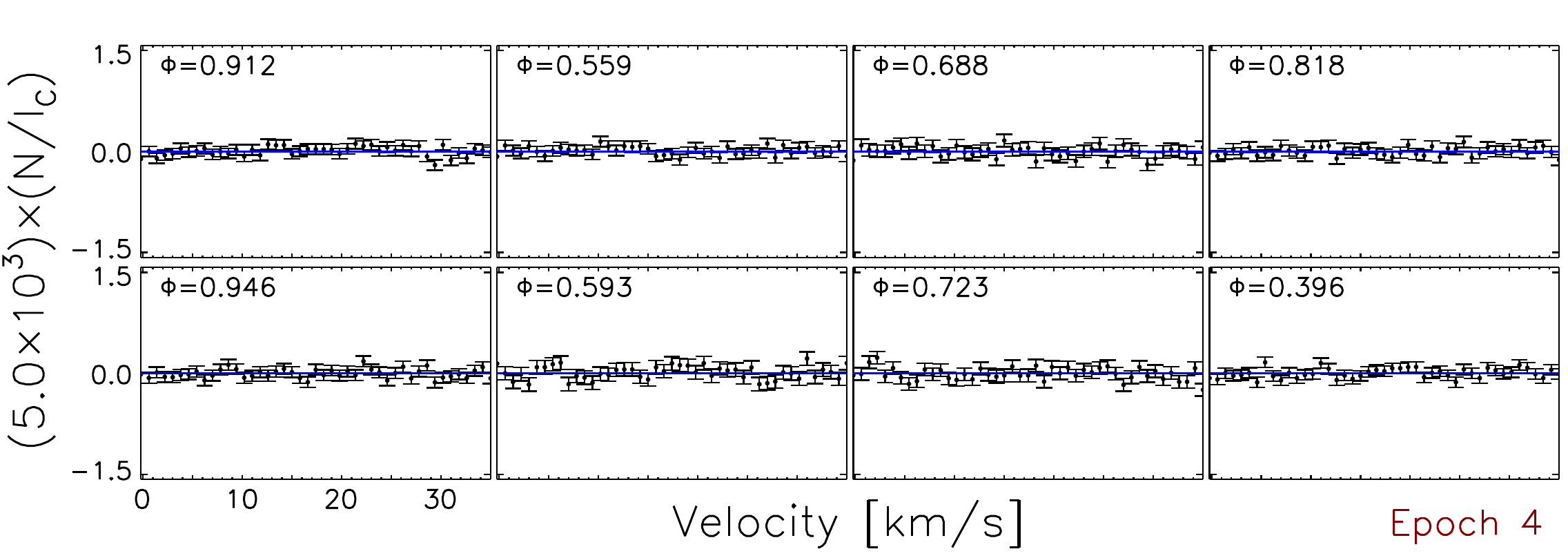}
\includegraphics[trim=0.5cm 0.8cm 1.0cm 0.2cm, clip=true, width=0.496\linewidth]{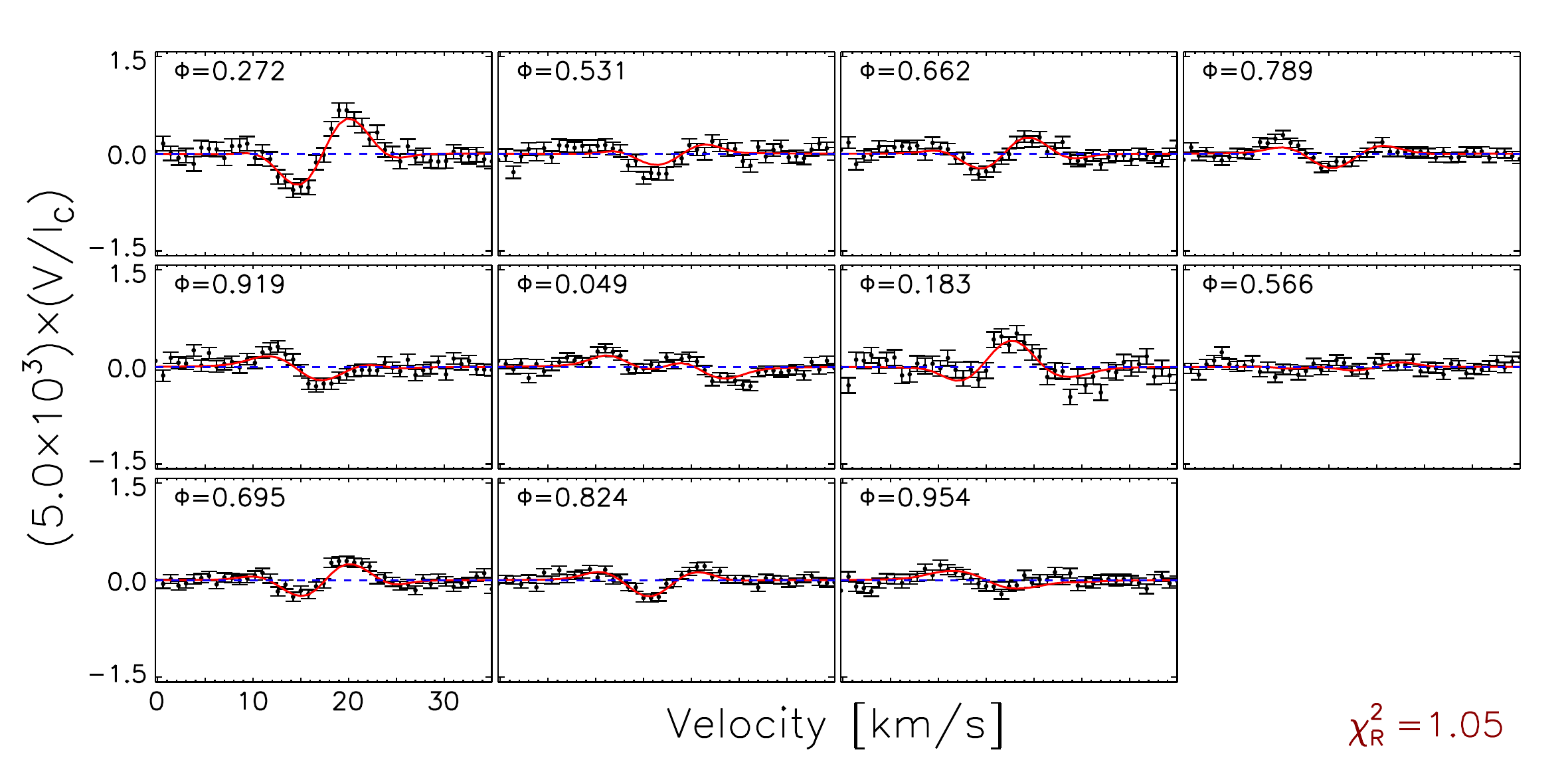}\hspace{2.5pt}\includegraphics[trim=0.5cm 0.8cm 1.0cm 0.2cm, clip=true, width=0.496\linewidth]{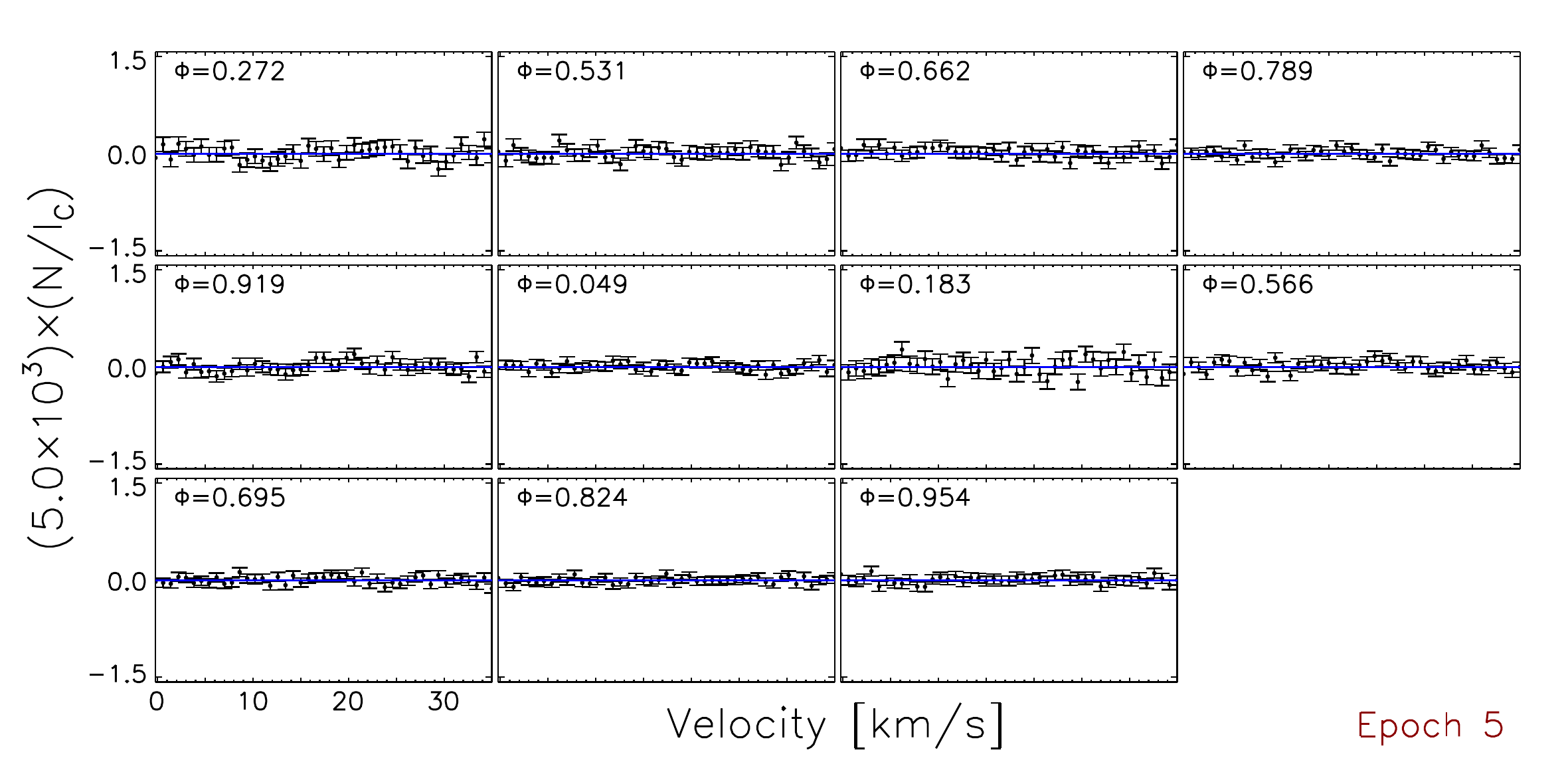}
\caption{Recovered LSD profiles from the spectropolarimetric observations of $\iota$~Hor. Each row contains a different observing epoch (Epochs 1 to 5), with individual observing phases ($\Phi$) shown in each sub-panel. The left column contains the ZDI fits (red) to the circularly polarized data (Stokes\,V, black), with the optimal reduced $\chi^2_{\rm R}$ achieved in each case. The corresponding diagnostic null (N) spectra are presented in the right column. Both profiles have been enhanced (by a factor of $5.0 \times 10^3$) and normalized to the continuum intensity ($I_{\rm C}$) for visualization purposes. The ZDI fits to the remaining epochs (6 to 18) are presented in Figs.~\ref{fig_5}-\ref{fig_7}.}\label{fig_4}
\end{figure*}

The full set of the large-scale ZDI magnetic field maps reconstructed consistently and using the same entropy-growth stopping criterion are shown in Figs.~\ref{fig_1} to \ref{fig_3}. These maps show the solutions for unconstrained $\ell$-modes (i.e. not pushing for an antisymmetric or symmetric large-scale field description). From a visual inspection of the maps, it is clear that the magnetic field evolves significantly over the course of the $\sim$\,3~year time-span (approximately $1077$~d which correspond to $\sim$\,$139$ stellar rotations, assuming $P_{\rm rot} = 7.73$~d). 

An epoch by epoch description of the evolution of the stellar large-scale magnetic field of \ihor~can be found in the Appendix~\ref{sec:ZDI-maps}. As illustrated in Fig.~\ref{fig_1}, the field is initially characterized by a dominant radial component ($B_{\rm R}$) with strengths ranging roughly from $-6.4$~G to $5.1$~G, accompanied by a significant negative azimuthal component ($B_{\rm A}$) and a weaker meridional field ($B_{\rm M}$). As the campaign progresses, notable changes occur: the field strength increases in some epochs (with $|B_{\rm R}|$ and $|B_{\rm A}|$ reaching up to approximately $12$~G and $16$~G, respectively), and the geometry of the field evolves. Several epochs show distinct polarity reversals in both $B_{\rm R}$, e.g., Epochs 8 (Fig.~\ref{fig_2})  and 14 (Fig.~\ref{fig_3}) and $B_{\rm A}$ (with multiple reversals noted throughout), suggesting the re-initiation of the magnetic cycle. The azimuthal field, in particular, transitions from a broad negative band to configurations featuring mix-polarity regions, with the mid-latitude bands either emerging or disappearing across the epochs.

A sample of the Stokes\,V time-series used to derive these maps are shown in Figs.~\ref{fig_4} (Epochs $1-5$), with the remaining ones presented in the Appendix \ref{sec:ZDI-fits} (Figs.~\ref{fig_5}-\ref{fig_7}). We also include the null (N) spectra which are employed to check for instrumental contributions to the polarization signature. As demonstrated in these spectral time-series, strong Zeeman signatures are detected at all observed epochs, with the amplitude and morphology of the Zeeman signatures varying over timescales on the order of one night. Throughout the campaign, the ZDI reconstructions consistently achieve good fits to the observed Stokes\,V LSD profiles, with $\chi^2_{\rm R}$ values typically near 1.0. Still, some epochs exhibit higher values which could be in part due to variations in the magnetic field within epochs, as well as observational phase gaps and worse S/N (which can affect the location of the optimal $\chi^2_{\rm R}$ value; see \citeads{2015A&A...582A..38A}). Nevertheless, the Stokes\,V time-series alone indicate changing, complex multipolar surface fields, which are further confirmed in the resulting ZDI magnetic field maps. 

As the ZDI reconstructions in Figs.~\ref{fig_1}-\ref{fig_3} are performed in an unconstrained manner, this leads to no magnetic field being recovered below $-60^\circ$ latitude, due to the star's inclination angle. In order to complete the ``missing'' field in the obscured hemisphere (assumed here to correspond to southern latitudes), we can push for antisymmetric and symmetric solutions consistent with the data by weighting against odd and even modes, respectively. We show the corresponding symmetric/antisymmetric maps in Appendix \ref{app:ZDI-maps}. We find that \ihor\ tends to host an antisymmetric large-scale field, as the maps derived from the unconstrained solutions have more power in the even spherical harmonic modes and tend to resemble the antisymmetric maps. 

Finally, note that the meridional field maps are more subject to cross-talk ---this is a well-known limitation that arises from the relatively weak contribution from this component to the circularly polarized Stokes\,V profiles in high inclination stars (see \citeads{1997A&A...326.1135D}, \citeads{2002A&A...388..868K}). Reconstructions based on time-series of all Stokes\,V, Q, and U profiles would help to address this limitation but Stokes Q and U signatures are typically an order of magnitude weaker in cool stars compared to Stokes\,V (e.g.~\citeads{2015ApJ...805..169R}, \citeads{2015MNRAS.449....8W}), hence the exposure times required would be too long to derive maps with high cadence employing all four Stokes profiles. As such, in the following sections, we will mainly focus on the properties of the radial and azimuthal components of the large-scale~field.

\subsection{The large-scale magnetic field of \ihor: Global properties}

\begin{figure}[!t]
\includegraphics[trim=0.6cm 0.4cm 1.7cm 1.0cm, clip=true, width=\linewidth]{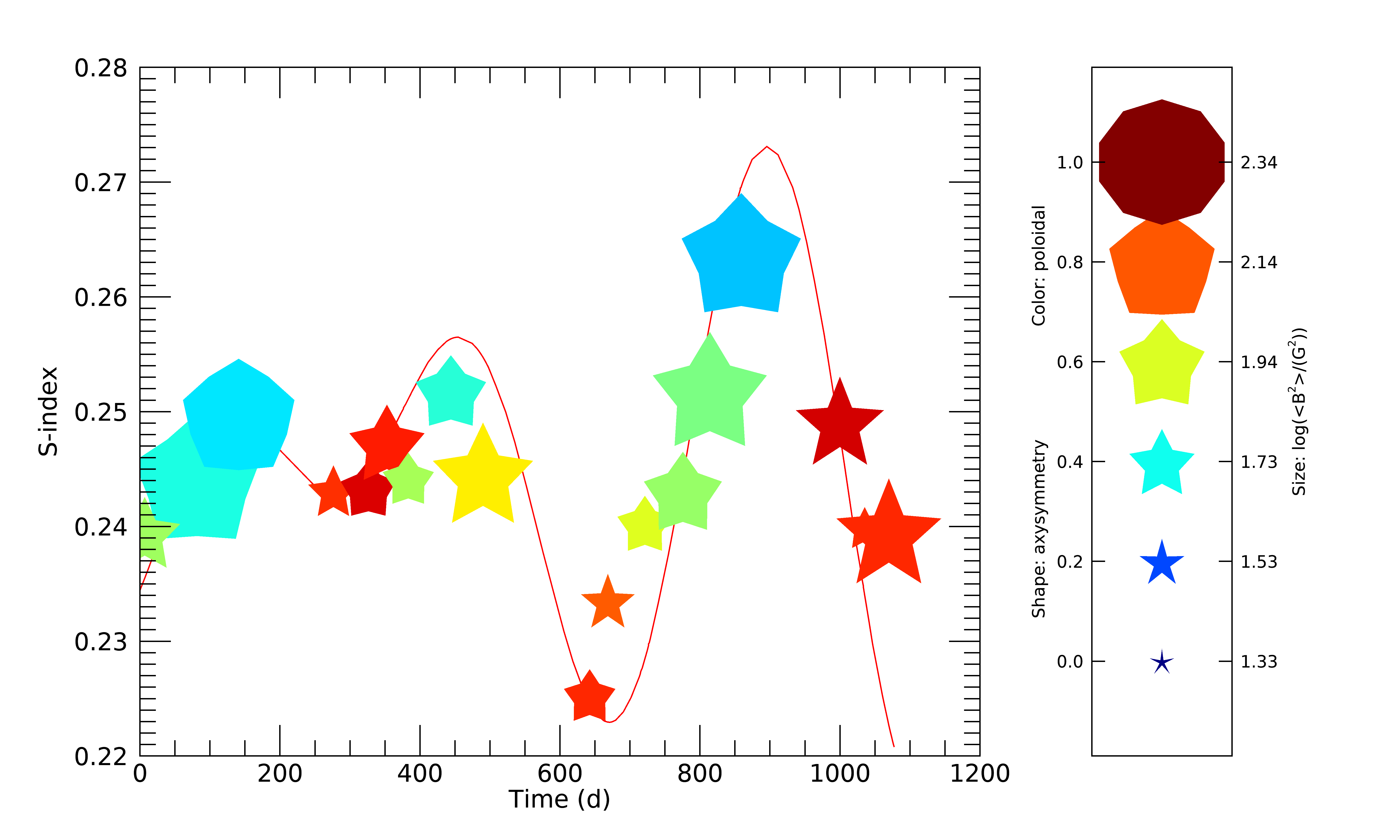}
\includegraphics[trim=0.6cm 0.4cm 1.7cm 1.0cm, clip=true,width=\linewidth]{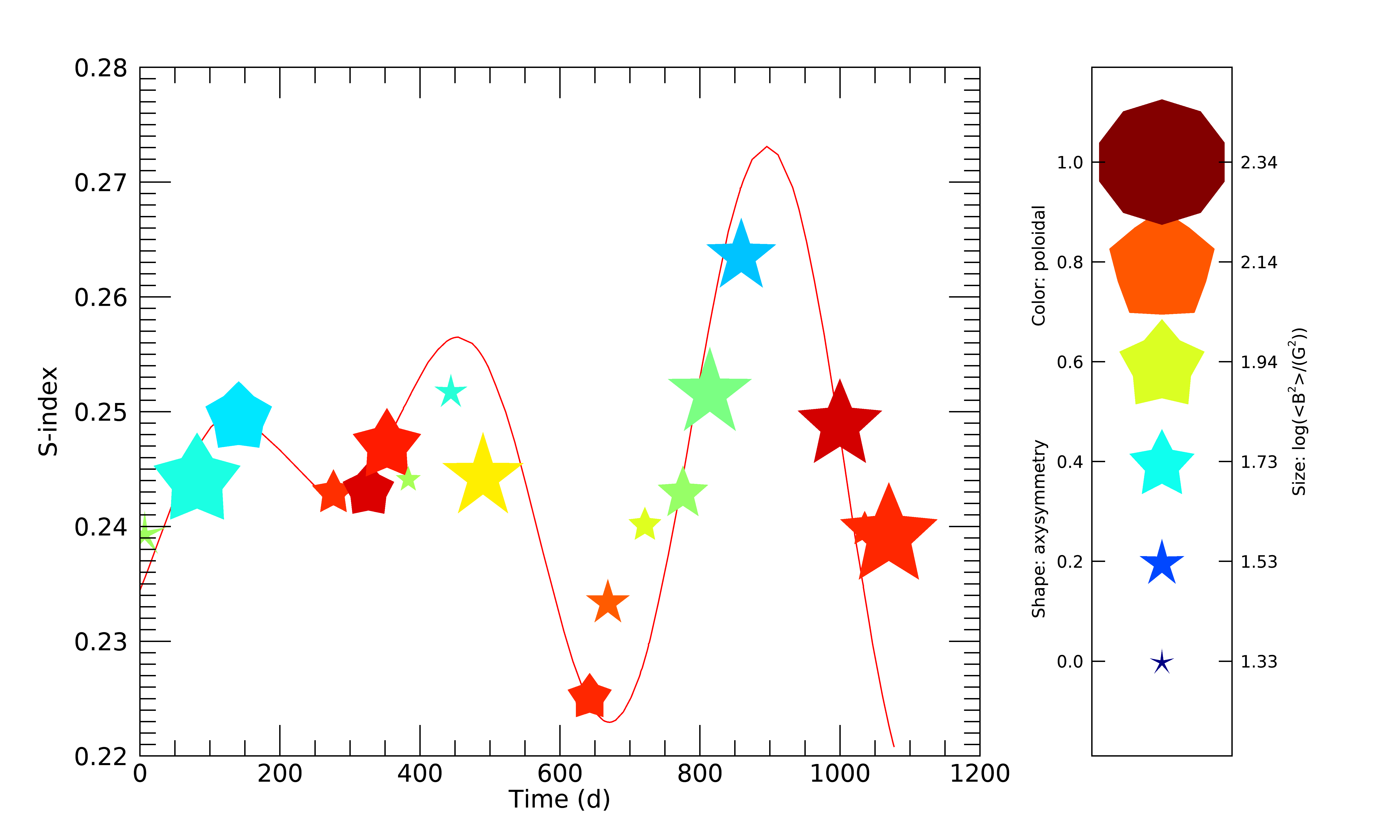}
\includegraphics[trim=0.6cm 0.4cm 1.7cm 1.0cm, clip=true,width=\linewidth]{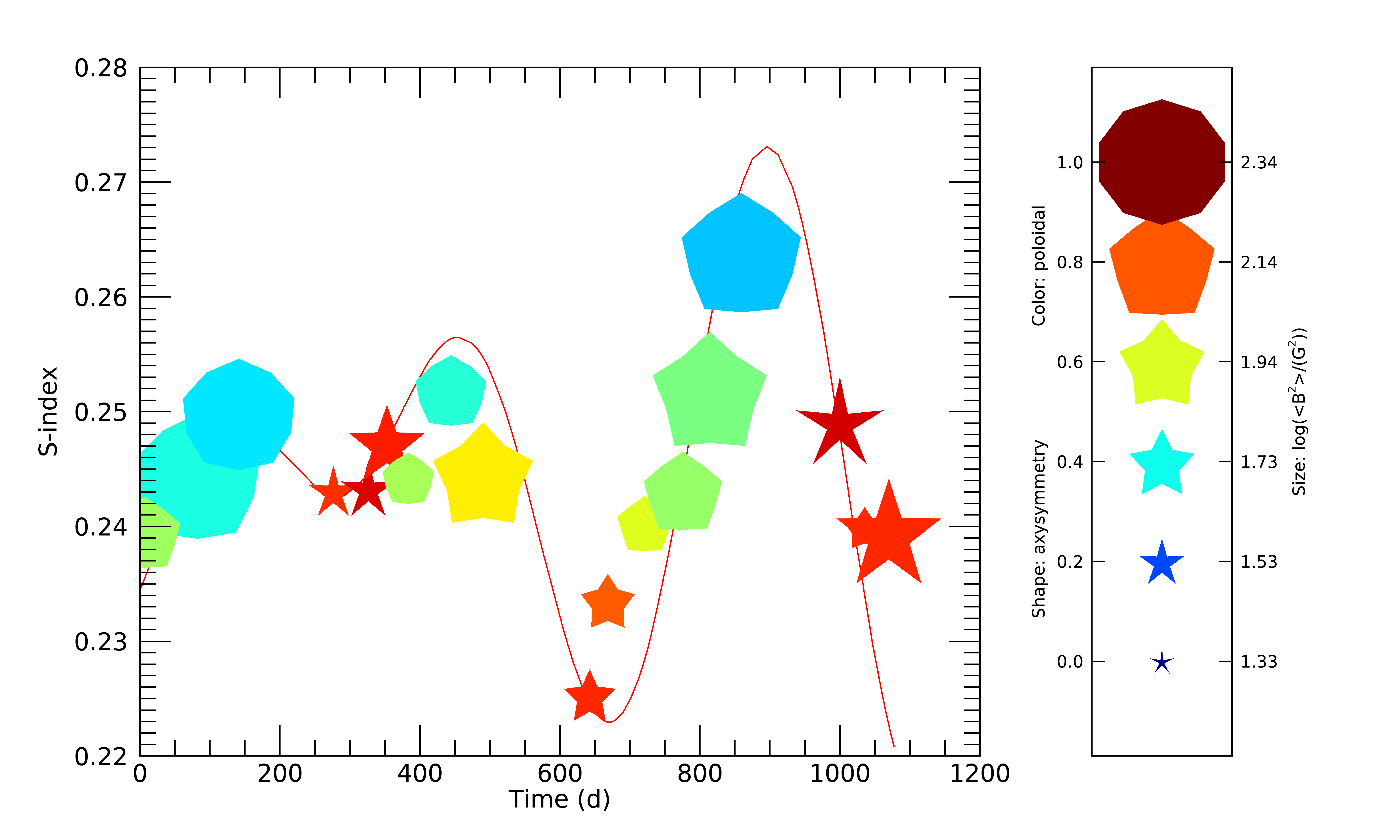}
\caption{The evolving global magnetic field properties for \ihor\ plotted as a function of the chromospheric S-index and time. The red solid line shows the best 2-period model fit to Ca\,II\,H\&K observations from \citetads{2023MNRAS.524.5725A}. Day "0" corresponds to ${\rm BJD} = 2457300.78580$. The symbol sizes scale with $\log B^2$, and symbols for poloidal- and toroidal-dominated fields range from red to blue. The degree of non-axisymmetry is represented by symbol shapes, ranging from circular for fully axisymmetric field and star-shaped symbols for non-axisymmetry. {\em Top:} the axisymmetry and $\log B^2_{\rm tot}$ for the total field and the {\em middle} and {\em bottom} panels show the axisymmetry and $\log B^2$ for the poloidal field and toroidal field components, respectively.}
\label{fig_confuso}
\end{figure}

\begin{figure}
\centering
\includegraphics[width=\linewidth]{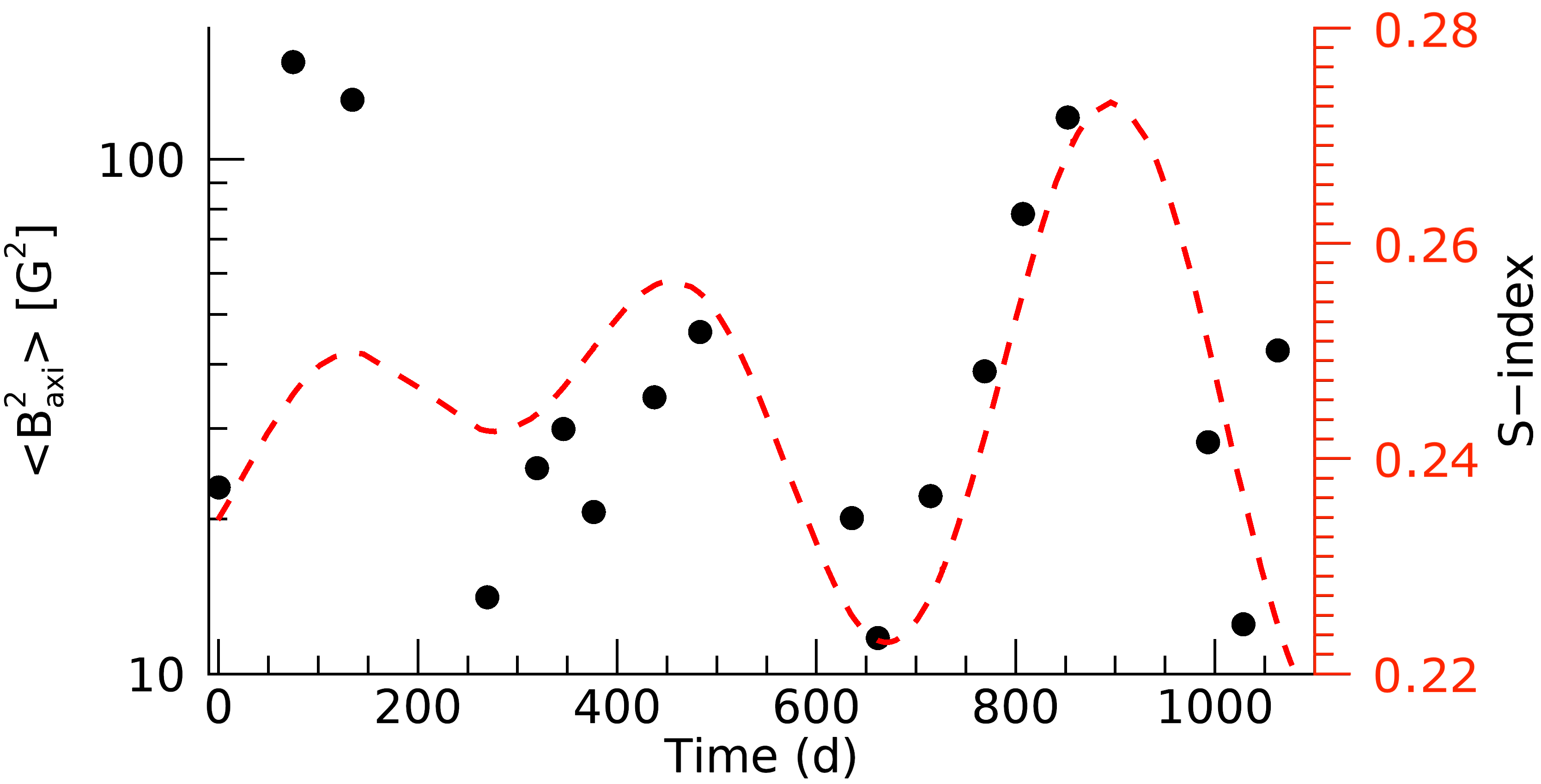}
\includegraphics[width=\linewidth]{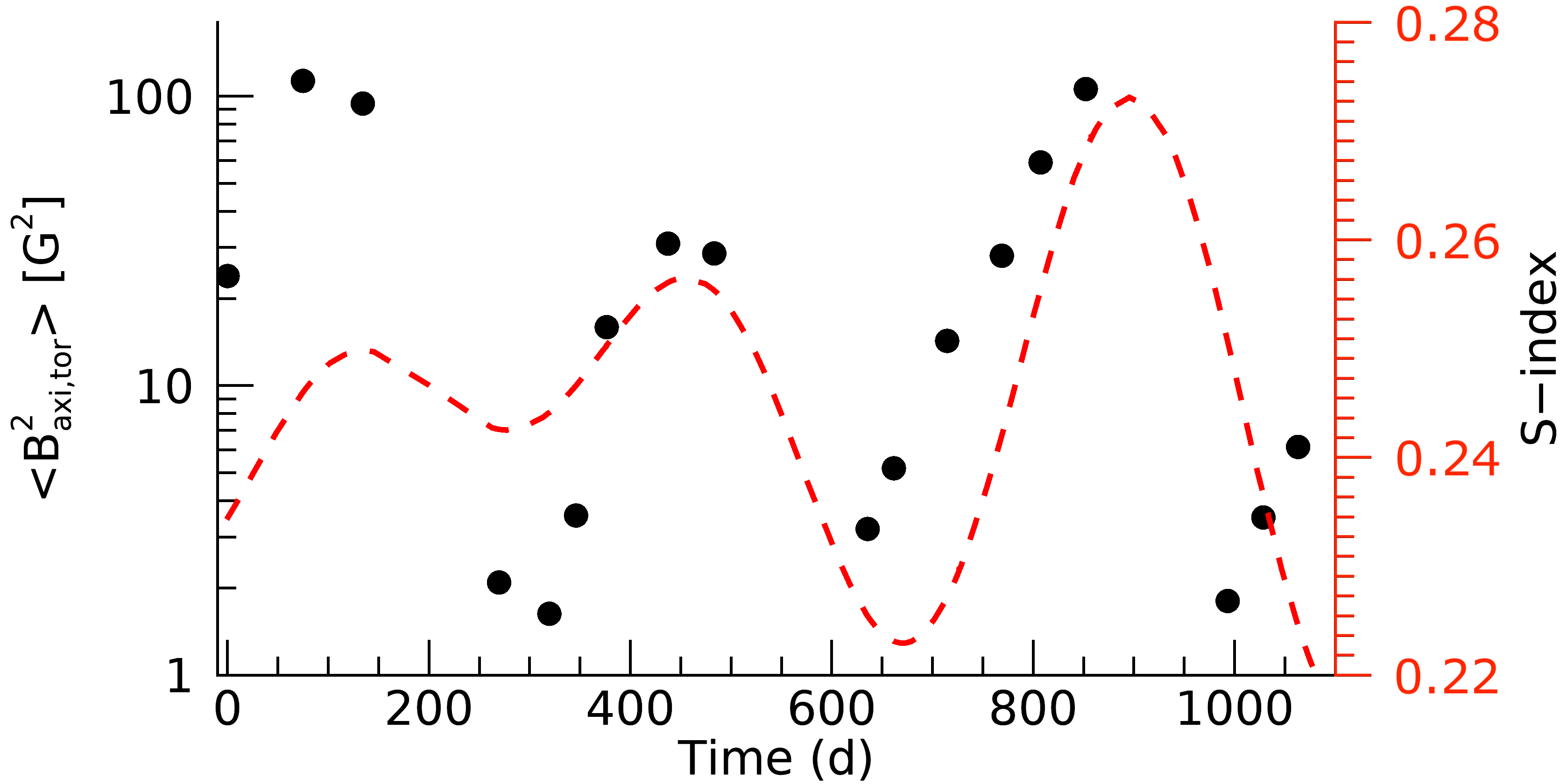}
\caption{The evolution of the axisymmetric total energy $\langle B^2_{\rm axi} \rangle$ and axisymmetric toroidal energy $ \langle B^2_{\rm tor,axi}\rangle$ most closely follow the evolution in the chromospheric activity (traced by the Ca\,II\,H\&K S-index; Red dashed line). See caption of Fig.~\ref{fig_confuso}.}
\label{fig_c2comp_tor}
\end{figure}

\noindent Figure~\ref{fig_confuso} traces how the ratio of poloidal to toroidal field changes over the different epochs covered in this study. There appears to be a dependence on chromospheric activity in the recovered total magnetic field. A comparison of the Figure~\ref{fig_confuso} middle and lower panels demonstrates that the large-scale toroidal field  changes more dramatically than the poloidal field in terms of both strength and axisymmetry. The poloidal field component stays predominantly non-axisymmetric, with only three epochs corresponding to  axisymmetry of 50\% or greater (see Table \ref{tab:ZDI-numbers}). On the other hand, the toroidal field component undergoes much larger changes in both axisymmetry, reaching over 80\% axisymmetry in half the epochs;  the epochs at which it is stronger and more axisymmetric appear closer in time to chromospheric activity maxima, while the minima are generally associated with weaker and non-axisymmetric toroidal fields.

Our dense time-sampling allows a detailed analysis of how different properties of the large-scale field evolve with activity cycle. This also allows a comparison with a similar analysis of the solar cycle conducted by \citetads{2021MNRAS.500.1243L} who identified which features of the large-scale field morphology could be recovered using ZDI techniques. They applied ZDI to magnetograms based on Solar Cycle\,23 and found that, while the size of the large-scale field tended to be underestimated in ZDI maps, changes in both axisymmetry and toroidal field could be used to detect and follow the solar cycle. They also observed that the strongest tracer of the solar activity cycle was the axisymmetric field; with the fraction of axisymmetry decreasing essentially linearly with increasing S-index (see \citeads{2021MNRAS.500.1243L}, Fig.\,8). \citetads{2024A&A...682A.156H} used a similar approach, applying ZDI to simulated  data based on dynamo models tracing different activity levels; they similarly find  that while weaker magnetic field strengths are poorly recovered, there is a sensitivity to the relative amounts of toroidal and poloidal field. On \ihor, we find no clear trend in the fraction of axisymmetry with S-index; though the total field reconstructed tends to be more axisymmetric during peaks of the chromospheric activity cycle, albeit with quite some scatter.

On the Sun, another tracer of the activity cycle is the fraction of the toroidal field, with $f_{\rm tor}$ in dipolar, quadrupolar and octupolar components all following the activity level of the Sun. On \ihor, the octupolar field $f_{\rm tor}$ component shows no trend with the activity cycle, but the  $f_{\rm tor}$ in the dipolar and quadrupolar components weakly trace the activity levels. An analysis of the time evolution of the real component of the large-scale field ($\alpha, \beta, \gamma$) also reveals a possible trend with $\gamma_{\ell=1,2}$, which traces the dipolar and quadrupolar components of the toroidal field. 

The magnetic energy $\langle B^2_{\rm total} \rangle$ shows an increase in its toroidal component which traces the peaks of the chromospheric cyclic activity, seen in Fig.~\ref{fig_confuso}. However, albeit with significant scatter, the strongest tracers of the chromospheric activity of \ihor~are the axisymmetric total energy $\langle B^2_{\rm axi} \rangle$, and the energy of its toroidal and toroidal axisymmetric component ($\langle B^2_{\rm tor,axi} \rangle$, see Fig.~\ref{fig_c2comp_tor}). We find Spearman correlation coefficients of $0.59$ for the axisymmetric magnetic energy vs the S-index evolution (given by the 2-period fit) and $0.54$ for the axisymmetric toroidal field energy vs the S-index fit. 
Their standard t-test significance p-values are $0.01$ (S-index vs $\langle B^2_{\rm axi}\rangle$) and $0.02$ (S-index vs $\langle B^2_{\rm axi,tor}\rangle$), which indicate a weak/moderate correlation between the involved quantities. We assess the significance of these correlations by calculating p-values after a two-sided permutation test which leads to similar results. This behaviour is drastically different from the Sun (see Fig. C2 in  \citeads{2021MNRAS.500.1243L}), where the non-axisymmetric energy $\langle B^2_{\rm naxi} \rangle$ is the stronger tracer of the underlying magnetic cycle. 

\begin{table*}
\centering
\small
\caption{Properties of the large-scale magnetic field of \ihor~obtained from the ZDI reconstructions.}
\label{tab:ZDI-numbers}
\begin{tabular}{c @{\hspace{0.2cm}\vline\hspace{0.2cm}} c c c @{\hspace{0.2cm}\vline\hspace{0.2cm}} c c @{\hspace{0.2cm}\vline\hspace{0.2cm}} c c c @{\hspace{0.2cm}\vline\hspace{0.2cm}} c c c @{\hspace{0.2cm}\vline\hspace{0.2cm}} c c c}
\toprule
\textbf{Epoch} & \textbf{$B^{2}$} & \textbf{$B^{2}_{\rm pol}$} & \textbf{$B^{2}_{\rm tor}$} & \textbf{$f_{\rm pol}$} & \textbf{$f_{\rm tor}$} & \textbf{$f_{\rm dip}$} & \textbf{$f_{\rm quad}$} & \textbf{$f_{\rm oct}$} & \textbf{$f_{\rm axisym}$} & \textbf{$f_{\rm axisym,pol}$} & \textbf{$f_{\rm axisym,tor}$} & \textbf{$f_{\rm dip,tor}$} & \textbf{$f_{\rm quad,tor}$} & \textbf{$f_{\rm oct,tor}$}\\

 & \textbf{[$\rm G^{2}$]} & \textbf{[$\rm G^{2}$]} & \textbf{[$\rm G^{2}$]} & [\%]  & [\%] & [\%] & [\%] & [\%] & [\%] & [\%] & [\%] & [\%] & [\%] & [\%] \\
\midrule
1  & 61.7  & 33.2  & 28.4   & 53.9 & 46.1 & 30.8 & 41.8 & 12.2 & 37.4 & 0.4  & 80.6 & 45.1 & 41.3 & 4.6 \\
2  & 217.2 & 89.8  & 127.4  & 41.3 & 58.7 & 37.3 & 25.8 & 12.3 & 71.2 & 47.5 & 87.8 & 49.0 & 22.7 & 7.7 \\
3  & 157.0 & 56.0  & 101.0  & 35.7 & 64.3 & 41.0 & 33.0 & 9.1  & 83.1 & 66.6 & 92.3 & 55.5 & 31.3 & 2.3 \\
4  & 38.8  & 32.6  & 6.2    & 84.0 & 16.0 & 5.1  & 27.4 & 30.0 & 36.3 & 39.9 & 17.5 & 4.9  & 7.4  & 21.3 \\
5  & 43.2  & 39.6  & 3.6    & 91.7 & 8.3  & 15.2 & 32.3 & 24.1 & 58.1 & 61.8 & 17.5 & 10.9 & 12.0 & 22.9 \\
6  & 68.3  & 58.5  & 9.7    & 85.7 & 14.3 & 11.0 & 39.8 & 28.8 & 43.8 & 46.8 & 26.3 & 9.1  & 41.2 & 19.5 \\
7  & 39.3  & 21.5  & 17.8   & 54.8 & 45.2 & 32.5 & 30.3 & 14.6 & 52.6 & 26.6 & 84.0 & 47.4 & 38.2 & 3.4 \\
8  & 61.1  & 26.0  & 35.1   & 42.6 & 57.4 & 37.2 & 29.5 & 14.0 & 56.4 & 17.5 & 85.2 & 50.0 & 37.7 & 4.4 \\
9  & 121.5 & 79.1  & 42.4   & 65.1 & 34.9 & 21.8 & 36.0 & 16.3 & 38.0 & 23.5 & 65.1 & 40.7 & 36.5 & 1.2 \\
10 & 39.6  & 33.5  & 6.1    & 84.6 & 15.4 & 12.6 & 33.6 & 26.2 & 50.8 & 53.4 & 36.1 & 26.8 & 28.0 & 8.2 \\
11 & 41.7  & 33.0  & 8.6    & 79.3 & 20.7 & 18.8 & 38.5 & 20.4 & 28.2 & 22.9 & 48.6 & 29.2 & 32.0 & 7.6 \\
12 & 42.9  & 25.9  & 17.0   & 60.4 & 39.6 & 24.8 & 39.2 & 16.0 & 51.7 & 34.3 & 78.2 & 44.1 & 42.7 & 5.1 \\
13 & 72.8  & 38.9  & 33.9   & 53.5 & 46.5 & 33.3 & 25.0 & 21.1 & 53.2 & 29.9 & 80.0 & 45.9 & 35.6 & 1.8 \\
14 & 167.0 & 84.4  & 82.6   & 50.5 & 49.5 & 30.8 & 23.5 & 17.7 & 46.9 & 24.1 & 70.2 & 37.9 & 26.6 & 13.3 \\
15 & 187.6 & 60.7  & 127.0  & 32.3 & 67.7 & 44.7 & 26.2 & 5.7  & 64.3 & 26.0 & 82.6 & 49.8 & 33.3 & 5.7 \\
16 & 91.8  & 84.8  & 7.0    & 92.4 & 7.6  & 27.1 & 27.1 & 25.5 & 30.7 & 32.3 & 11.5 & 9.2  & 25.8 & 24.3 \\
17 & 31.0  & 26.1  & 4.9    & 84.1 & 15.9 & 34.9 & 20.2 & 13.1 & 40.3 & 38.3 & 50.7 & 25.5 & 27.8 & 11.6 \\
18 & 135.4 & 114.3 & 21.1   & 84.4 & 15.6 & 22.4 & 34.0 & 15.1 & 31.4 & 32.7 & 24.4 & 17.3 & 24.2 & 11.1 \\
\bottomrule
\end{tabular}
\end{table*}

\subsection{Stellar butterfly diagrams}\label{sec:B-fly}

\begin{figure}[!h]
\centering
\includegraphics[trim=0.1cm 0.0cm 0.1cm 0.0cm, clip=true,width=0.495\textwidth]{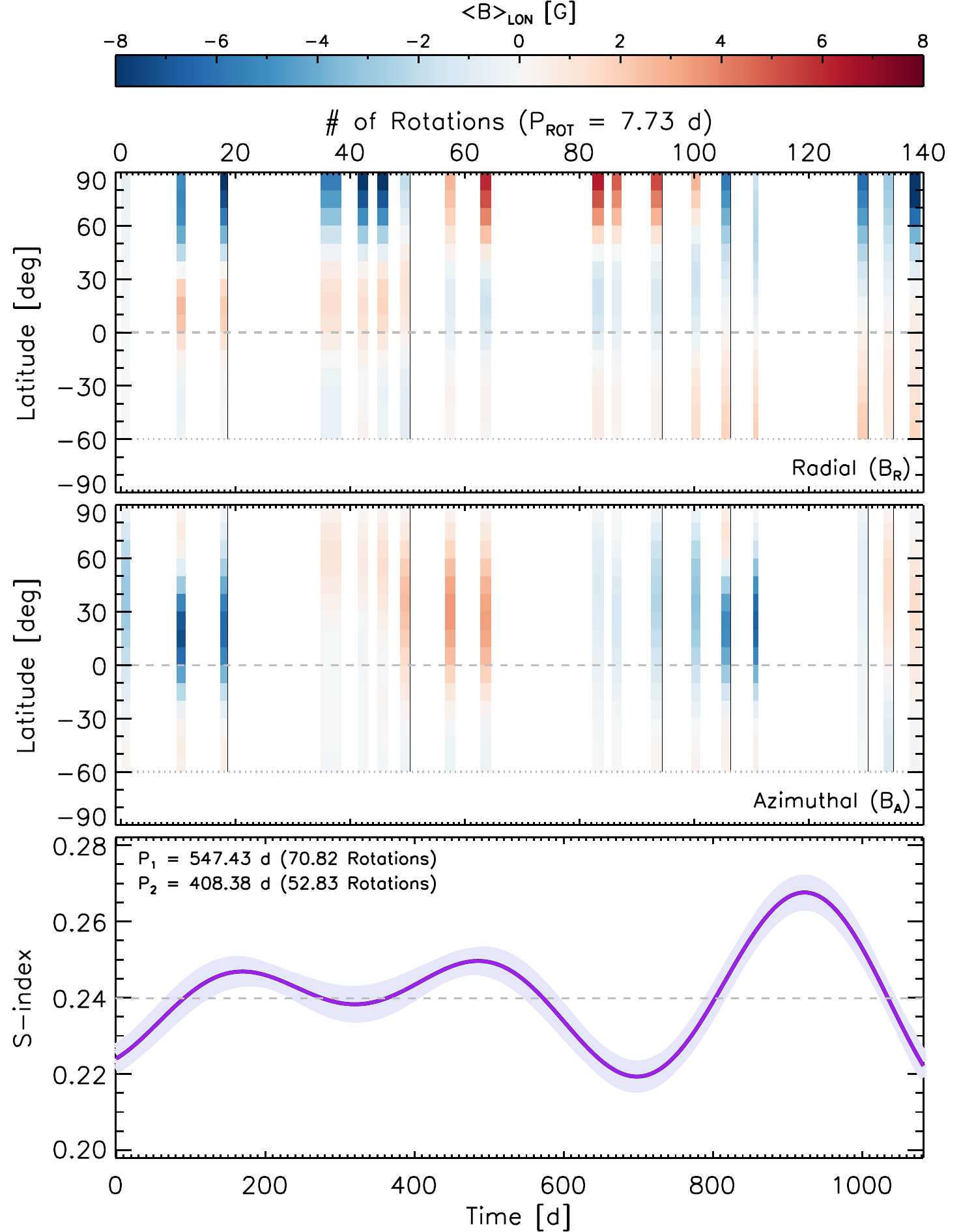}
\caption{``Butterfly diagrams'' of the radial (top) and azimuthal (middle) large-scale magnetic field of \ihor. These synoptic maps display the longitudinally-averaged magnetic field strength (color bar, in Gauss), extracted over $10^{\circ}$ latitudinal strips on the individual ZDI epochs (Figs.~\ref{fig_1}~to~\ref{fig_3}). The bottom panel shows the best 2-period fit to the Ca~II H\&K S-index variations obtained by \citetads{2023MNRAS.524.5725A}. The shaded region encapsulates the 1-$\sigma$ period and amplitude uncertainties. Day "0" corresponds to ${\rm BJD} = 2457300.78580$ (2015-10-05 06:51:33.12 UTC) and a secondary x-axis indicates the number of stellar rotations covered by the campaign (assuming $P_{\rm rot} = 7.73$~d).}
\label{fig_butterfly}
\end{figure}

Based on the 18 different ZDI reconstructions of \ihor~(Figs.~\ref{fig_1}~to~\ref{fig_3}) we can retrieve a counterpart of the well-known solar magnetic butterfly diagrams (e.g. \citeads{2015LRSP...12....4H}, \citeads{2018A&A...609A..56C}). On each ZDI reconstruction, we consider $10^{\circ}$ latitudinal strips\footnote[3]{This value was chosen based on the approximate maximum latitudinal resolution expected for the ZDI reconstructions. With a rotational velocity $v\sin(i) \simeq 6.5$~km~s$^{-1}$ and a HARPSpol velocity spacing $\Delta v = 0.8$~km~s$^{-1}$ ($R \simeq 110000$), the extracted LSD profiles of \ihor~encompass $v\sin(i)/\Delta v \simeq 8.125$ resolution elements from the line center to the wing. Therefore, the $90^{\circ}$ pole-to-equator angle is mapped by ZDI in latitudinal bins of $90^{\circ}/8.125 \simeq 11^{\circ}$ \citepads{2016LNP...914.....R}. This estimate neglects the variations due to the phase coverage of the individual ZDI maps (fairly consistent among all of our epochs) as well as the possible rotational smearing during the exposures (negligible in our observations).} over which longitudinally-averaged values of the radial and azimuthal magnetic field components are extracted. Since the averaging procedure is performed over the signed magnetic field, the resulting ``meridional\footnote[4]{Do not confuse with the meridional component of the magnetic field (not included in the butterfly diagram analysis).} snapshots'' (covering the dates comprising each ZDI epoch) capture the dominant field polarity at different latitudes. By plotting all the retrieved snapshots as a function of time, we can construct the butterfly diagrams of \ihor~for $B_{\rm R}$ and $B_{\rm A}$, which are presented in Fig.~\ref{fig_butterfly}. We also include the corresponding number of stellar rotations since the beginning of the campaign (with $P_{\rm rot} = 7.73$~d), as well as, the 2-period fit to the Ca~II H\&K S-index variations characterizing the evolution of the chromospheric activity \citepads{2023MNRAS.524.5725A}.

Notice that the width of each snapshot (i.e.~vertical slices) in the diagram is not the same among the various epochs, illustrating the number of consecutive rotations used for retrieving each ZDI map. While ZDI studies commonly achieve better phase coverage by combining data from multiple rotations (even non-consecutive ones), it is clear that this would clearly impact the capability to reconstruct a Butterfly diagram. Ideally, each ZDI map should be recovered over a single rotation, minimizing in this way the evolution the field can have in between epochs. As such, wider snapshots in Fig.~\ref{fig_butterfly} indicate epochs with more uncertainty in terms of the long-term magnetic field evolution. 

\begin{figure}[!h]
\centering
\includegraphics[trim=0.2cm 0.1cm 0.2cm 0.2cm, clip=true,width=0.495\textwidth]{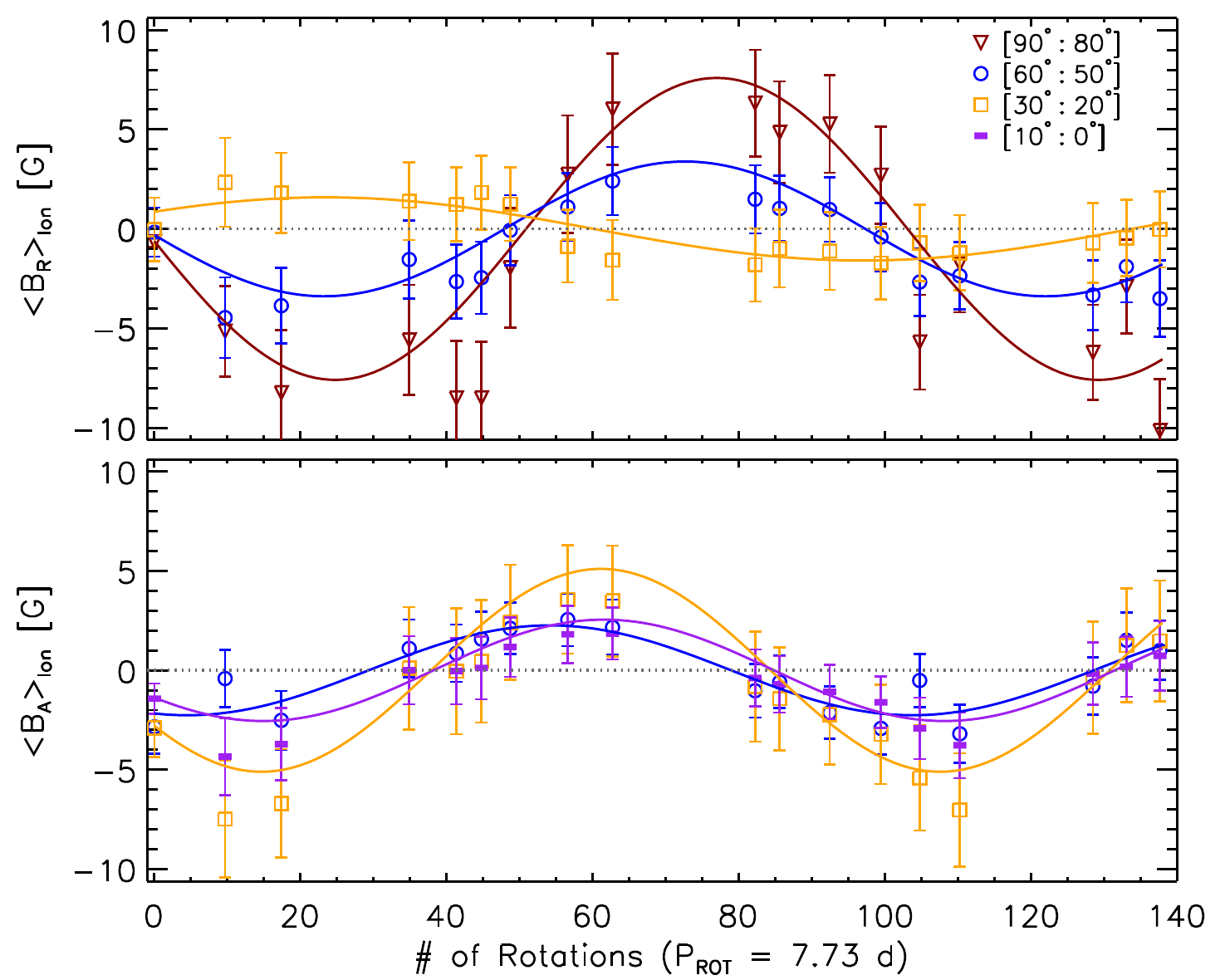}
\caption{Temporal profiles of the longitudinally-averaged radial (top) and azimuthal (bottom) large-scale magnetic field of \ihor. Symbols/colors denote the range of latitudes of extraction, with error bars indicating the standard deviation of each component within the associated latitude strip as well as with longitude. Solid lines (with corresponding colors) show the resulting best-fit solution for a single period model (3 free parameters) at a given latitude bin (see text for details).}
\label{fig_B-cycle}
\end{figure}

As expected from the individual ZDI maps, multiple polarity reversals of both components of the large-scale field of \ihor~are evident in Fig.~\ref{fig_butterfly}. Around the visible pole, $B_{\rm R}$ flips sign two times (between rotations 48.64\,-\,56.51  and 99.46\,-\,104.76) while, closer to the equator, three polarity reversals are seen in $B_{\rm A}$ (between rotations 17.56\,-\,33.91, 62.58\,-\,82.03, and 128.35\,-\,133.00). At face value, the large-scale magnetic field does not show an immediate correspondence with the double periodicity observed in the chromospheric activity of the star. However, the magnetic field evolution appears to be more regular compared to the Ca~II H\&K S-index. To better quantify the magnetic cycle timescales, we analyze the temporal evolution of the longitudinally-averaged magnetic field components at different locations on the stellar surface. 

Figure~\ref{fig_B-cycle} shows temporal profiles for $\left<B_{\rm R}\right>_{\rm LON}$ and $\left<B_{\rm A}\right>_{\rm LON}$, extracted at four different $10^{\circ}$ latitude strips including the visible pole ($90^{\circ}$) and the stellar Equator ($0^{\circ}$). The error bars indicate the absolute variation (standard deviation\footnote[5]{This information was extracted from the individual ZDI maps.}) of each component with longitude and within the corresponding latitudinal bin.  To determine the periodicity at each latitudinal band, we consider a single-period sinusoidal model $\left<B\right>_{\rm LON}(t) = B^{0}\sin(\omega^{\rm B}t + \phi^{\rm B})$, composed of 3 free parameters: amplitude ($B^{0}$), period ($P^{\rm B} = 2\pi/\omega^{\rm B}$), and normalized phase ($\phi^{\rm B}$), and find the best-fit to the individual temporal profiles of $\left<B_{\rm R}\right>_{\rm LON}$ and $\left<B_{\rm A}\right>_{\rm LON}$ (solid lines in Fig.~\ref{fig_B-cycle}). Similar fits were obtain for all the other latitude bins down to the visibility limit ($-60^{\circ}$). The resulting model parameters with their formal uncertainties are listed in Table~\ref{tab:B-cycle_Pars}.   

Several aspects of Fig.~\ref{fig_B-cycle} and Table~\ref{tab:B-cycle_Pars} are noteworthy. First of all, both field components reveal a time-scale of approximately $100$~rotations for the completion of the magnetic cycle (consistent with the evolution observed in the ZDI maps and in Fig.~\ref{fig_butterfly}). There are however some deviations in the determined polarity reversal time-scale, with larger values (by up to $\sim$\,$60\%$) identified in $\left<B_{\rm R}\right>_{\rm LON}$ and smaller ones (by $\sim$\,$20\%$) in the evolution of $\left<B_{\rm A}\right>_{\rm lon}$. In the case of the radial component, these deviations appear at mid-latitudes ($20^{\circ}$\,--\,$40^{\circ}$) in both hemispheres (see Table~\ref{tab:B-cycle_Pars}). For instance, evolution at $30^{\circ}$ (Fig.~\ref{fig_B-cycle}, top panel, orange) displays an associated reversal time-scale of $147.2\pm24.8$ rotations ($\sim$\,$1137.9$~d), whereas a value of $154.5\pm14.9$ rotations ($\sim$\,$1194.3$~d) emerges at $-30^{\circ}$. Note that these longer polarity reversal time-scales exceed our observational window of $\sim$\,$139$ rotations ($\sim$\,$1077.0$~d), which is reflected by their relatively large uncertainties ($\sim$\,$20\%$) even when the field variance in the individual epochs is small. In terms of the amplitude of variability, there is a clear trend of decreasing $B^{0}_{\rm R}$ with latitude, starting with a maximum value close to $7.5$~G for the visible pole, down to sub-Gauss amplitudes at the stellar equator and in the partially obscured hemisphere. With the exception of latitude $40^{\circ}$ (which displays the largest uncertainty in the period determination), the magnetic cycle phase in $\left<B_{\rm R}\right>_{\rm LON}$ does not show too much variation staying within the $0.4-0.6$ range for all latitudes.   

The situation is slightly different for the azimuthal component (Table~\ref{tab:B-cycle_Pars}, Fig.~\ref{fig_B-cycle}, bottom panel). The majority of latitudes display $P_{A}^{B}$ values around $90$~rotations, with $80.9\pm7.7$~rotations ($\sim$\,$622.9$~d) at $-30^{\circ}$, appearing as the shortest time-scale for a polarity reversal in this component. While some longer values ($\gtrsim 105$~rotations) emerge at high latitudes ($70^{\circ}$\,--\,$90^{\circ}$) and at $-20^{\circ}$, these display the largest uncertainties in both, $P_{A}^{B}$ (up to $\sim$\,$20$\%) and in $\phi_{\rm A}^{\rm B}$ (up to $\sim$\,$50\%$). The cycle amplitude in this component shows a maximum value of $B^{0}_{\rm A} \sim 5.1$~G around $30^{\circ}$, declining in strength down to sub-Gauss levels at very high latitudes ($80^{\circ}$\,--\,$90^{\circ}$) as well as in the obscured hemisphere ($\leq -20^{\circ}$). As evidenced by the best-fit curves plotted in Fig.~\ref{fig_B-cycle}, there is a small phase difference between the radial and azimuthal large-scale magnetic field at the latitudes that show the largest cycle amplitudes (i.e.,~at $90^{\circ} - 60^{\circ}$ for $\left<B_{\rm R}\right>_{\rm LON}$ and $30^{\circ} - 0^{\circ}$ for $\left<B_{\rm A}\right>_{\rm LON}$). From the best-fit parameters presented in Table~\ref{tab:B-cycle_Pars}, this phase difference is on the order of $0.1$, which corresponds to $\sim$\,$10$~rotations ($77.3$~d) assuming a mean cycle period of $\sim$\,$100$~rotations for both components.     

\begin{table*}
\centering
 \caption{Best-fit parameters of the single-period sinusoidal models describing the evolution of the magnetic cycle in \ihor.}
 \label{tab:B-cycle_Pars}
 \begin{tabular}{l c c c c c c}
  \toprule
  Latitude Bin & \multicolumn{3}{c}{$\left<B_{\rm R}\right>_{\rm lon}$} & \multicolumn{3}{c}{$\left<B_{\rm A}\right>_{\rm lon}$}\\
  
\hspace{0.8cm}[$^{\circ}$] & $B^{0}_{\rm R}$ [G] & $P^{\rm B}_{\rm R}$ [rot] & $\phi^{\rm B}_{\rm R}$ & $B^{0}_{\rm A}$ [G] & $P^{\rm B}_{\rm A}$ [rot] & $\phi^{\rm B}_{\rm A}$ \\
   \midrule
  $[90:80]$ & $7.59 \pm 0.96$ & $104.5 \pm 2.6$  & $0.513 \pm 0.005$ & $ 0.21 \pm 0.10$ & $109.2 \pm 19.2$ & $0.828 \pm 0.459$\\[2pt]
  $[80:70]$ & $6.90 \pm 0.91$ & $103.7 \pm 2.8$  & $0.511 \pm 0.009$ & $ 0.82 \pm 0.20$ & $107.0 \pm 10.6$ & $0.797 \pm 0.255$\\[2pt]
  $[70:60]$ & $5.28 \pm 0.82$ & $101.0 \pm 3.5$  & $0.506 \pm 0.022$ & $ 1.38 \pm 0.33$ & $106.7 \pm 9.9$  & $0.793 \pm 0.087$\\[2pt]
  $[60:50]$ & $3.39 \pm 0.71$ & $98.8  \pm 5.3$  & $0.515 \pm 0.044$ & $ 2.25 \pm 0.45$ & $99.3  \pm 7.8$  & $0.705 \pm 0.081$\\[2pt]
  $[50:40]$ & $1.62 \pm 0.65$ & $98.9  \pm 10.5$ & $0.563 \pm 0.099$ & $ 3.49 \pm 0.61$ & $94.4  \pm 5.6$  & $0.633 \pm 0.060$\\[2pt]
  $[40:30]$ & $1.17 \pm 0.54$ & $158.6 \pm 36.8$ & $0.083 \pm 0.012$ & $ 4.68 \pm 0.84$ & $93.2  \pm 4.6$  & $0.605 \pm 0.050$\\[2pt]
  $[30:20]$ & $1.59 \pm 0.65$ & $147.2 \pm 24.8$ & $0.591 \pm 0.120$ & $ 5.10 \pm 1.04$ & $93.0  \pm 4.5$  & $0.592 \pm 0.048$\\[2pt]
  $[20:10]$ & $1.58 \pm 0.65$ & $133.5 \pm 20.3$ & $0.574 \pm 0.112$ & $ 4.72 \pm 1.06$ & $92.7  \pm 4.5$  & $0.586 \pm 0.048$\\[2pt]
  $[10:0]$ & $1.34 \pm 0.56$ & $122.6 \pm 18.7$ & $0.557 \pm 0.112$ & $ 3.91 \pm 0.91$ & $92.8  \pm 4.7$  & $0.585 \pm 0.049$\\[2pt]
  $[0:-10]$ & $0.90 \pm 0.40$ & $103.7 \pm 13.9$ & $0.515 \pm 0.105$ & $ 2.55 \pm 0.58$ & $93.3  \pm 4.8$  & $0.590 \pm 0.050$\\[2pt]
 $[-10:-20]$ & $0.34 \pm 0.30$ & $100.0 \pm 19.8$ & $0.563 \pm 0.192$ & $ 1.29 \pm 0.37$ & $94.8  \pm 6.6$  & $0.606 \pm 0.070$\\[2pt]
 $[-20:-30]$ & $0.09 \pm 0.18$ & $159.1 \pm 30.0$ & $0.607 \pm 0.632$ & $ 0.34 \pm 0.18$ & $106.5 \pm 21.4$ & $0.703 \pm 0.227$\\[2pt]
 $[-30:-40]$ & $0.69 \pm 0.13$ & $154.5 \pm 14.9$ & $0.546 \pm 0.045$ & $ 0.30 \pm 0.15$ & $80.9  \pm 7.7$  & $0.020 \pm 0.065$\\[2pt]
 $[-40:-50]$ & $0.46 \pm 0.09$ & $139.9 \pm 13.0$ & $0.496 \pm 0.036$ & $ 0.46 \pm 0.12$ & $87.0  \pm 4.7$  & $0.074 \pm 0.052$\\[2pt]
 $[-50:-60]$ & $0.41 \pm 0.09$ & $137.9 \pm 13.9$ & $0.475 \pm 0.041$ & $ 0.42 \pm 0.08$ & $86.3  \pm 4.4$  & $0.076 \pm 0.051$\\[2pt]
  \bottomrule
 \end{tabular}
\tablefoot{The parameters $B^{0}$ (Amplitude), $P^{\rm B}$ (Period), $\phi^{\rm B}$ (Normalized Phase) describe the temporal profiles of $\left<B_{\rm R}\right>_{\rm LON}$ and $\left<B_{\rm A}\right>_{\rm LON}$ extracted at the indicated $10^\circ$ latitudinal bins. For all cases the phase of the fit has been adjusted so that the amplitude takes positive values.
} 
\end{table*}

\subsection{Large-scale flows}\label{sec:LS_flows}

Apart from the analysis of the magnetic cycle period, we can use the butterfly diagrams to put constraints on the large-scale photospheric flows on \ihor. To our knowledge, this is the first time such an estimate has been carried out on a star different from the Sun. 

From each butterfly diagram, we trace the average position of the magnetic regions of a given polarity as a function of time. The calculation is performed on each hemisphere and separately for $\left<B_{\rm R}\right>_{\rm LON}$ and $\left<B_{\rm A}\right>_{\rm LON}$. The extracted pairs of positions/times (rotations) are then fitted by a simple linear function to estimate the speed of the large-scale flow on each field component. This methodology is inspired in similar analyses performed on solar data, retrieving the so-called 'rush-to-the-poles' and equator-ward flux migration of the solar magnetic field (e.g.~\citeads{2001AJ....122.2115L}, \citeads{2007ApJ...670L..69S}, \citeads{2022MNRAS.510.1331M}). From the behavior observed in the Butterfly diagrams, the analysis that follows assumes that the pole-ward flux transport dominates the evolution of the radial field, whereas the equator-ward drift manifests more strongly in the azimuthal component, being the precursor of spot/facula. Note that while both components are treated in a similar manner, only the pole-ward motion is expected to trace the stellar meridional circulation (\citeads{2021SCPMA..6439601C}, \citeads{2023SSRv..219...77H}) whereas the equator-ward drift would be the result of consecutive emerging magnetic regions as the cycle progresses (\citeads{1955epds.book.....W}, \citeads{2011SoPh..273..221H}).  

\begin{figure}
\centering
\includegraphics[trim=0.0cm 0.0cm 0.0cm 0.0cm, clip=true,width=0.495\textwidth]{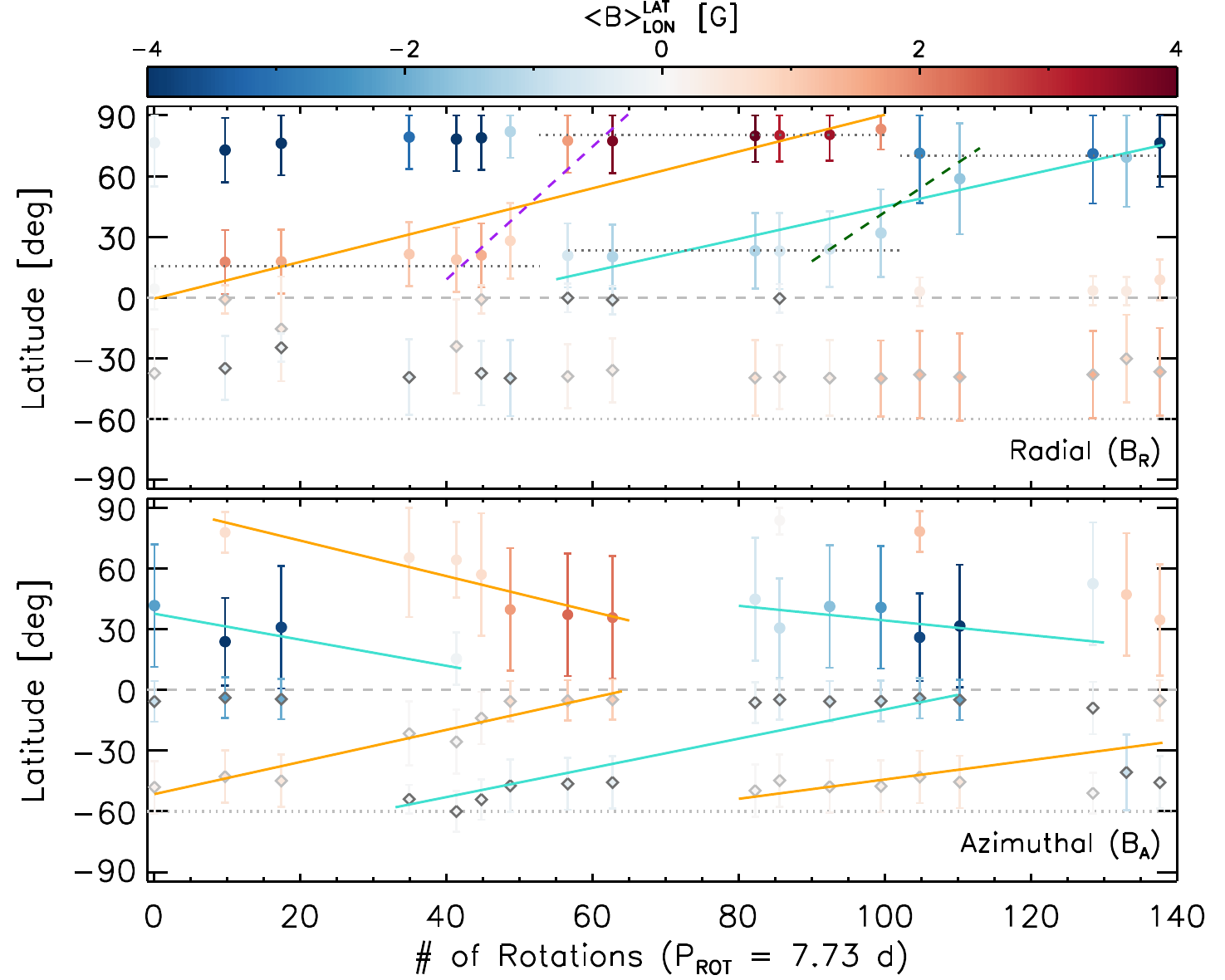}
\caption{Analysis of the large-scale flows on \ihor~based on the magnetic butterfly diagrams (Fig.~\ref{fig_butterfly}). The top and bottom panels contain results for the radial and azimuthal components, respectively. Individual points indicate the field-weighted mean latitude hosting each polarity on a given epoch, with error bars denoting its standard deviation within the averaged latitudinal bin. Filled colored circles and gray open diamonds separate locations within the Northern and Southern hemispheres, respectively. The latitudinally-averaged magnetic field strength, $\left<B\right>^{\rm LAT}_{\rm LON}$, is denoted by the color bar. Lines (straight and dashed) correspond to different fits aimed at estimating the pole-ward and equator-ward migration of the large-scale field. The line colors indicate which polarity of the magnetic field the fits are tracing (Positive: Orange, Magenta; Negative: Cyan, Green). Dotted lines in $B_{\rm R}$ show the best-fit two-latitude models (slope = 0.0) for each polarity used for statistical model comparison (see Section~\ref{sec:LS_flows} for details).}
\label{fig_B-flows}
\end{figure}

The results are shown in Fig.~\ref{fig_B-flows} for the radial (top) and azimuthal (bottom) components. The individual points show the field-weighted mean latitude of a particular polarity of the field at each epoch. Only the visible hemisphere is considered for the radial field, as the remaining region lacked sufficient information for identifying any trends in latitude for this component. The color bar indicates the latitudinally-averaged value of the magnetic field on each location. As this averaging procedure takes place over the Butterfly diagram values (averaged over longitude) we label it as  $\left<B\right>^{\rm LAT}_{\rm LON}$. In principle, the weighting procedure could also incorporate temporal information (i.e. number of rotations), taking into account the widths of the individual ZDI epochs (see Sect.~\ref{sec:B-fly}). However, most of our maps have similar temporal coverage (between $\sim1.0 - 1.9$ rotations) with only Epoch 4 largely deviating from this range ($3.48$). As such, we did not include any temporal uncertainty in our analysis. The different lines (solid, dashed) in Fig.~\ref{fig_B-flows} correspond to the various fits considered to the data with their colors matching the polarity of the field (Positive: Orange, Magenta; Negative: Cyan, Green).  

For the radial magnetic field (Fig.~\ref{fig_B-flows}, top) two sets of linear fits are employed in the flow analysis. The first one (solid line) considers all the points of a given polarity, from low latitude regions to the pole, before the polarity reversal. The second case (dashed line) considers only 2-3 epochs before and after the polarity reversal at the visible pole. These two cases are selected to obtain two 'limits' on the pole-ward flow speed (slope), being the latter faster than the former one. The second approach should be closer to the methodology employed on the Sun based on magnetogram data (see e.g. \citeads{2022MNRAS.510.1331M} and references therein). The first method however allows us to set the lowest speed compatible with the observed evolution of the radial field in the Butterfly diagrams (with the added benefit of having more data points for the fit). In this way, we obtain pole-ward flow speeds between $19.2 \pm 1.2$~m~s$^{-1}$ (slow) and $69.2 \pm 8.6$~m~s$^{-1}$ (fast) for the positive polarity. Similarly, the fits to the negative polarity regions yield limit speeds of $16.9 \pm 1.6$~m~s$^{-1}$ (slow) and $51.3 \pm 11.2$~m~s$^{-1}$~(fast). 

For completeness, we also consider a two constant-latitude model (i.e., no flow velocity) per polarity (dotted lines in Fig.~\ref{fig_B-flows}, top panel) to perform a statistical model comparison with the pole-ward flow models using the Bayesian Information Criterion ($\rm BIC$; \citeads{1978AnSta...6..461S}). For the positive polarity, the best-fit zero-slope model yield weighted-mean latitudes of $\lambda_1^+ = 15.5 \pm 5.5^{\circ}$ and $\lambda_2^+ = 80.4 \pm 5.3^{\circ}$, with a break at $t_1 = 52.64$~rotations and an associated BIC of $102.8$. Similarly, for the negative polarity we obtain $\lambda_1^- = 23.1 \pm 7.3^{\circ}$ and $\lambda_2^- = 70.2 \pm 10.9^{\circ}$, with $t_2= 102.11$~rotations and a BIC value of $94.6$. Computing the BIC for the slow pole-ward linear models (2 free parameters) yields $107.7$ and $94.5$ for the positive ($N=13$ points) and negative ($N=11$ points) polarities, respectively. This analysis indicates that the two constant-latitude model is statistically slightly preferred against the slow pole-ward fit in the positive polarity ($\Delta\rm BIC < 0$), whereas the opposite situation occurs for the negative one. However, as mentioned earlier, the slow pole-ward drift has been taken as a coarse lower limit to the flow speed but admitting a physically-motivated preference for the faster speed determined before and after field polarity reversal at the pole (magenta and green fits in Fig.~\ref{fig_B-flows}). Repeating the two-latitude fit for the 5 points involved in the fast positive polarity\footnote[6]{The negative polarity fast pole-ward model does not have sufficient points to perform a meaningful comparison.} pole-ward model yields a BIC value of $42.1$, slightly less favored than the corresponding pole-ward model with a BIC of $41.7$. Note that, in general, these results make both models (pole-ward flow vs two-latitude) statistically comparable, so that another possible interpretation of the data is for the star to have two separate active latitudes. We will however keep the interpretation based on the meridional flow transport, motivated by the known solar phenomenology and the assumption that the dynamo operating in Sun-like stars like \ihor~should behave similarly as in the Sun.     

As can be seen in Fig.~\ref{fig_B-flows} (bottom), we consider several ``tracks'' for fitting the evolution of the azimuthal component. These were constructed by taking all the locations of a given polarity with the requirement of having more than three consecutive epochs and a maximum of one epoch gap per track. This is the reason why some points are not included in the fitting analysis (e.g., negative polarity in the southern hemisphere between rotations $0-20$ and positive polarity in the northern hemisphere between rotations $100-138$). Similar to the radial component, the fitted slope yields the mean equator-ward drift speed compatible with the observations. In this way, we obtain speed values of $15.1 \pm 4.0$~m~s$^{-1}$ for the positive polarity and $12.1 \pm 3.3$~m~s$^{-1}$ for the negative one. These values correspond to the average between both hemispheres, employing the three identified tracks for each polarity. Similarly, the average speeds on the northern and southern hemisphere (regardless of the field polarity) yield $13.2 \pm 3.3$~m~s$^{-1}$ and $14.0 \pm 4.1$~m~s$^{-1}$, respectively. This suggest that our derived equator-ward drift speeds are robust even when employing data from the partially obscured hemisphere. Finally, given that our flow speed and drift rate determinations are inspired by the known solar phenomenology, we replicate in Sect.~\ref{sec:Mer-Equ} the methodology applied on \ihor~on solar magnetic field observations as a way to assess the performance of our ZDI-based procedure.

\section{Discussion}\label{sec:discussion}

We have presented a comprehensive characterization of the magnetic cycle of the young planet-host Sun-like star \ihor~based on spectropolarimetry and ZDI. In the following sub-sections we place our results in the context of other cool stars with information on their magnetic cycles as well as with the Sun.

\subsection{The large-scale field of \ihor~among cool stars}\label{sec:B-comparison}

The $B^2_{\rm tor}$ vs $B^2_{\rm pol}$ range recovered in the \ihor\ maps is similar to what has been observed for solar-type stars with similar ages and rotation periods. In terms of their morphologies, solar-type stars in their first 250~Myr host stronger axisymmetric large scale fields \citepads{2016MNRAS.457..580F}, and as they age (between 250-650 Myr)  they show significantly weaker large-scale non-axisymmetric field morphologies that are similar to those found for \ihor\ \citepads{2018MNRAS.474.4956F}. In the latter study, the authors note the difficulty of finding trends in magnetic geometry in the 250-650 Myr range and suggest this is due to intrinsic variability as the stars follow longer term cycles. Our maps with dense time sampling of \ihor's large-scale field over the course of its cycle further supports this conclusion. As with many of the stars studied in \citetads{2018MNRAS.474.4956F}, \ihor's field also tends to have a significant toroidal component for much of its cycle and displays a tendency towards non-axisymmetry. 

The dense time-sampling of \ihor\ maps presented here also allows a detailed comparison with the solar activity cycle. \citetads{2021MNRAS.500.1243L} carried out an analysis of which features of the solar activity cycle would be recovered using ZDI techniques by applying ZDI-like diagnostics to magnetograms acquired during Solar Cycle 23. While there is a strong trend in the solar activity cycle with the fraction of axisymmetric field recovered; with axisymmetry decreasing essentially linearly with increasing S-index (see \citeads{2021MNRAS.500.1243L}, Fig.\,8), on \ihor\ the trend is not as clear. Indeed, the fraction of axisymmetry tends to increase during activity maxima albeit with quite some scatter. The clearest trends with activity cycle found on \ihor\ correspond to $B^2_{\rm tor}/B^2_{\rm pol}$, with both dipolar and quadrupolar components showing increases during the chromospheric activity maxima traced with the chromospheric S-index. Likewise, we see that the axisymmetric total energy $\langle B^2_{\rm axi} \rangle$, and energy in the toroidal axisymmetric component $\langle B^2_{\rm tor,axi} \rangle$ roughly follow the evolution of the chromospheric activity (with significant scatter).

\subsection{Activity and magnetic cycles connection}\label{sec:cycles}

We can compare the cycle periods obtained for the magnetic field components with the behaviour observed in the coronal and chromospheric activity of the star (\citeads{2019A&A...631A..45S}, \citeads{2023MNRAS.524.5725A}). While the X-ray observations indicate a coronal cycle of $P^{\rm cyc}_{\rm X} \sim 1.6$~yr ($\sim75.9$~rotations), the long-term evolution of the S-index is more consistent with the superposition of two periodicities around $P^{\rm cyc}_{S1} = 1.49$~yr ($\sim70.7$~rotations) and $P^{\rm cyc}_{S2} = 1.09$~yr ($\sim51.7$~rotations). Interestingly, the mean magnetic cycle period of $\sim$\,$100$ rotations would match the solar 2:1 relation (i.e. two activity cycles for one magnetic cycle) with the secondary chromospheric cycle ($P^{\rm cyc}_{S2}$). While uncertain for the reasons exposed in Sect.~\ref{sec:B-fly}, the second magnetic cycle period in $B_{\rm R}$ (around $150$ rotations) would also roughly follow the solar expectation with the coronal ($P^{\rm cyc}_{\rm X}$) and the primary chromospheric cycle ($P^{\rm cyc}_{S1}$). Note also that the $\sim20\%$ shorter magnetic cycle period identified in $B_{\rm A}$ is also in agreement with the behaviour observed in the solar toroidal magnetic field \citepads{2022ApJ...927L...2L}. 

As with the global magnetic field properties (Sect.~\ref{sec:B-comparison}), our results for \ihor~also align well with the activity and magnetic cycle properties established by ZDI studies of other cool stars. Note, however, that none of the previous investigations has been able to resolve the magnetic cycle as we have done for \ihor~(i.e. retrieve a magnetic Butterfly diagram), so that comparisons can only be performed on the most general aspects of the~cycles.  

The studies of Boro Saikia~et~al. (\citeyearads{2016A&A...594A..29B}, \citeyearads{2018A&A...620L..11B}) showed that 61~Cyg~A displays a Sun-like behavior in its magnetic cycle, where the polarity reversals occur in phase with the observed chromospheric and coronal activity cycle of $7.3\pm0.1$~yr (\citeads{1995ApJ...438..269B}, \citeads{2012A&A...543A..84R}). This was the first stellar case on which such a cycle coincidence could be confirmed.

The star $\epsilon$~Eri has been found to follow two chromospheric cycles of $2.95$~yr and $12.7$~yr (\citeads{1995ApJ...438..269B}; \citeads{2000ApJ...544L.145H}; \citeads{2013ApJ...763L..26M}).  A coronal counterpart in X-rays to the shorter chromospheric cycle has only recently been confirmed \citepads{2020A&A...636A..49C}. On the other hand, the ZDI maps obtained by Jeffers~et~al.~(\citeyearads{2014A&A...569A..79J}, \citeyearads{2017MNRAS.471L..96J}) suggest polarity reversals on time-scales that are much shorter than these cycles; two polarity reversals have occurred which may be associated with the chromospheric activity minimum, but longer and denser monitoring is still required.  

The planet-hosting star $\tau$~Boo was initially believed to show a $1$:$3$ relation between the magnetic and chromospheric cycle periods ($360$ vs $116$ days, \citeads{1997ApJ...474L.119B}; \citeads{2000ApJ...531..415H}; \citeads{2013MNRAS.435.1451F}). However, more densely sampled ZDI reconstructions discussed by \citetads{2018MNRAS.479.5266J} showed polarity reversals in the radial field in phase with the shorter chromospheric cycle (similar to what we observe in \ihor). Although X-ray observations of the star carried out by \citetads{2012AN....333...26P} found no evidence of such a short coronal activity cycle, a re-analysis of the data performed by \citetads{2017A&A...600A.119M} did find such periodicity albeit with a relatively small maximum to minimum amplitude ($L_{\rm X}^{\rm max}/L_{\rm X}^{\rm min} \simeq 1.6$). As such, the authors concluded that $\tau$~Boo would also display a solar-like behavior on its activity and magnetic cycles.  

Long-term monitoring of $\kappa$ Ceti's chromospheric activity showed a high variability exhibiting multiple cycles of $22.3$\,yr and $5.7$\,yr (\citeads{1995ApJ...438..269B}, \citeads{2018A&A...616A.108B}). A recent investigation suggests a shorter period of  $3.1$\,yr \citepads{2022A&A...658A..16B}. This short period is suspected to be a harmonic of the $5.7\,(6)$\,yr period, but opens the possible scenario of two activity cycles with a 2:1 ratio similar to the Hale and the Schwabe cycles, even though showing a complex temporal evolution, unlike the Sun. The ZDI analysis by \citetads{2022A&A...658A..16B} reports a radial field for $\kappa$ Ceti with considerable evolution during $6$\,yr of yearly follow-up. The large-scale magnetic field reconstructions show a polarity reversal of the radial and azimuthal components, with a magnetic cycle period of approximately $10\pm2$\,yr.

Finally, ZDI studies of the G dwarfs $\chi^1$\,Ori (G0V, $P_{\rm rot}~\simeq~5.0$~d), HD~9986 (G5V, $P_{\rm rot} \simeq 21.0$~d) and HD~56124 (G0V, $P_{\rm rot} \simeq 20.7$~d) report relatively short magnetic cycle periods ---between $3-10$~yr; \citeads{2016A&A...593A..35R}, \citeads{2022A&A...659A..71W},  \citeads{2025A&A...693A.269B}--- albeit with significant uncertainties given the sparseness of the ZDI reconstructions. In all cases, however, it does not seem to be a clear connection between the chromospheric S-index variability and the evolution of the large-scale field, with some of these targets originally classified as ``variable'' by the Mount Wilson survey \citepads{1995ApJ...438..269B}. From our analysis of \ihor, it is possible that a denser sampling of the large-scale field is needed to properly capture the behaviour and relations between these rapidly evolving cycles.  

\subsection{Meridional circulation and equator-ward drift}\label{sec:Mer-Equ}

As mentioned in Sect.~\ref{sec:LS_flows}, there are no estimates of the large-scale photospheric flow properties for stars other than the Sun. Therefore, we will compare our \ihor~findings with established solar large-scale flow values, focusing particularly on spatio-temporal speed averages, which should be more directly comparable to those derived from stellar ZDI data.

\begin{figure}
    \centering
    \includegraphics[width=0.5\textwidth]{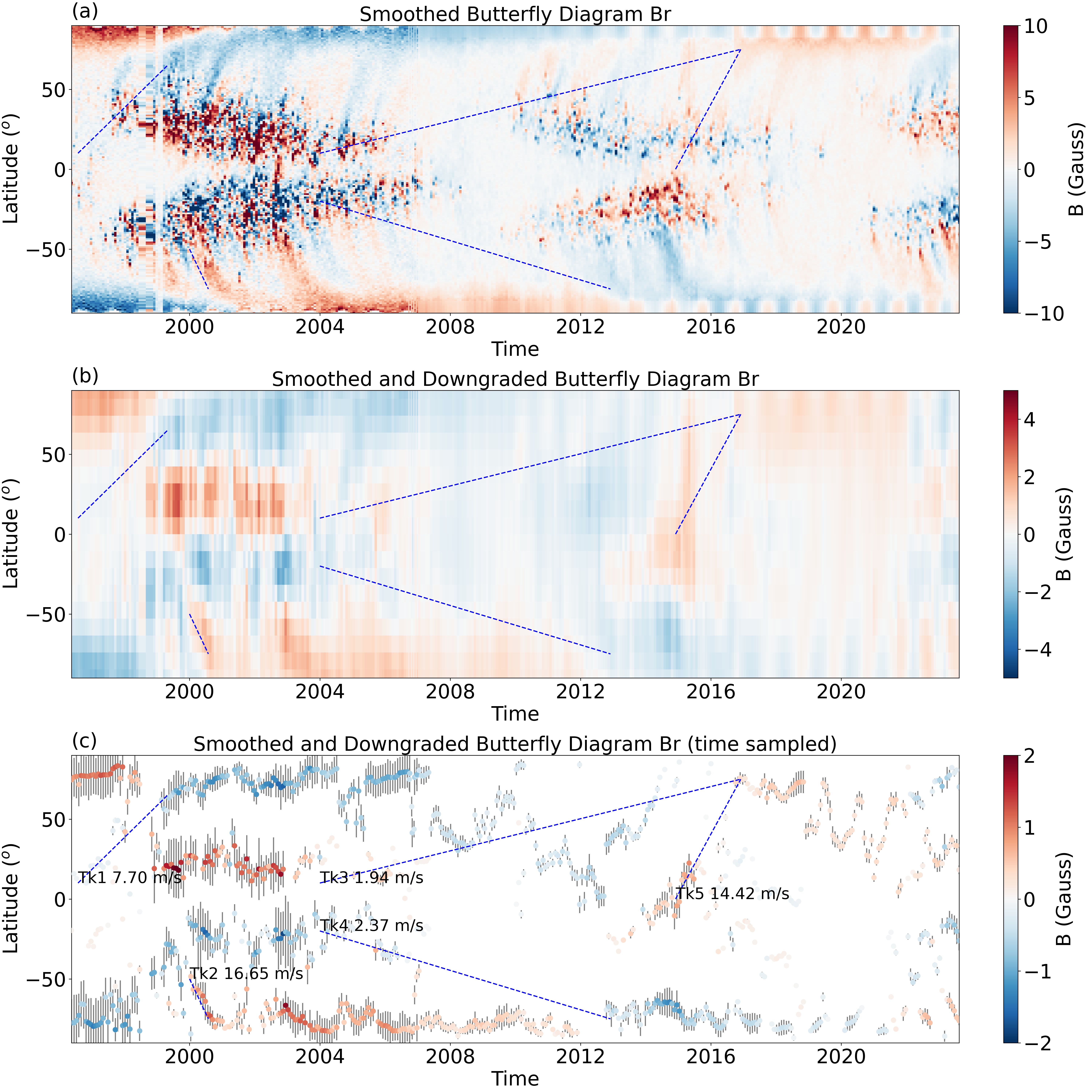}
    \caption{Solar magnetic butterfly diagrams of the radial field ($B_{\rm r}$) from MDI and GONG data. (a) Smoothed butterfly diagram of $B_{\rm r}$, constructed from Carrington maps and combined datasets from SOHO/MDI and GONG. (b) Resolution-downgraded butterfly diagram, with latitudinal resolution reduced to match our observations of \ihor~($10^{\circ}$ per pixel). (c) Field-weighted butterfly diagram, where the location of each polarity is indicated, with error bars representing the standard deviation of field strength-weighted latitudes. Linear fits with the corresponding speeds are indicated.}
    \label{fig:br}
\end{figure}

\begin{figure}
    \centering
    \includegraphics[width=0.5\textwidth]{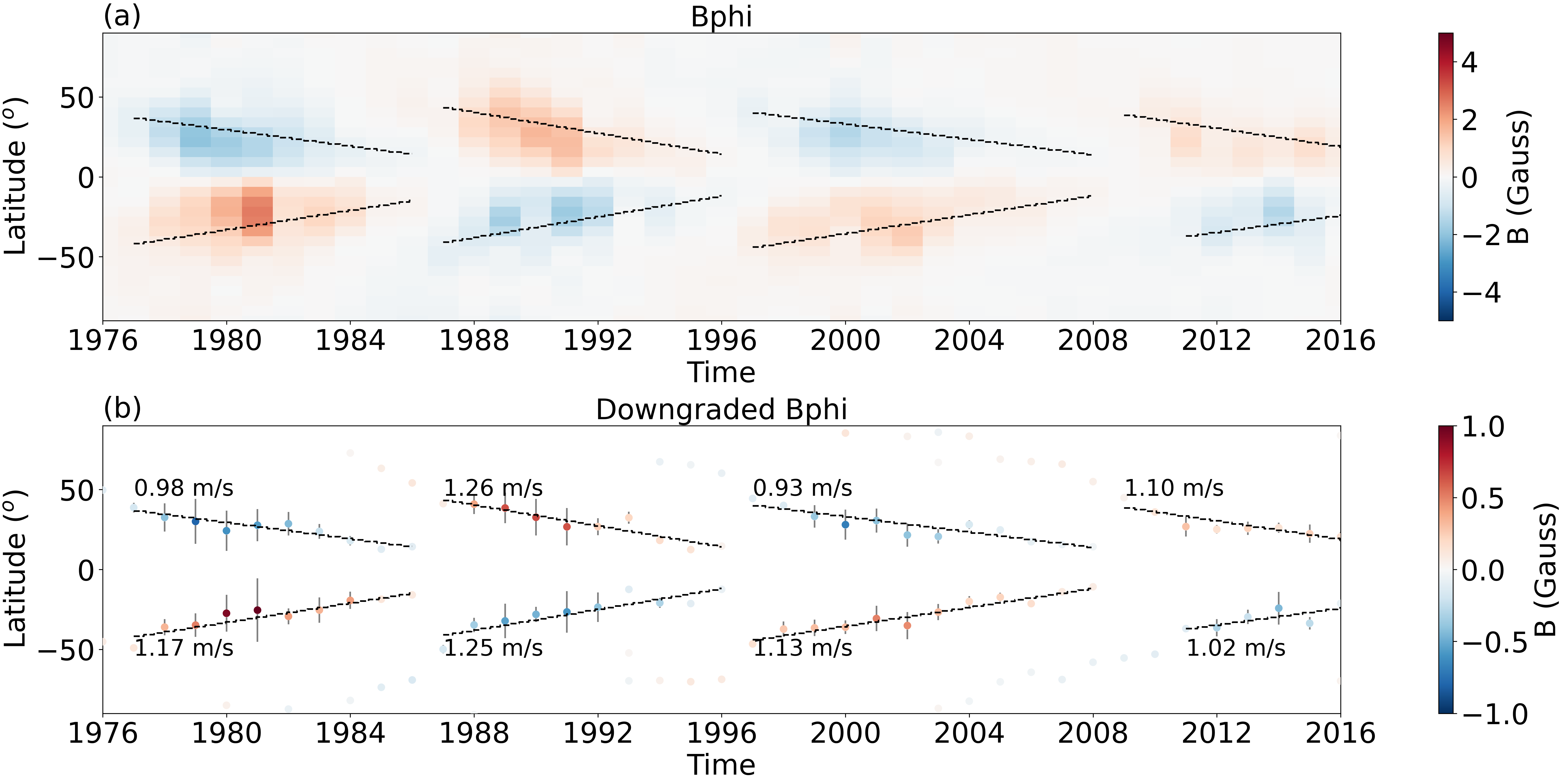}
    \caption{Analysis of the solar equator-ward drift based on the azimuthal magnetic field ($B_{\phi}$). (a) Reconstructed butterfly diagram for $B_{\phi}$ taken from \citetads{2018A&A...609A..56C}. (b) Downgraded butterfly diagram with half the original latitudinal resolution used for the analysis. The drift speeds of different polarity regions are determined via linear fits and are indicated by dashed~lines.}
    \label{fig:bphi}
\end{figure}

Including errors, our analysis revealed a pole-ward flux transport speed within $15 - 78$~m~s$^{-1}$ and an equator-ward drift in range of $9 - 19$~m~s$^{-1}$ for \ihor. Given the relatively fast activity and magnetic cycles on the star, it is not surprising that these speeds appear larger than the nominal solar values, namely a mean meridional flow speed of $\sim$\,$10$~m~s$^{-1}$ \citepads{2022LRSP...19....3H} as well as a $\sim$\,$0.61$~m~s$^{-1}$ average equator-ward drift rate\footnote[7]{Typically expressed as $1.6^{\circ}$\,yr$^{-1}$, which we have converted to m~s$^{-1}$ assuming the fiducial value of the solar radius ($R_{\odot} = 6.95\times10^{8}$~m).} \citepads{2001AJ....122.2115L}.

As mentioned earlier, the values listed above correspond to global averages, as both of these solar large-scale motions are known to vary as a function of latitude as well as with the activity cycle \citepads{2015LRSP...12....4H}. As the methodologies involved in their determination on the Sun are very different compared to our procedure based ZDI data (see \citeads{2022LRSP...19....3H}), we will emulate the analysis over stellar-like solar observations as a way to improve the comparability of the results. Furthermore, this will allow us to assess the robustness of the procedure described in Sect.~\ref{sec:LS_flows}, as this is the first time this determination is carried out in the literature.

As with \ihor, we begin constructing solar magnetic butterfly diagrams. The Carrington maps of the radial field are obtained from the Michelson Doppler Imager (MDI) project onboard the Solar and Heliospheric Observatory (SOHO) spacecraft (\citeads{1991AdSpR..11d.113S}, \citeads{1995SoPh..162....1D}, \citeads{1995SoPh..162..129S}) for the period between 1976 to 2006 and the Global Oscillation Network Group (GONG; \citeads{1987S&T....74..470H}, \citeads{1988AdSpR...8k.117H}) for the 2006 to 2023 period. The resolution of the MDI maps is 3600x1080~pixels while that of the GONG map is 360x180~pixels. In order to make the maps more comparable to what we get from the ZDI stellar observations, we applied a Gaussian smooth filter to lower the resolution. The widths of the Gaussian filters were set to 401 pixels and 81 pixels for MDI and GONG maps, respectively. The two datasets were then combined together to create the butterfly diagram of the radial field as shown in Fig.~\ref{fig:br} (a). We downgraded the resolution of the original butterfly diagram in the latitudinal direction to 18 (i.e. $10^{\circ}$ per pixel) in order to match our observations of $\iota$\,Hor, and the results are shown in Fig.~\ref{fig:br} (b). Figure~\ref{fig:br} (c) contains the field-weighted butterfly diagram, where the location of each point indicates the dominant latitude of hosting each polarity (analogous to Fig.~\ref{fig_B-flows} for \ihor). The error bars represent the standard deviation of the field strength-weighted latitudes. 

Figure~\ref{fig:br}\,(c) is used to estimate the solar pole-ward flux transport speeds where five trends are identified and plotted. These trends try to replicate the two limiting cases we considered on \ihor, with a fast lane following a given polarity surge to the pole right before its reversal (TK1, TK2 and TK5) and the slow lane starting from a given polarity region at mid-latitudes and following it to the polar region right after the reversal (TK3 and TK4). The tracks TK2 and TK5 roughly represent the accepted value for the meridional flow speed of the Sun, which varies around $10$~m~s$^{-1}$ to $20$~m~s$^{-1}$ (\citeads{2022LRSP...19....3H}). The speeds of the other three trends are much lower and they help to gauge the values obtained from our observations of $\iota$\,Hor, which are heavily limited by the gaps between epochs.  

For the solar azimuthal field, we employ the butterfly diagram previously reconstructed by \citetads{2018A&A...609A..56C} which is shown in Fig.~\ref{fig:bphi} (a). The diagram has a time resolution of one year and a latitudinal resolution of six degrees, which is downgraded by a factor of two to carry out our stellar-like analysis. We tracked the equatorial motions of each polarity in the field-weighted butterfly diagram as shown in Fig.~\ref{fig:bphi}\,(b). Similar to \ihor, we applied linear fits to the different polarity regions and obtain the equator-ward drift speeds. The results are labeled beside the corresponding trends and they own an average speed of around $1.1$~m~s$^{-1}$. 

Based on the solar tests we see that the field-weighted latitudinal fitting procedure can roughly capture the accepted large-scale flow speeds on the Sun, with results falling within their nominal ranges. As mentioned earlier, for the pole-ward flux transport the solar analysis suggests that the slow tracks found in Sect.~\ref{sec:LS_flows} most likely underestimate the flow speed in \ihor. This is because the analogous solar tracks, 'TK3' and 'TK4', yield values close to one order of magnitude below the standard $10-20$~m~s$^{-1}$ range. In this way, the fast speeds resulting from the analysis on \ihor~should be closer to the actual meridional flow on the star, which would surpass the solar values by factors between $4-7$ approximately. Similarly, while the solar equator-ward drift rate derived from our field-weighted butterfly diagram analysis appears slightly larger than global average obtained by \citet{2001AJ....122.2115L}, it falls right within the standard range measured at various latitudes and activity states (see e.g.~\citeads{2003ApJ...589..665H}). Note also that the linear fits applied here are a crude approximation of this solar observable, which can be better represented by a decaying exponential function as shown by \citetads{2011SoPh..273..221H}. These results corroborate the robustness of our estimates on the equator-ward migration drift obtained for \ihor, placing it at about $17 - 22$ times faster than its solar counterpart.

\section{Summary}\label{sec:summary}

This paper presented our investigation of the magnetic cycle in the young solar analogue \ihor, based on 18 Zeeman-Doppler Imaging maps acquired over nearly three years. Our analysis showed clear changes in the large-scale surface magnetic field, including multiple polarity reversals and cyclic variations in both the poloidal and toroidal components. Dense ZDI sampling allowed us, for the first time, to obtain stellar magnetic butterfly diagrams tracking the latitudinal migration of magnetic features and to derive initial estimates of large-scale flow properties on a star beyond the Sun (under the assumption of a similar flow phenomenology). These results provide useful benchmarks for models of dynamo processes in young solar-type stars.

Our analysis revealed a magnetic cycle intricately linked to the star's activity. The ZDI reconstructions indicate that \ihor~undergoes a full magnetic cycle over roughly 100 rotations ($\sim$\,$2.1$~yr), though latitudinal variations are evident---some mid-latitude regions exhibit polarity reversal time-scales that extend up to about 150 rotations in the radial component, while the azimuthal component oscillates on a slightly shorter time-scale of around 90 rotations. In comparison, the chromospheric activity traced by the Ca~II H\&K S-index displays dual periodicities of approximately 70.7 rotations ($\sim$\,$1.49$~yr) and 51.7 rotations ($\sim$\,$1.09$~yr), suggesting a 2:1 correspondence for the shorter cycle, mirroring the relationship observed in the Sun. On the other hand, the coronal activity cycle traced by the X-ray emission indicates a periodicity of about 75.5 rotations ($\sim$\,$1.49$~yr), which could be linked to the longer magnetic variability-timescale identified in the butterfly diagrams of the radial magnetic field.

In a broader perspective, this work demonstrates how intensive spectropolarimetric monitoring can reveal the temporal evolution of stellar magnetism. By mapping variations in field strength, geometry, and drift speeds, we can connect observed magnetic behavior to the underlying dynamo action. This approach can help advance the understanding of magnetic cycles in stars like \ihor~and lays the groundwork for future comparative studies with the solar cycle.

\begin{acknowledgements}

We would like to thank the anonymous referee for their constructive feedback, which helped improve the quality of this article. Based on observations collected at the European Organisation for Astronomical Research in the Southern Hemisphere under ESO programmes 096.D-0257, 097.D-0420, 098.D-0187, 099.D-0236, 0100.D-0535, and 0101.D-0465. Data were obtained from the ESO Science Archive Facility under request number jalvarad.212739. This work has made use of the VALD database, operated at Uppsala University, the Institute of Astronomy RAS in Moscow, and the University of Vienna. This research has made use of the SIMBAD database, operated at CDS, Strasbourg, France. J.D.A.G. would like to thank R. Cameron and M. Sch\"{u}ssler for providing the solar azimuthal magnetic field reconstruction. This research has made use of NASA's Astrophysics Data System. J.D.A.G. and E.M.A.G. were partially supported by HST GO-15299 and GO-15512 grants. E.M.A.G and K.P.\ acknowledge support from the German \textit{Leibniz-Gemeinschaft} under project number P67/2018 and support from the European Research Council (ERC) under grant agreement 101170037 (Evaporator). J.S.F. acknowledges support from the Agencia Estatal de Investigaci\'on of the Spanish Ministerio de Ciencia, Innovaci\'on y Universidades through grant PID2022-137241NB-C42.

\end{acknowledgements}

\section*{Data availability}
The raw data associated to this project can be obtained from the ESO archive (\url{http://archive.eso.org}). The reduced spectropolarimetric sequences can be downloaded from the Polarbase repository (\url{https://www.polarbase.ovgso.fr}). The three sets of ZDI maps can be found in machine-readable format in the Zenodo repository with \href{https://doi.org/10.5281/zenodo.17251923}{DOI: 10.5281/zenodo.17251923}.

\bibliographystyle{aa}
\bibliography{Biblio} 

\begin{appendix} 
\section{A ZDI view of the evolution of the surface magnetic field}\label{sec:ZDI-maps}

\subsection*{Epoch 1 (10.2015, Campaign day: 0, Rotation: 0 - fixed)}

The large-scale field of this epoch is dominated by the radial component ($B_{\rm R}$), displaying values between $-6.4$~G to $5.1$~G. The field is mostly concentrated in four regions with alternating polarity, covering from $-30^{\circ}$ up to $60^{\circ}$ in latitude and varying extensions in longitude ($\sim$\,$30^{\circ}$ up to $90^{\circ}$). The pole accessible to the observations (North) displays a weak radial field of negative polarity. A band of negative (East-West direction) azimuthal field ($B_{\rm A}$) covers the visible hemisphere of the star, with a maximum field strength of $5.8$~G. The magnitude of the meridional field ($B_{\rm M}$) remains below $30$\% of the peak value associated with the radial component. The ZDI map reproduces the Stokes\,V LSD profiles down to an optimal $\chi^2_{\rm R} = 1.10$.  

\vspace{-0.25cm}
\subsection*{Epoch 2 (12.2015, Campaign day: 74.75, Rotation: 9.67)}

This epoch shows some important evolution in the large-scale field with respect to the previous visit. The field strength increases reaching $\pm$11~G and $-12.0$~G for $B_{\rm R}$ and $B_{\rm A}$, respectively. The radial field displays a negative polarity region towards higher latitudes and the pole, with patches of positive polarity located at lower latitudes ($\pm30^{\circ}$). The band of negative polarity in $B_{\rm A}$, observed in Epoch~1, appears stronger and more concentrated in latitude ($0^{\circ}$ to $30^{\circ}$), while two weaker bands of positive polarity (West-East direction) in $B_{\rm A}$ can be seen in the remaining latitudes. The magnitude of $B_{\rm M}$ reaches up to $70$\% of the maximum radial field albeit much of this increase is associated with cross-talk. A $\chi^2_{\rm R} = 1.55$ is achieved by this ZDI reconstruction.

\vspace{-0.25cm}
\subsection*{Epoch 3 (02.2016, Campaign day: 135.76, Rotation: 17.56)}

The third map of the campaign reveals similarities with the reconstruction from Epoch 2. The maximum field strength remains close to the $\pm10$~G level for $B_{\rm R}$ and around $-11$~G in $B_{\rm A}$. A strong negative radial field is established at the north pole with several weaker mixed-polarity regions distributed close to the equator. The azimuthal field preserves the geometry from Epoch~2, although some weakening is visible in the high-latitude positive $B_{\rm A}$ band. The maximum strength of $B_{\rm M}$ decreases down to $40$\% of the peak radial field. The ZDI fit to the observations in this epoch reaches a $\chi^2_{\rm R} = 1.75$.  

\vspace{-0.25cm}
\subsection*{Epoch 4 (06.2016, Campaign day: 262.14, Rotation: 33.91)}

Some weakening of the field in all components is observed in this epoch, with values down to $\pm7$~G and $\pm4$~G for $B_{\rm R}$ and $B_{\rm A}$, respectively. While the geometry in the radial field largely resembles the previous visit, the strong mid-latitude ring of azimuthal field is not visible anymore whereas the high-latitude one strengthens. By examining the ZDI distributions obtained in Epochs 3 and 5, we believe that the weakening of the field could be partially introduced by the relatively poor phase coverage we had in this visit (no observations within $0 < \Phi \lesssim 0.4$, see Fig.~\ref{fig_4}). This also generates significant cross-talk in $B_{\rm M}$, which reaches up to $65$\% of the maximum value for $B_{\rm R}$. The Stokes\,V LSD profiles are reproduced at the $\chi^2_{\rm R} = 1.20$ level in this epoch.  

\vspace{-0.25cm}
\subsection*{Epoch 5 (08.2016, Campaign day: 319.04, Rotation: 41.27)}

The ZDI observations in this visit reveal an almost identical $B_{\rm R}$ distribution as to Epoch~3, and to the geometry retrieved in Epoch~4 for the $B_{\rm A}$ component. The magnitude of the radial field appears within $-8~{\rm G} < B_{\rm R} < 4~{\rm G}$, while for the azimuthal component ranges between $\pm3$~G. The meridional component stays below $40$\% of the maximum radial magnetic field. The associated fit to the Stokes\,V LSD profiles yields a $\chi^2_{\rm R} = 1.05$ for this ZDI reconstruction. 

\vspace{-0.25cm}
\subsection*{Epoch 6 (09.2016, Campaign day: 345.01, Rotation: 44.63)}

\begin{figure*}
\includegraphics[trim=0.5cm 0.8cm 1.0cm 1.2cm, clip=true, width=0.496\linewidth]{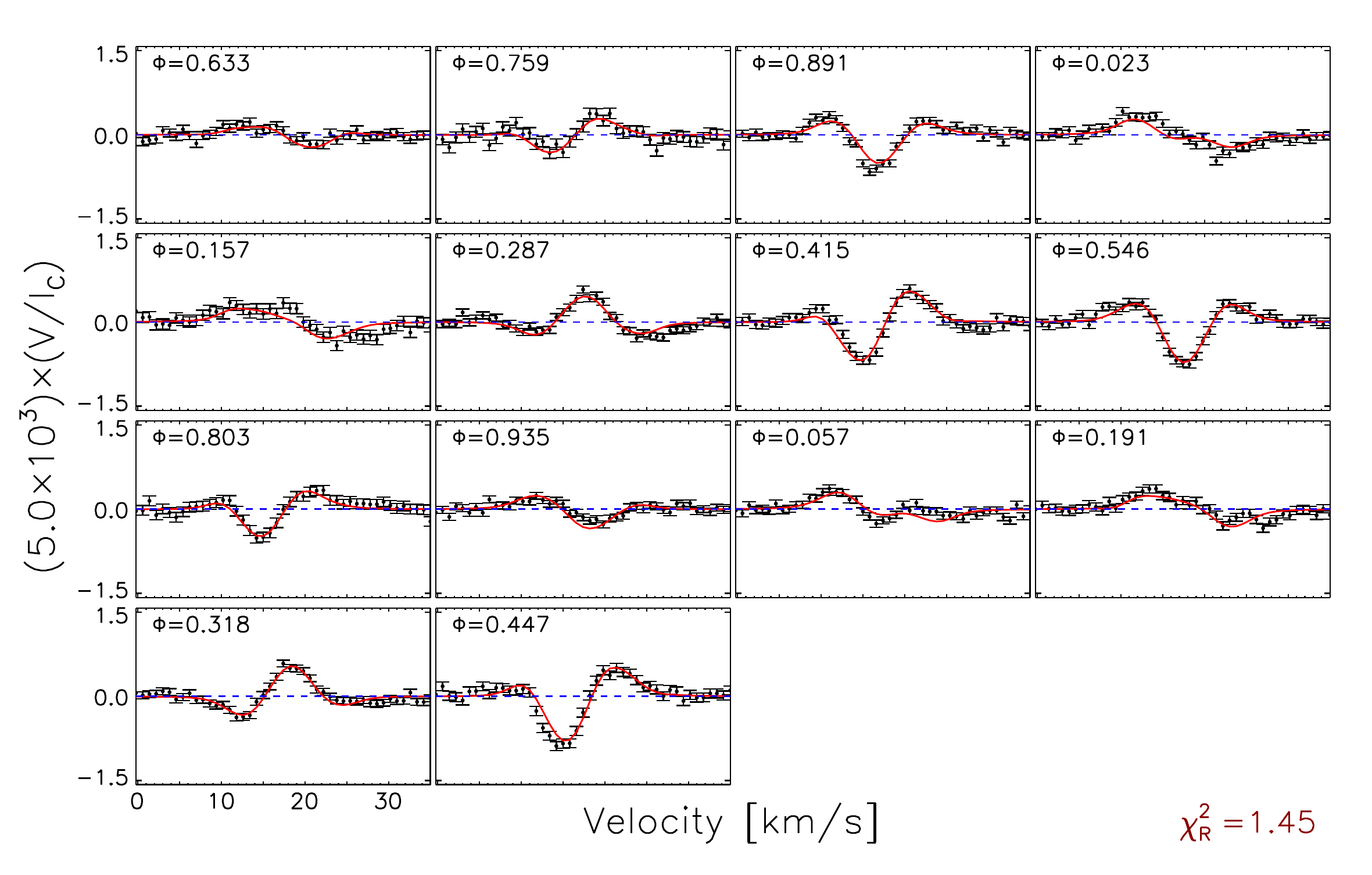}\hspace{2.5pt}\includegraphics[trim=0.5cm 0.8cm 1.0cm 1.2cm, clip=true, width=0.496\linewidth]{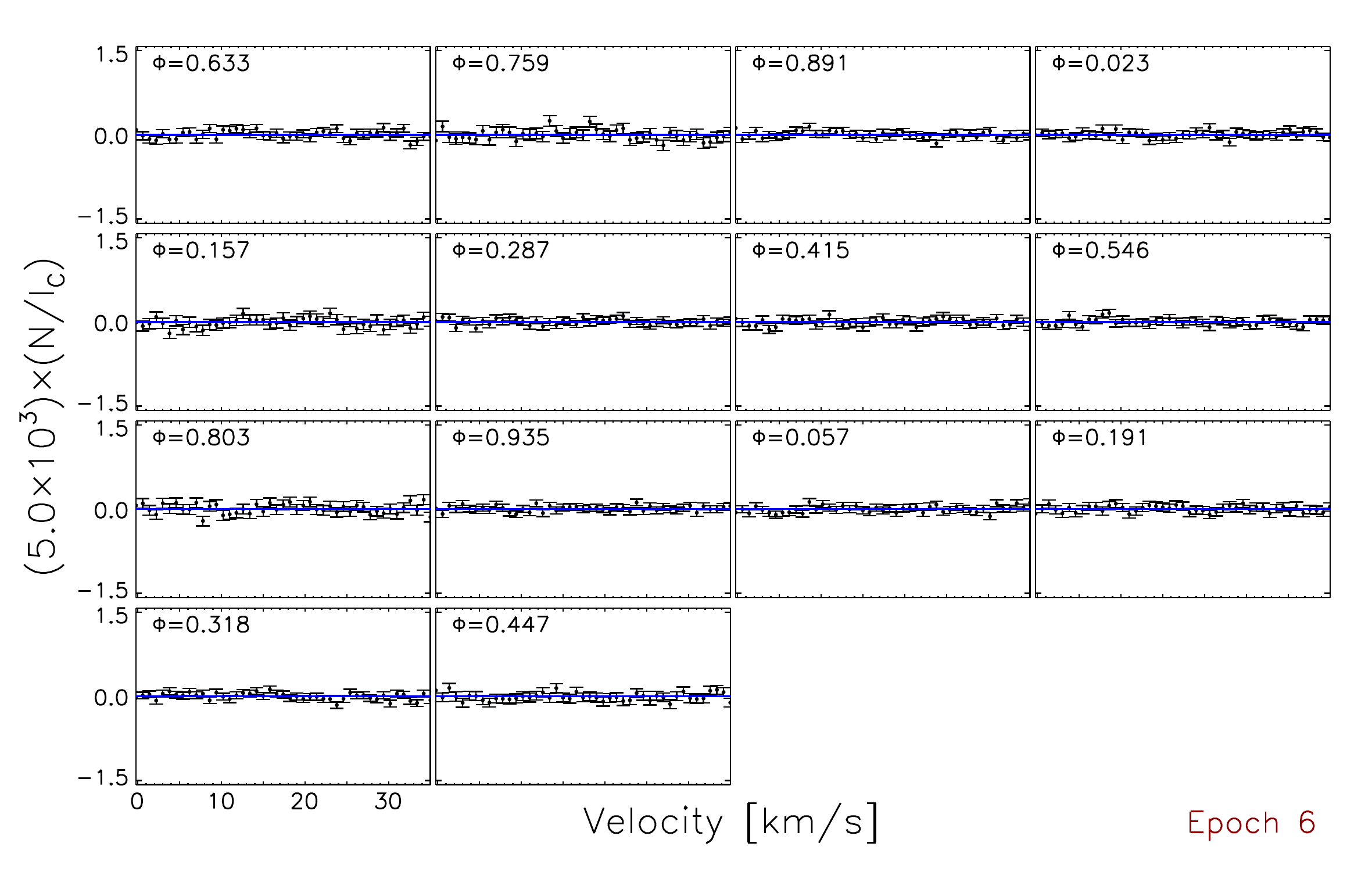}
\includegraphics[trim=0.5cm 0.8cm 1.0cm 0.2cm, clip=true, width=0.496\linewidth]{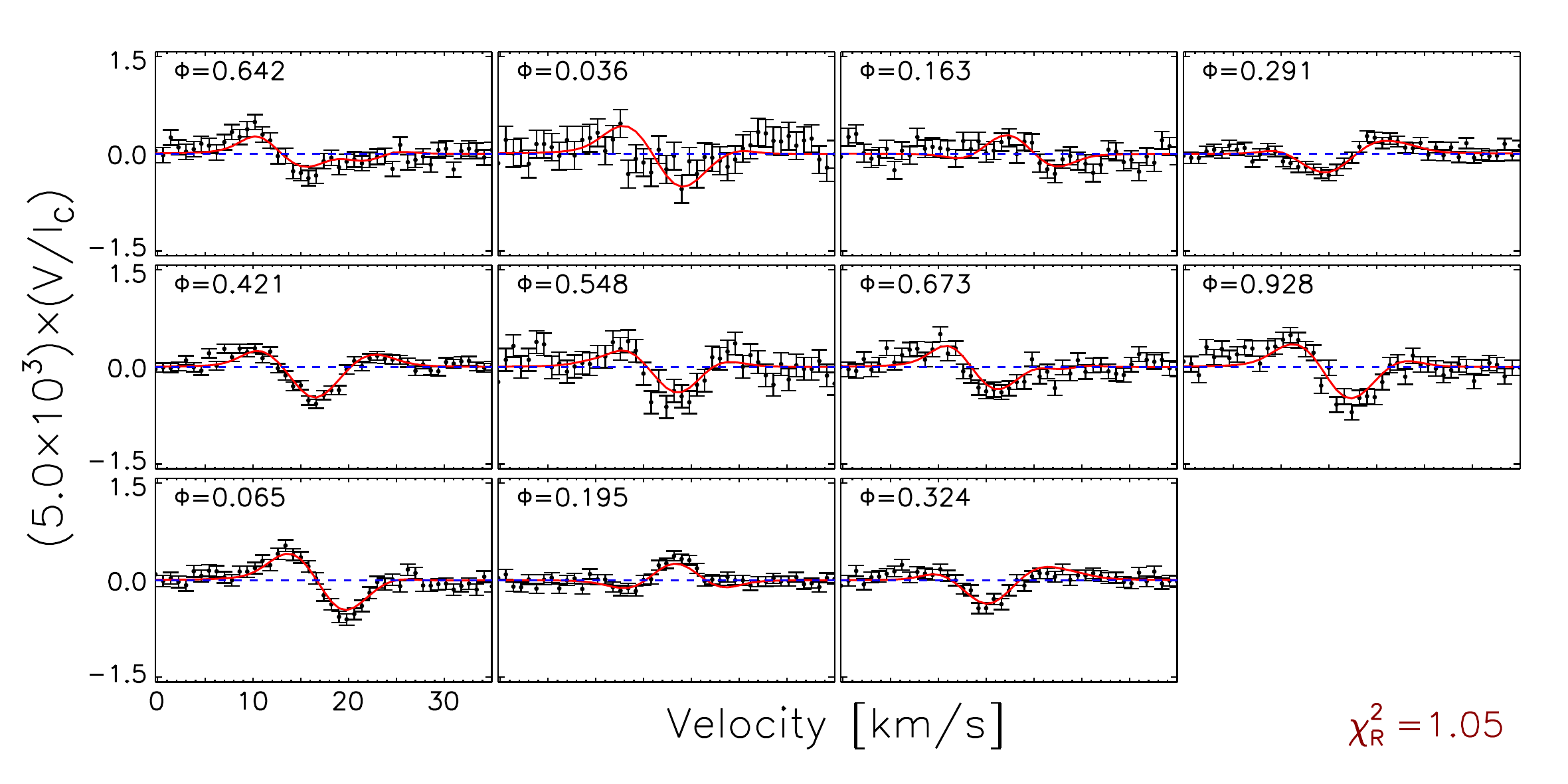}\hspace{2.5pt}\includegraphics[trim=0.5cm 0.8cm 1.0cm 0.2cm, clip=true, width=0.496\linewidth]{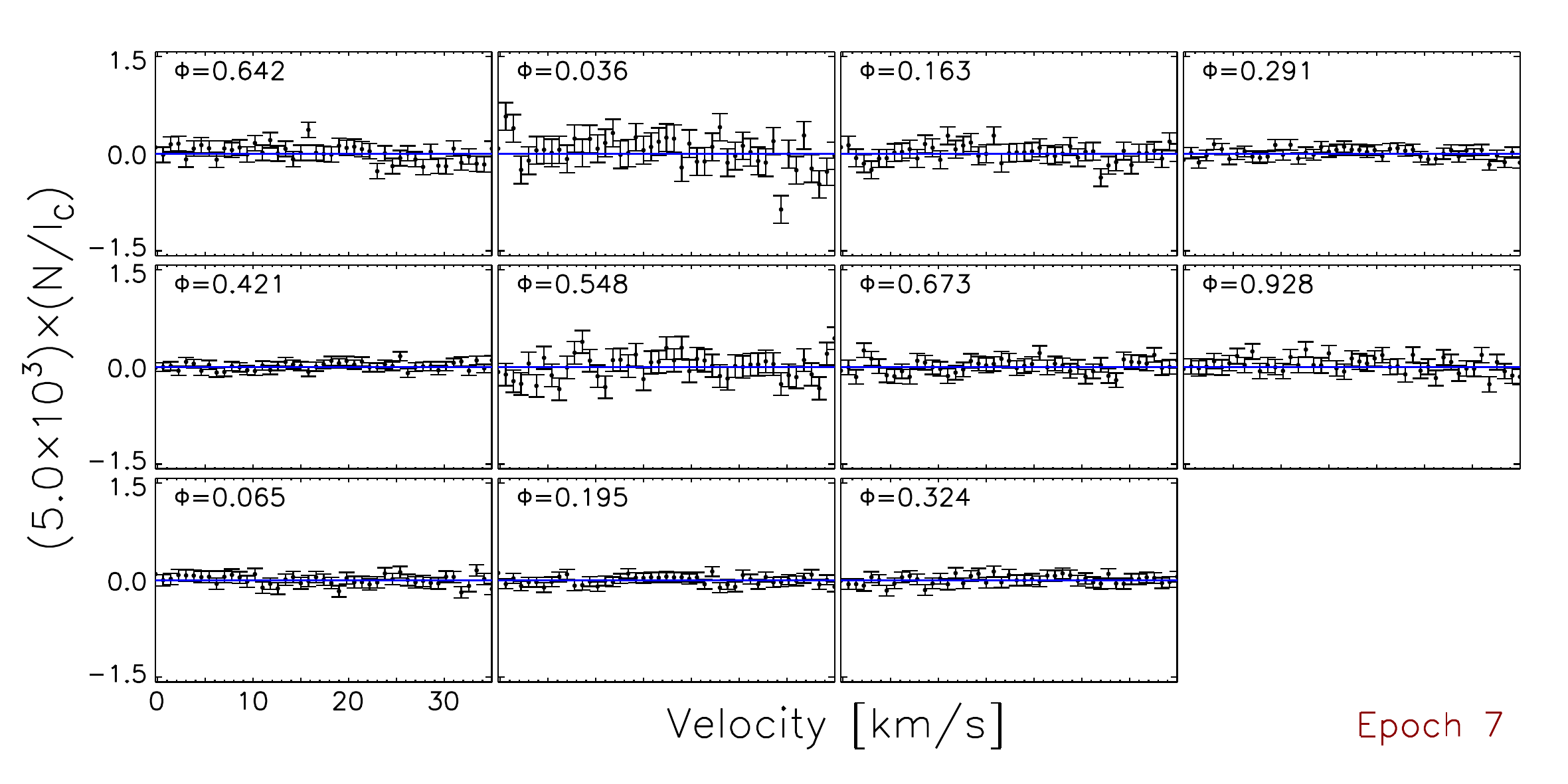}
\includegraphics[trim=0.5cm 0.8cm 1.0cm 0.2cm, clip=true, width=0.496\linewidth]{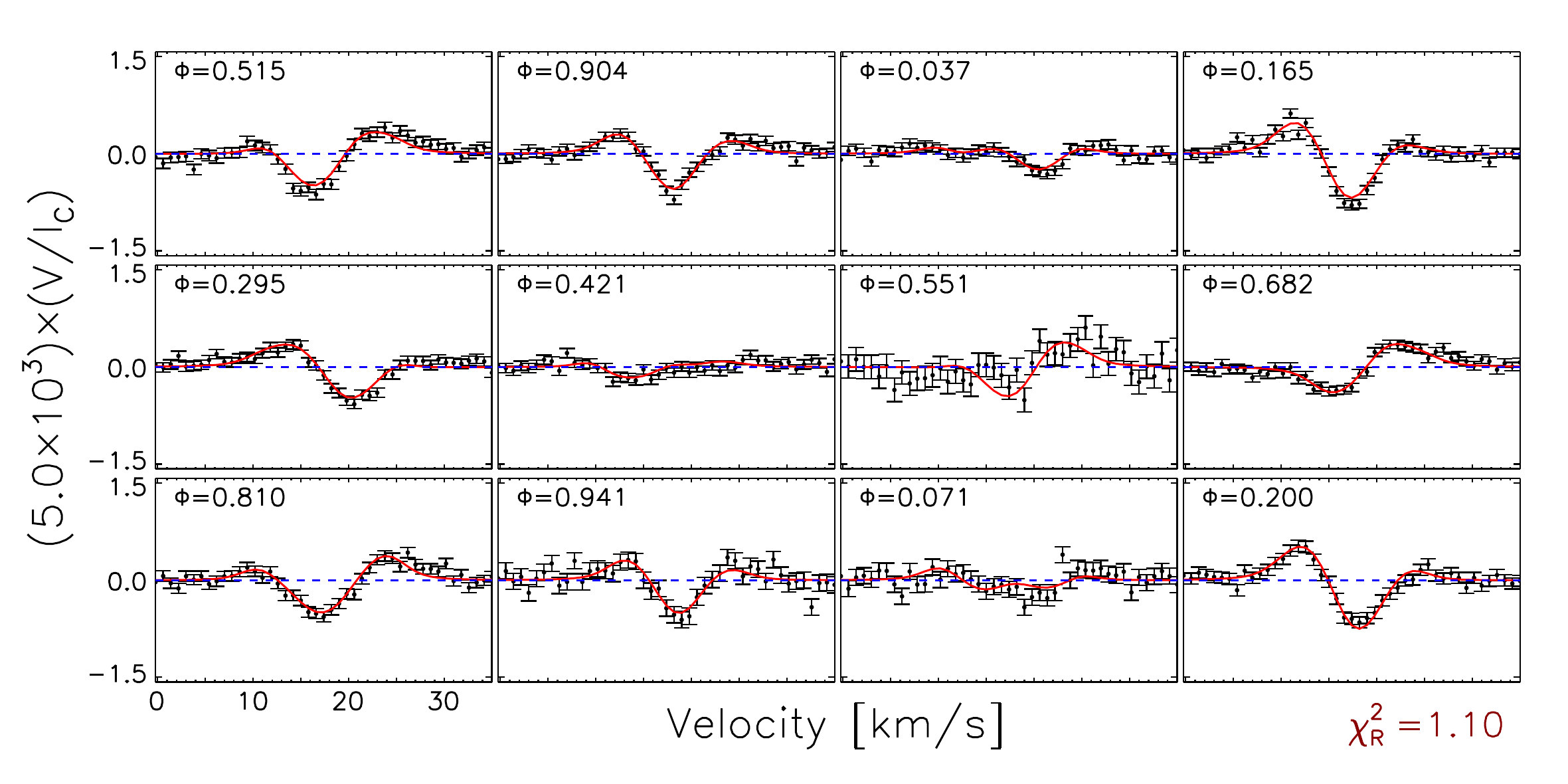}\hspace{2.5pt}\includegraphics[trim=0.5cm 0.8cm 1.0cm 0.2cm, clip=true, width=0.496\linewidth]{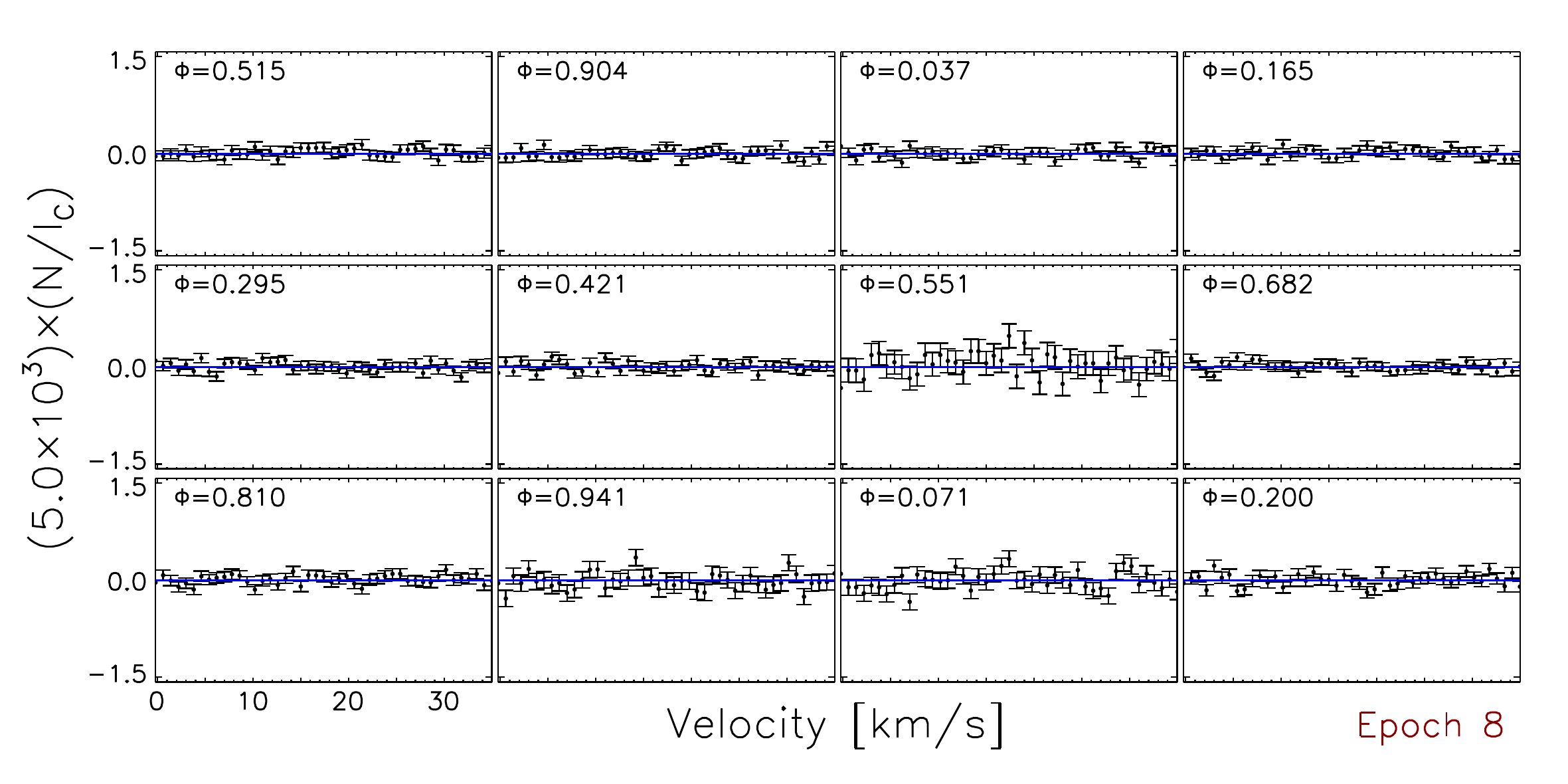}
\includegraphics[trim=0.5cm 6.1cm 1.0cm 0.2cm, clip=true, width=0.496\linewidth]{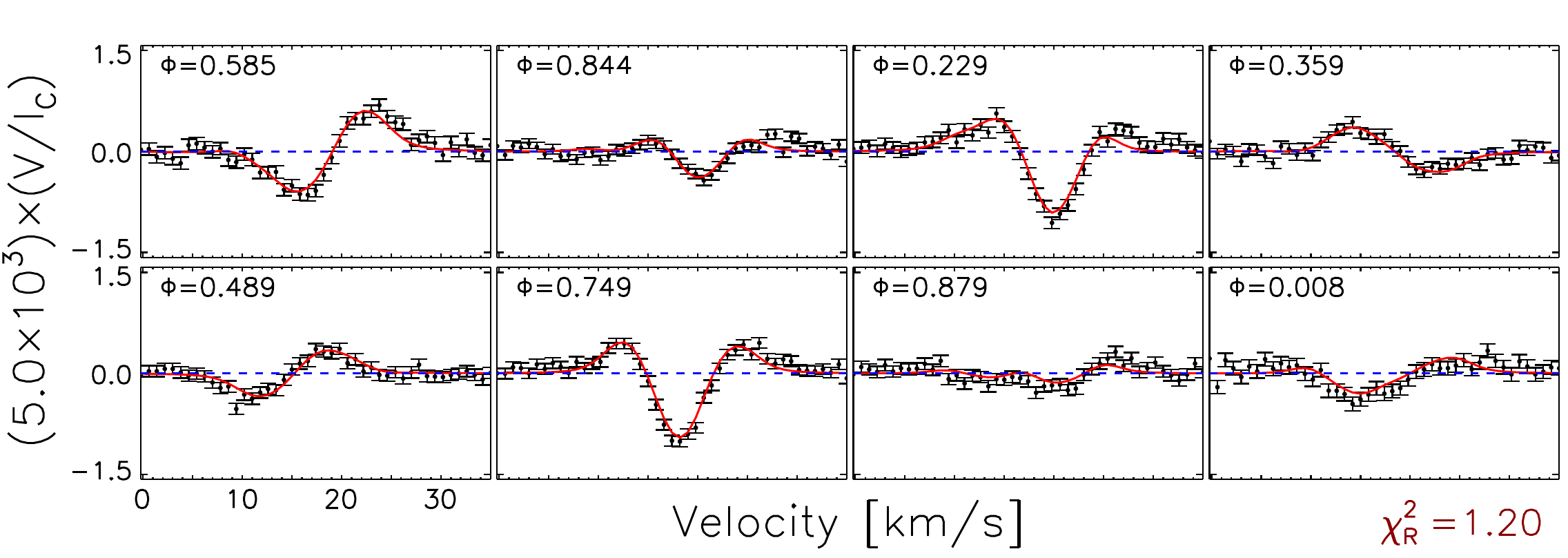}\hspace{2.5pt}\includegraphics[trim=0.5cm 6.1cm 1.0cm 0.2cm, clip=true, width=0.496\linewidth]{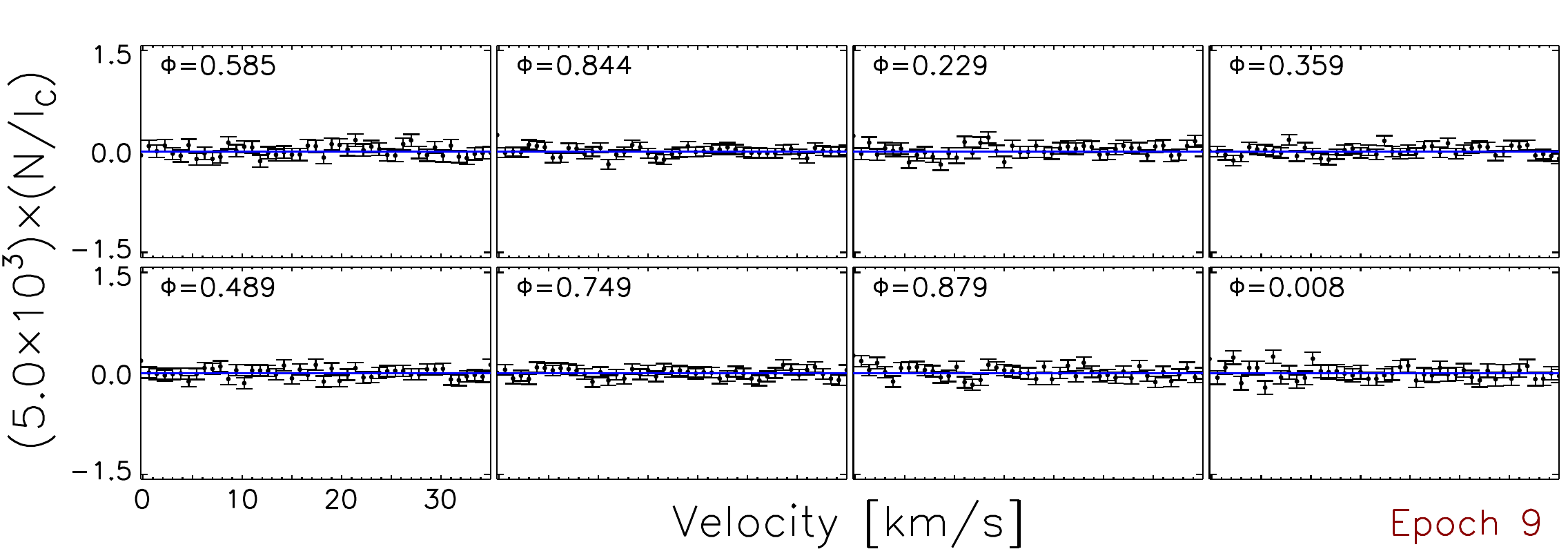}
\includegraphics[trim=0.5cm 0.8cm 1.0cm 0.2cm, clip=true, width=0.496\linewidth]{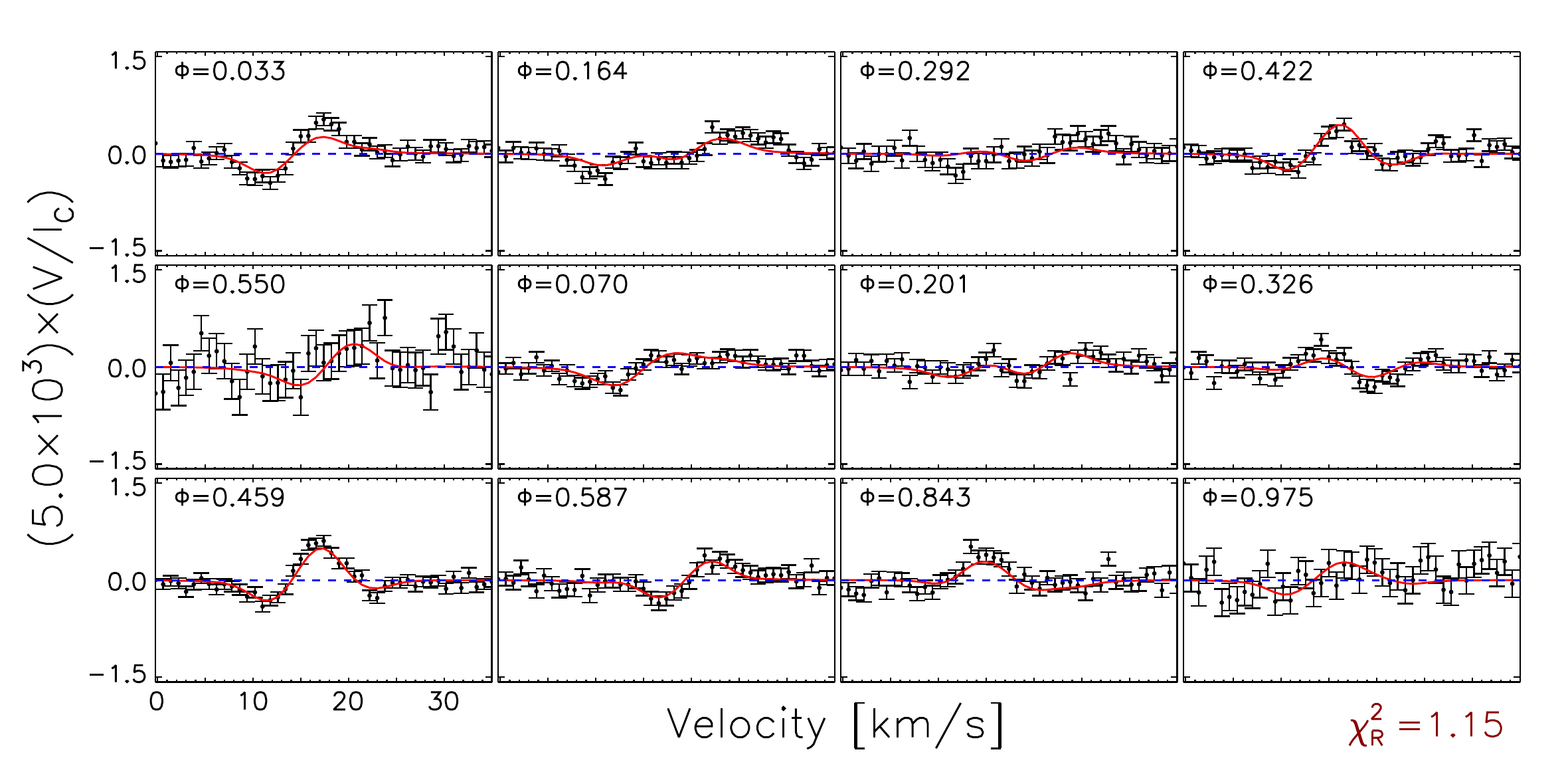}\hspace{2.5pt}\includegraphics[trim=0.5cm 0.8cm 1.0cm 0.2cm, clip=true, width=0.496\linewidth]{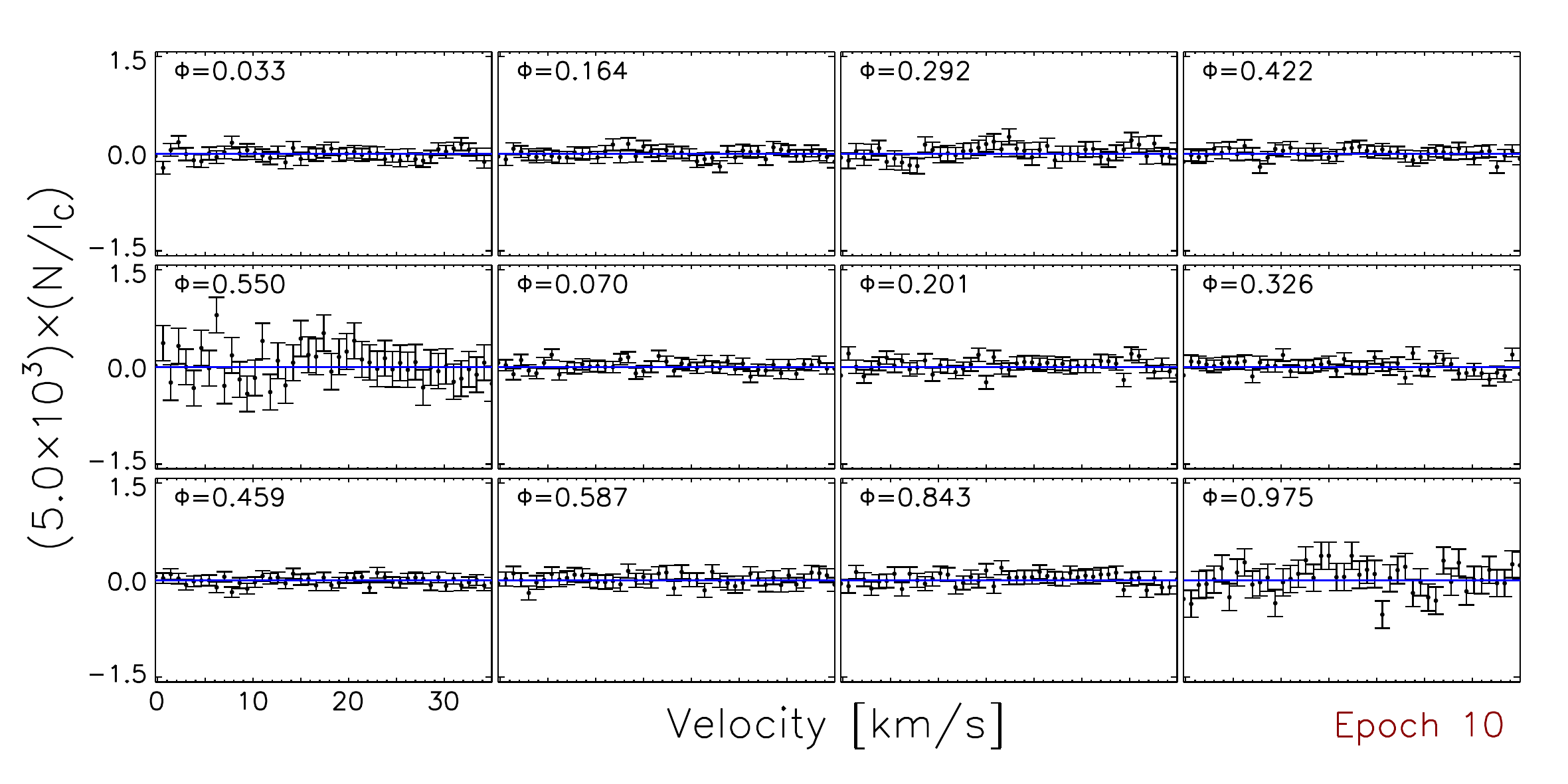}
\caption{Recovered LSD profiles from the spectropolarimetric observations of $\iota$~Hor (Epochs 6 to 10). See caption of Fig.~\ref{fig_4}.}\label{fig_5}
\end{figure*}

The geometry of the large-scale field does not show major evolution with respect to Epoch 5 ($\sim$\,$3.5$ stellar rotations separate both visits). The radial component strengthens to $\pm9$~G while a high-latitude positive band of azimuthal field starts to form, with a field magnitude up to $4$~G. The meridional component resembles closely the one observed in the previous visit. The large-scale field reconstruction reproduces the LSD profiles at the $\chi^2_{\rm R} = 1.45$ level. Notice that with 14 Stokes\,V profiles, this epoch has more exposures than any other visit of the campaign (Fig.~\ref{fig_5}, top).     

\vspace{-0.25cm}
\subsection*{Epoch 7 (10.2016, Campaign day: 376.00, Rotation: 48.64)}

As can be seen in the top-panel of Fig.~\ref{fig_2}, the distribution of the large-scale field is not too dissimilar from the previous epoch of observations. An important weakening of the radial component takes place ($|B_{\rm R}| < 5$~G) while the opposite occurs to the positive band of the azimuthal field, which is now fully formed at mid-latitudes ($B_{\rm A}^{\rm max} \simeq 6.0$~G). The magnitude of the meridional component remains below $35\%$ of the radial field. While the ZDI maps are similar to the first observed epoch of the campaign (separated by roughly one year), the reversal of the dominant polarity is $B_{\rm A}$ is now evident. The magnetic field reconstruction reproduces the Stokes\,V LSD profiles down to a $\chi^2_{\rm R} = 1.05$. 

\vspace{-0.25cm}
\subsection*{Epoch 8 (12.2016, Campaign day: 436.86, Rotation: 56.51)}

This epoch reveals the first polarity reversal in the radial component of the large-scale magnetic field. The polar and equatorial regions are dominated by the positive and negative polarity in $B_{\rm R}$, respectively, with field strength values ranging between $\pm5$~G. The ring-like structure of positive polarity in $B_{\rm A}$ persists and strengthens up to $7$~G. The peak magnitude of $B_{\rm M}$ reaches close to $50\%$ of the radial field, indicating that some cross-talk is present in the reconstruction. A $\chi^2_{\rm R} = 1.10$ results from the model-to-data comparison for this visit.   

\vspace{-0.25cm}
\subsection*{Epoch 9 (02.2017, Campaign day: 483.78, Rotation: 62.58)}

A global increase in the field strength is observed during this epoch, with little variation in terms of geometry with respect to the previous visit. The radial component reaches values close to $\pm10$~G, while the band of positive field in $B_{\rm A}$ peaks at $9$~G. The retrieved values in $B_{\rm M}$ remain within $50\%$ of the radial field strength. The ZDI solution provides an LSD Stokes\,V fit at the $\chi^2_{\rm R} = 1.20$ level.

\vspace{-0.25cm}
\subsection*{Epoch 10 (07.2017, Campaign day: 634.11, Rotation: 82.03)}

As can be seen in Fig.~\ref{fig_2}, while the topology of the radial field persists compared to the last two epochs, a dramatic variation in the azimuthal field is observed in this visit. Not only the magnitude of the field decreases ($|B_{\rm A}| < 3.5$~G) but also the mid-latitude band of positive azimuthal field is almost completely replaced by relatively weak mix-polarity regions in $B_{\rm A}$. Still, an overall dominance of the negative polarity is seen, indicating in this way the onset of the second polarity reversal of this component over the course of the cycle. The radial component also shows some minor weakening, with $|B_{\rm R}| \simeq \pm7.0$~G. The meridional component remains below $50\%$ of the radial field strength. The final ZDI reconstruction provides a $\chi^2_{\rm R} = 1.15$ to the spectropolarimetric data set.

\vspace{-0.25cm}
\subsection*{Epoch 11 (08.2017, Campaign day: 661.11, Rotation: 85.53)}

Being only $3.5$ stellar rotations apart, the large-scale field in this epoch retains most of the properties observed in the previous visit. The field strengths in all components display similar values as in Epoch 10, with just a subtle increase in the surface coverage of the azimuthal field with negative polarity. An optimal $\chi^2_{\rm R} = 0.90$ is achieved from the associated ZDI fit of the observations in this visit.

\vspace{-0.25cm}
\subsection*{Epoch 12 (09.2017, Campaign day: 713.10, Rotation: 92.25)}

Minor evolution of the large-scale magnetic field is observed in this epoch. The radial and azimuthal components remain within $B_{\rm R} \simeq \pm5.8$~G and $-3.9$~G $\leq B_{\rm A} \leq 0.8$~G. While the geometry of $B_{\rm R}$ stays largely unaltered, the distribution of $B_{\rm A}$ displays a clear establishment of a negative polarity band at mid-latitudes (analogous to the one observed in Epochs 7 - 9). The meridional component stays under $45\%$ of the radial field peak value. This ZDI reconstruction achieves a $\chi^2_{\rm R} = 1.05$ in comparison with the Stokes\,V LSD profiles. 

\vspace{-0.25cm}
\subsection*{Epoch 13 (11.2017, Campaign day: 768.82, Rotation: 99.46)}

The distribution of the large-scale field continues to be dominated by a negative band of azimuthal field, which now covers an entire hemisphere. The radial component displays a similar configuration to the last visit, with a positive field located at the visible pole and latitudes close to the visibility limit. Still, it is possible to observe a small increase in the coverage of negative polarity regions at mid-latitudes. The magnitude of the field ranges between $\pm6$~G for $B_{\rm R}$ and $-7.0$~G $\leq B_{\rm A} \leq 1.4$~G. The meridional component reaches up to $25\%$ of the maximum radial field strength. The ZDI reconstruction reproduces the LSD Stokes\,V profiles down to a $\chi^2_{\rm R} = 1.30$.  

\vspace{-0.25cm}
\subsection*{Epoch 14 (12.2017, Campaign day: 809.83, Rotation: 104.76)}

Close to $105$ stellar rotations since the beginning of the campaign ($\sim2.22$~yr), we detect the second polarity reversal in the radial component of the large-scale magnetic field. As such, the visible pole returns to a negative polarity in the radial field, while latitudes below the equator and close to the visibility limit display positive $B_{\rm R}$ values. On the azimuthal component, it is possible to see the emergence of a positive polarity region close to the polar region, while the negative band of the azimuthal field, persistent in the last two epochs, strengthens and is pushed to lower latitudes. This global behavior is very similar to what was observed during Epochs 1-3, providing further evidence of the re-initiation of the magnetic cycle. The radial field takes values between $-7.7$~G and $8.7$~G, whereas the azimuthal field varies between $-13.0$~G and $4.4$~G. The meridional field shows a slight increase in magnitude, reaching up to $40\%$ of the maximum azimuthal field. The ZDI fit of the spectropolarimetric data for this epoch achieves a $\chi^2_{\rm R} = 1.30$.      

\vspace{-0.25cm}
\subsection*{Epoch 15 (02.2018, Campaign day: 854.75, Rotation: 110.58)}

After approximately $5$ rotations, the large-scale magnetic field in Epoch 15 roughly matches the configuration observed in the previous visit. This includes the geometry of each of the field components, as well as their magnitudes (i.e., $B_{\rm R} \simeq \pm7.5$~G, $-16.4$~G $\leq B_{\rm A} \leq 2.0$~G). Only minimal variation is observed in the retrieved $B_{\rm M}$ distribution. The ZDI map achieves a $\chi^2_{\rm R} = 1.20$ when compared with the observed Stokes\,V LSD profiles.

\vspace{-0.25cm}
\subsection*{Epoch 16 (06.2018, Campaign day: 992.12, Rotation: 128.35)}

\begin{figure*}
\includegraphics[trim=0.5cm 6.1cm 1.0cm 1.2cm, clip=true, width=0.496\linewidth]{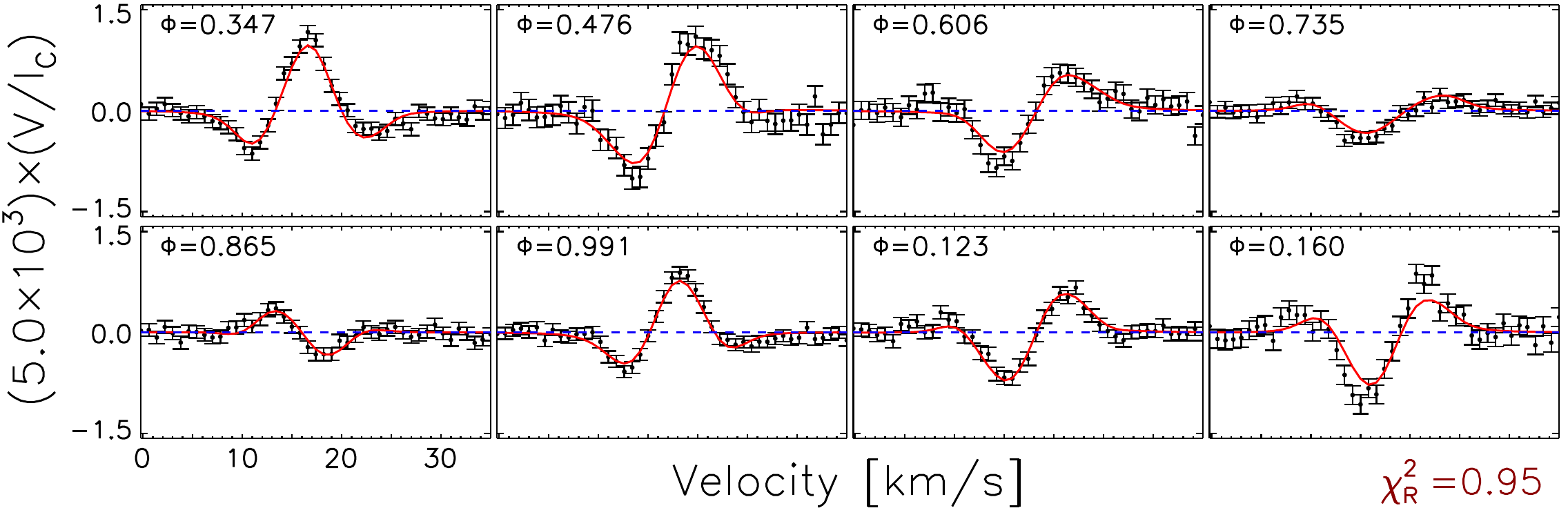}\hspace{2.5pt}\includegraphics[trim=0.5cm 6.1cm 1.0cm 1.2cm, clip=true, width=0.496\linewidth]{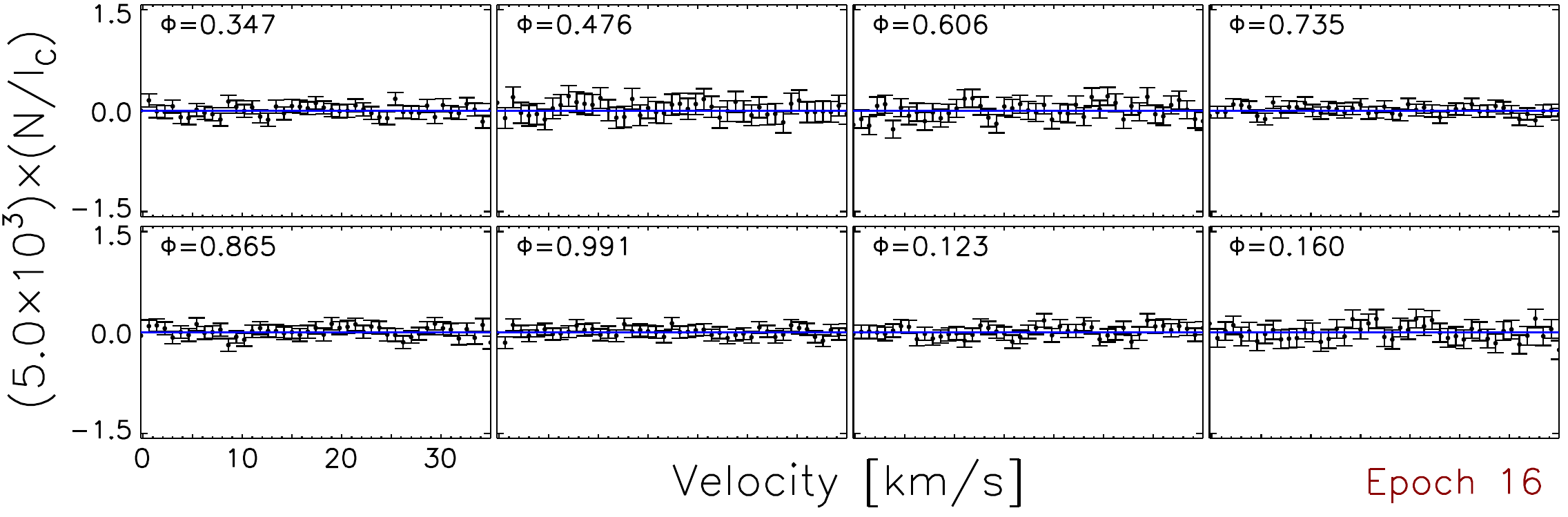}
\includegraphics[trim=0.5cm 6.1cm 1.0cm 0.2cm, clip=true, width=0.496\linewidth]{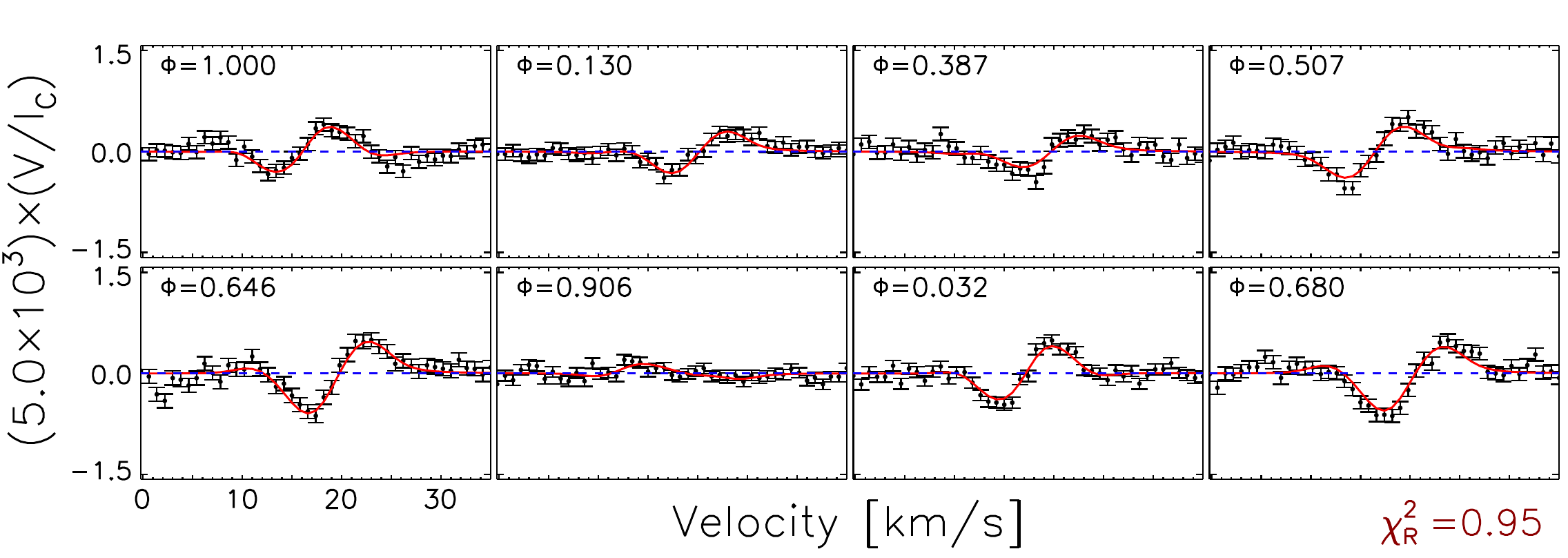}\hspace{2.5pt}\includegraphics[trim=0.5cm 6.1cm 1.0cm 0.2cm, clip=true, width=0.496\linewidth]{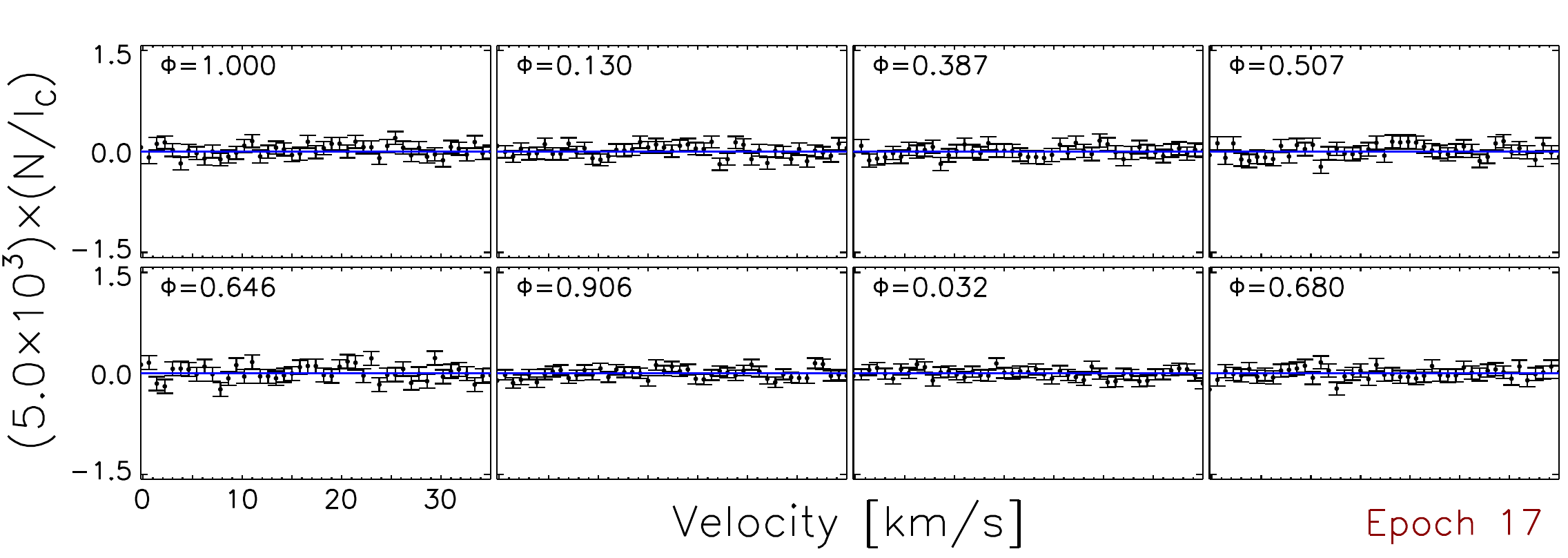}
\includegraphics[trim=0.5cm 0.8cm 1.0cm 0.2cm, clip=true, width=0.496\linewidth]{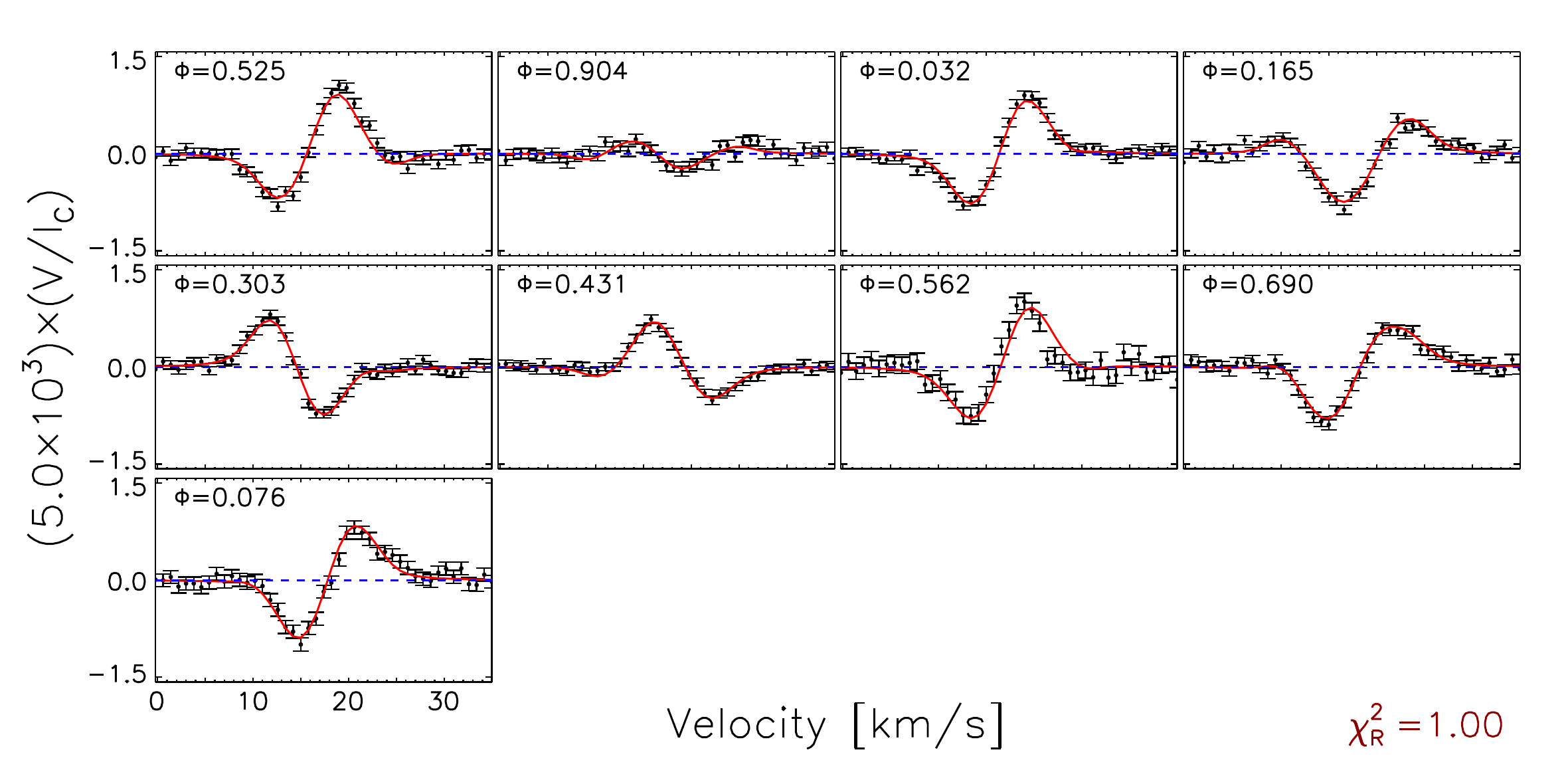}\hspace{2.5pt}\includegraphics[trim=0.5cm 0.8cm 1.0cm 0.2cm, clip=true, width=0.496\linewidth]{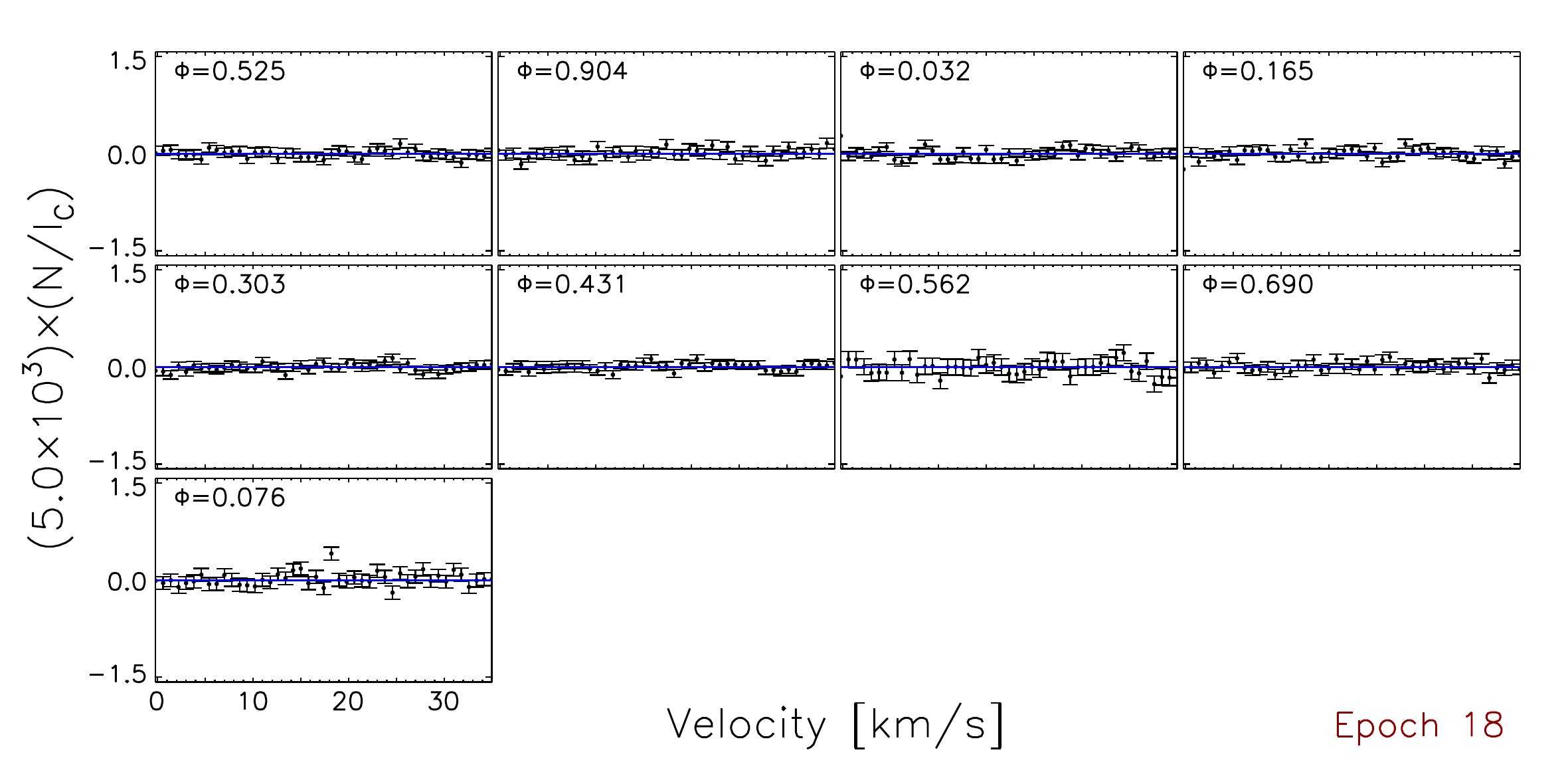}
\caption{Recovered LSD profiles from the spectropolarimetric observations of $\iota$~Hor (Epochs 16 to 18). See caption of Fig.~\ref{fig_4}.}\label{fig_7}
\end{figure*}

The ZDI maps of this epoch reveal the almost complete disappearance of the mid-latitude band of the azimuthal field of negative polarity. As such, the magnitude of this component dramatically decreases, going down to $-4.4$~G $\leq B_{\rm A} \leq 1.9$~G ($\sim$\,$12$~G of variation in comparison with the previous visit). In contrast, the radial component shows similar features as in the previous visit, with a distribution tending to be more dipolar displaying a subtle increase in field strength ($-8.9$~G $\leq B_{\rm R} \leq 7.8$~G). The meridional field stays below $40\%$ of the maximum radial field. As indicated in Fig.~\ref{fig_7} (top panel), this reconstruction reproduces the observed LSD Stokes\,V profiles down to a $\chi^2_{\rm R} = 0.95$. 

\vspace{-0.25cm}
\subsection*{Epoch 17 (08.2018, Campaign day: 1028.09, Rotation: 133.0)}

Relatively close in time to the previous epoch ($<5$~stellar rotations), the large-scale magnetic field maps of this visit indicate a new reversal in the dominant polarity of the azimuthal component (from negative to positive). This reversal is the third one registered in $B_{\rm A}$ during the campaign (with the first two detected between Epochs 3-4 and 9-10), indicating that the next magnetic cycle is well underway during these observations of our program. While the geometry of $B_{\rm R}$ closely resembles the configuration observed in Epoch~16, a global reduction of the field strength is observed with $-5.5$~G $\leq B_{\rm R} \leq 3.5$~G and $-1.5$~G $\leq B_{\rm A} \leq 3.3$~G. The meridional field reaches only up to $35\%$ of the maximum radial magnetic field. The LSD Stokes\,V profiles are reproduced down to an optimal $\chi^2_{\rm R} = 0.95$ by these ZDI reconstructions.

\vspace{-0.25cm}
\subsection*{Epoch 18 (09.2018, Campaign day: 1063.07, Rotation: 137.5)}

In this last observing epoch the ZDI maps do not show any major changes in the geometry or polarity dominance with respect of the behavior observed in the previous two visits. The radial field configuration roughly matches the one observed in Epoch 16 ($\sim$\,10 rotations before), whereas the azimuthal one is more similar to the reconstruction in Epoch 17 ($\sim$\,4.5 rotations before). Still, the dominant polarity in the azimuthal field appears more concentrated in a band-like structure, clearly resembling the behavior observed throughout the campaign. Some strengthening of the field is observed in all components with  
$|B_{\rm R}| \leq \pm12$~G and $-3.8$~G $\leq B_{\rm A} \leq 7.7$~G. Some cross-talk is visible in the meridional component, which reaches up to $50$\% of the maximum radial field strength. These ZDI maps reproduce the observed LSD profiles down to a $\chi^2_{\rm R} = 1.00$. 

\section{Spectropolarimetric data and ZDI fits (Epochs 6-18)}\label{sec:ZDI-fits}

\begin{figure*}
\includegraphics[trim=0.5cm 0.8cm 1.0cm 1.2cm, clip=true, width=0.496\linewidth]{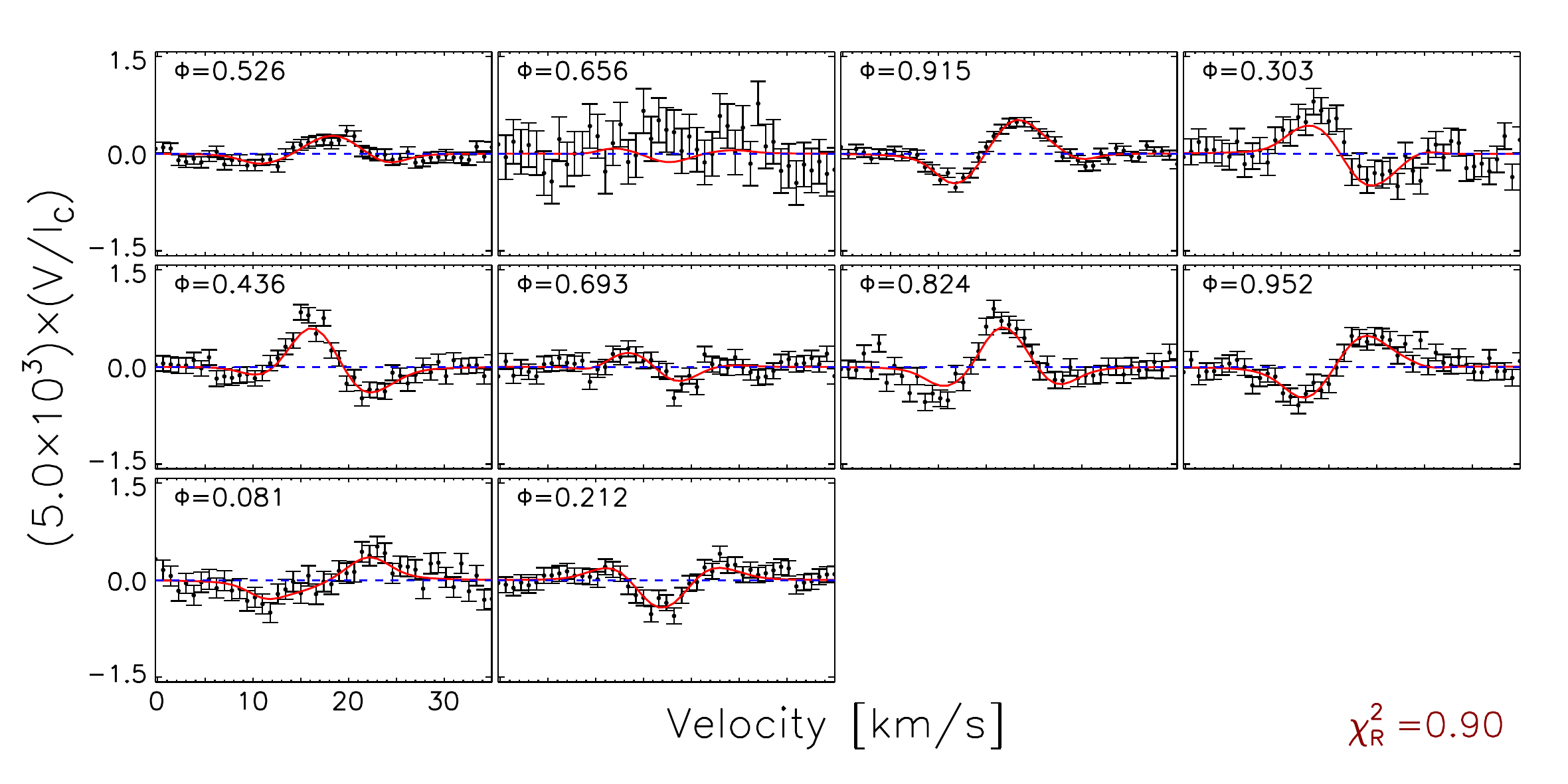}\hspace{2.5pt}\includegraphics[trim=0.5cm 0.8cm 1.0cm 1.2cm, clip=true, width=0.496\linewidth]{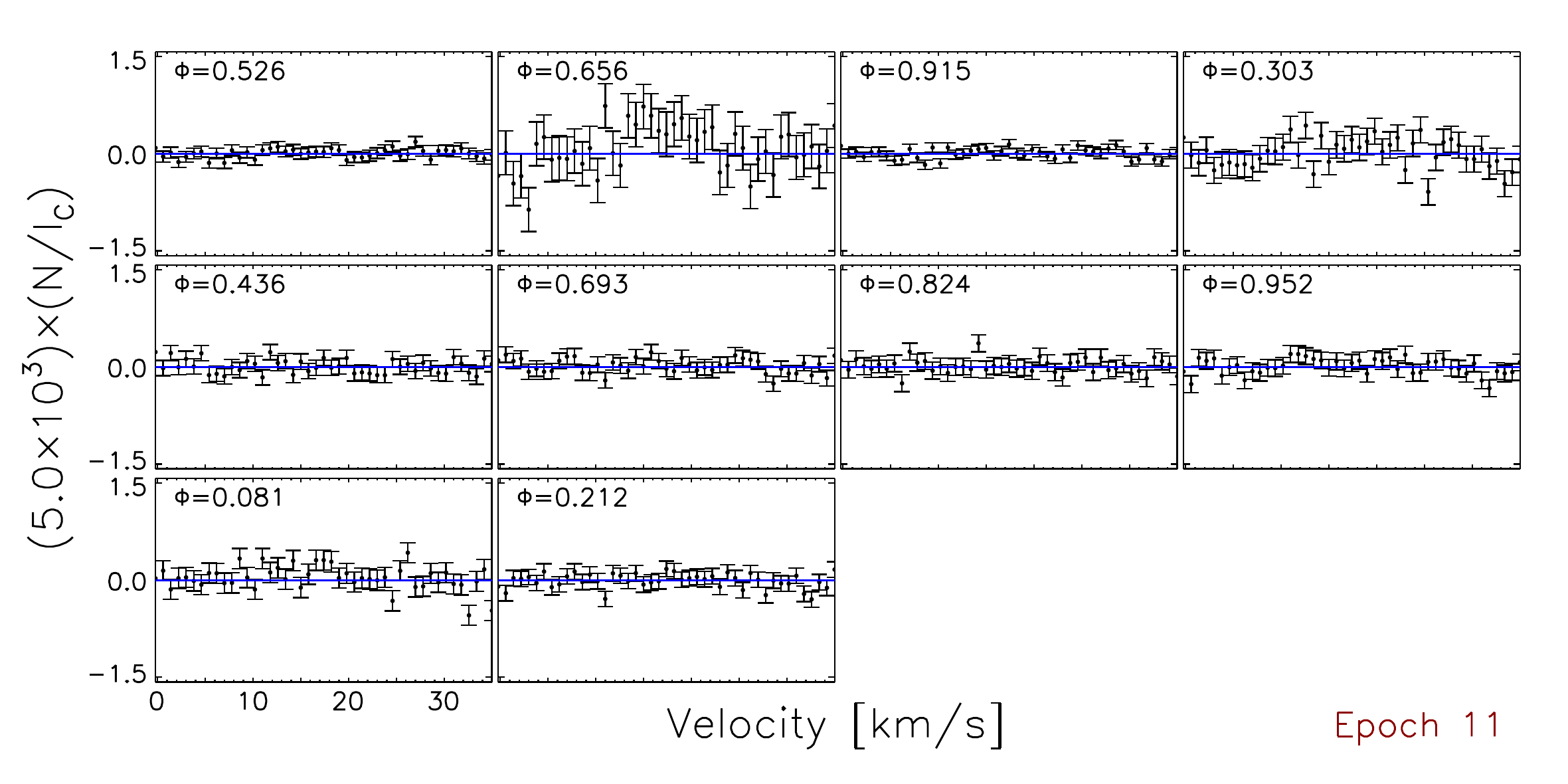}
\includegraphics[trim=0.5cm 0.8cm 1.0cm 0.2cm, clip=true, width=0.496\linewidth]{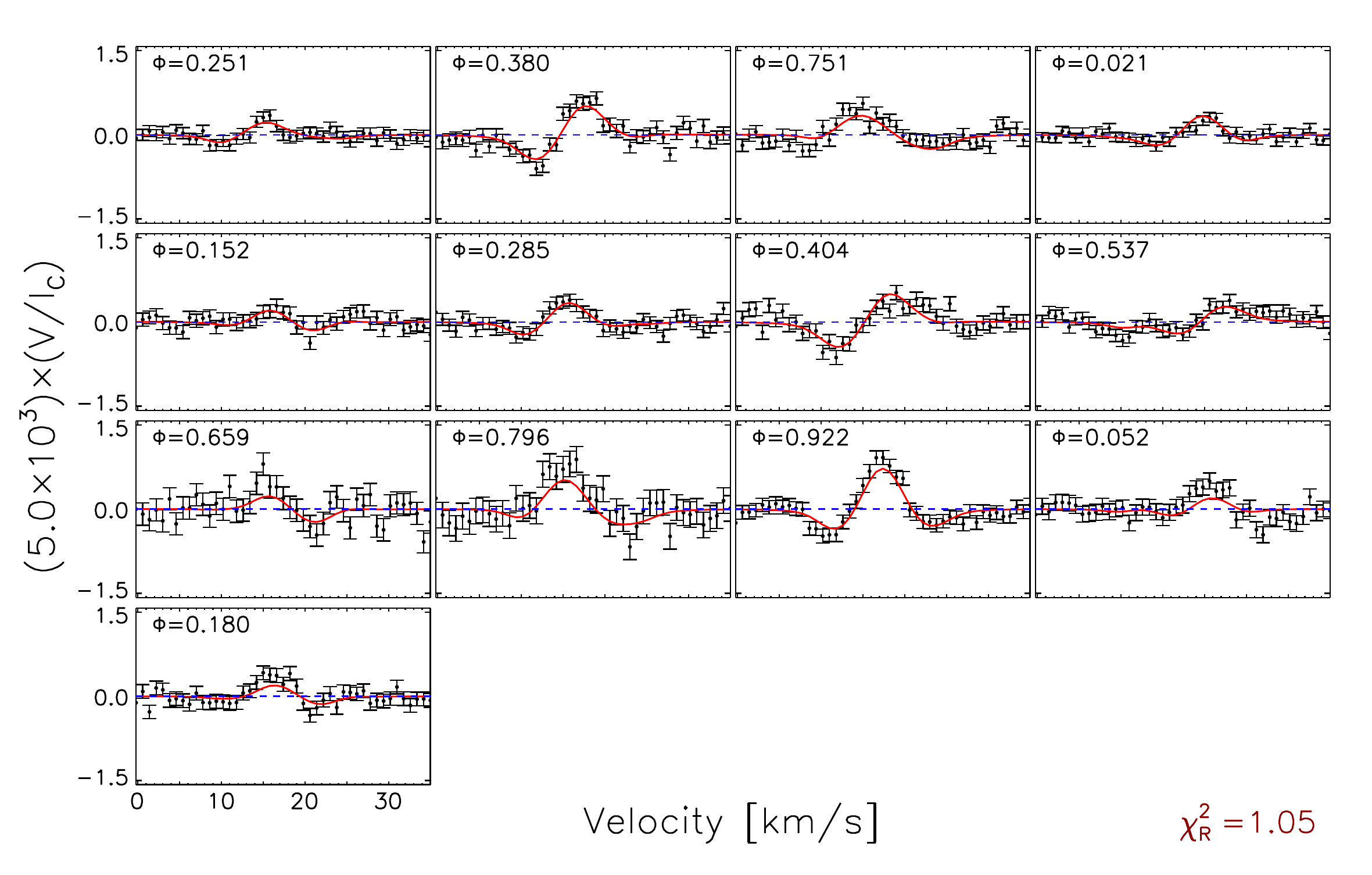}\hspace{2.5pt}\includegraphics[trim=0.5cm 0.8cm 1.0cm 0.2cm, clip=true, width=0.496\linewidth]{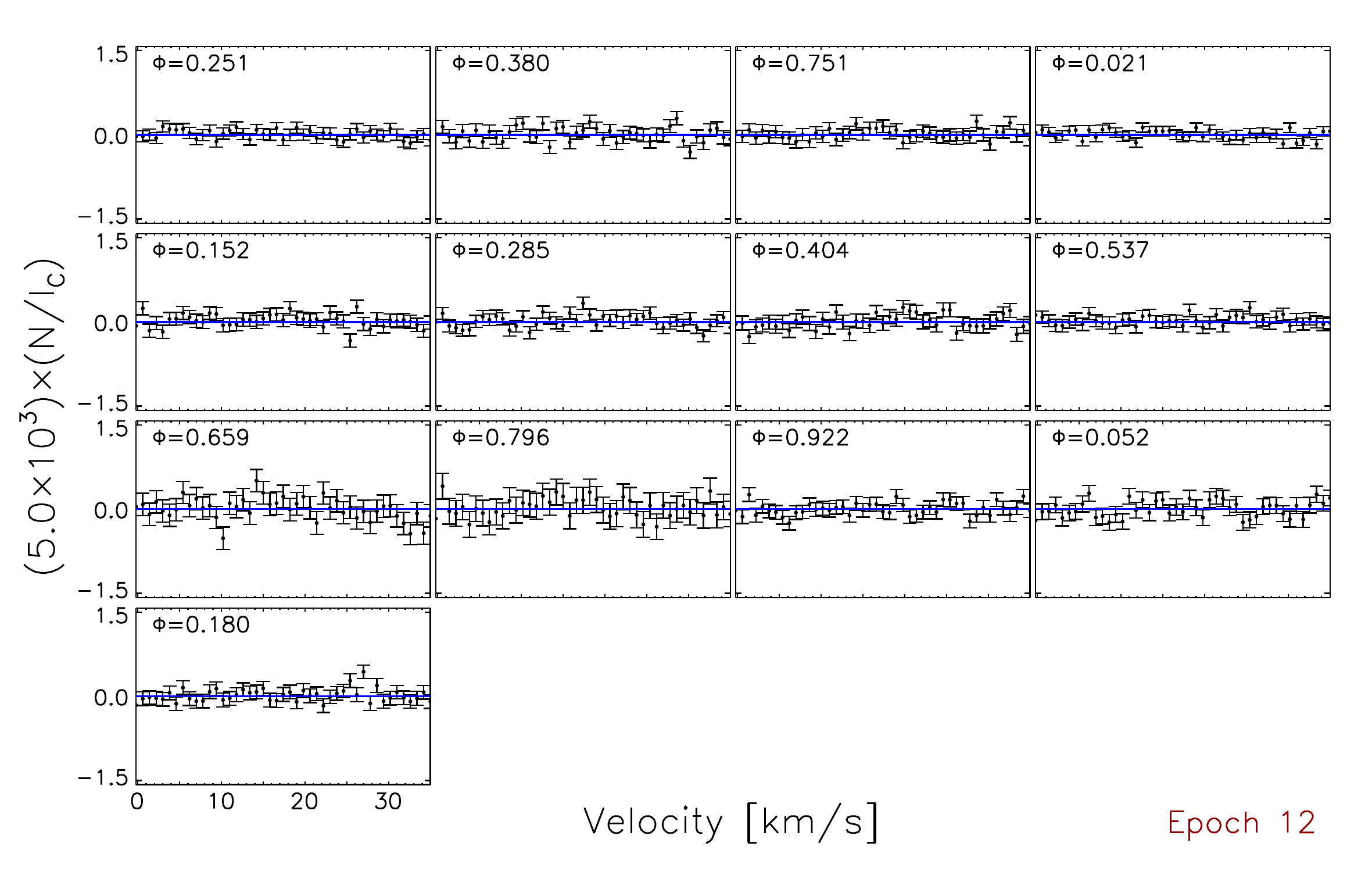}
\includegraphics[trim=0.5cm 0.8cm 1.0cm 0.2cm, clip=true, width=0.496\linewidth]{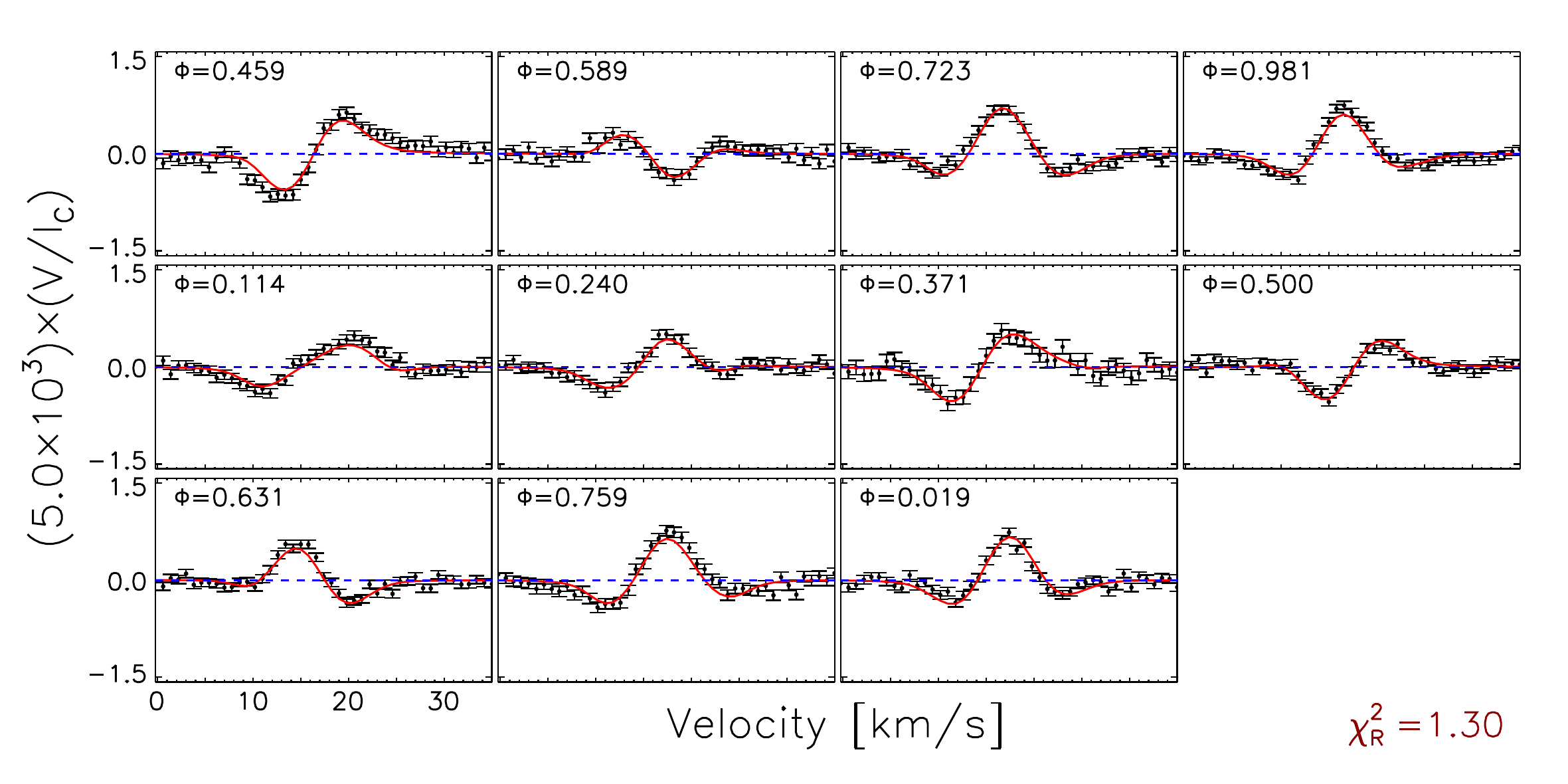}\hspace{2.5pt}\includegraphics[trim=0.5cm 0.8cm 1.0cm 0.2cm, clip=true, width=0.496\linewidth]{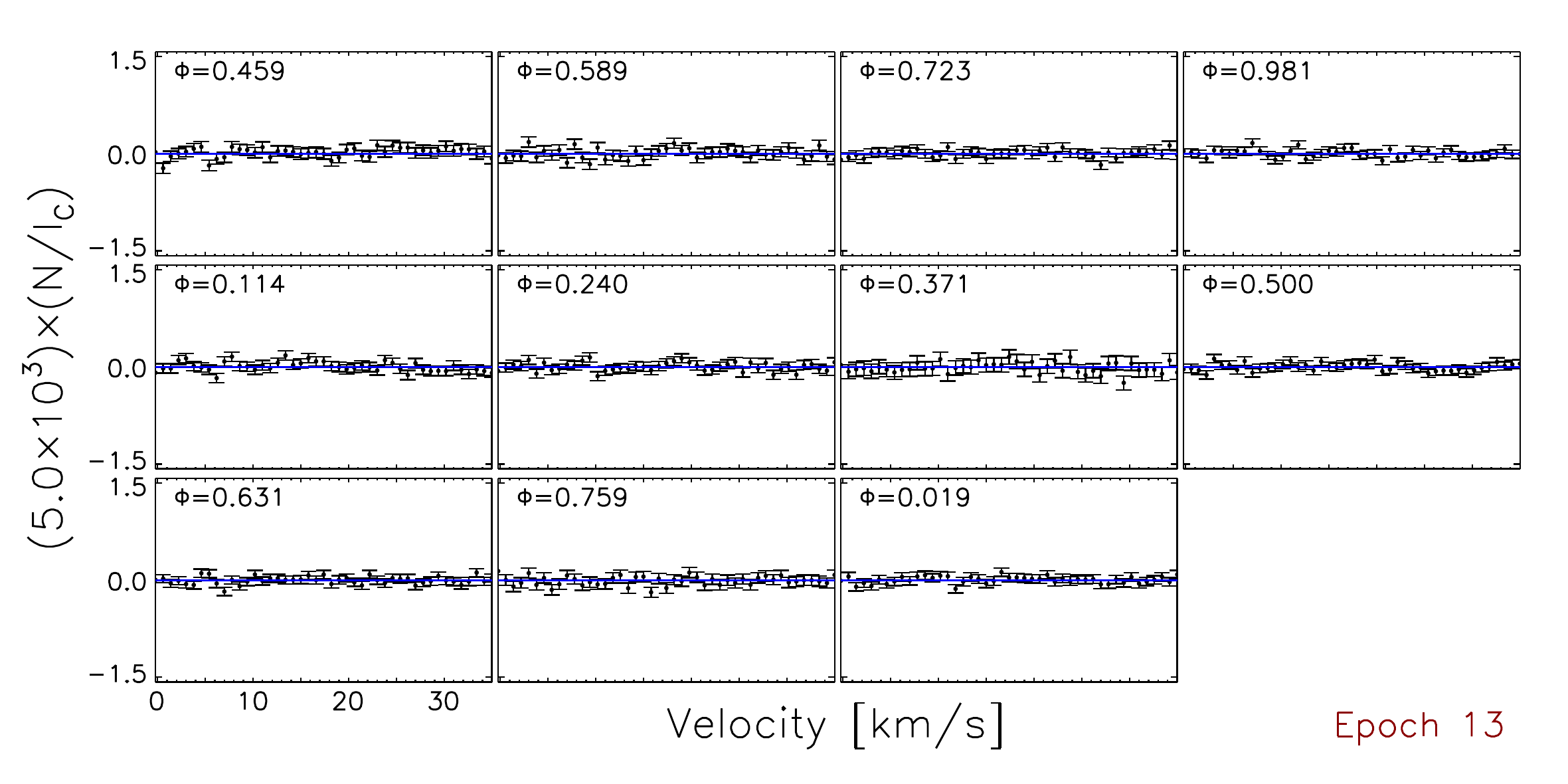}
\includegraphics[trim=0.5cm 6.1cm 1.0cm 0.2cm, clip=true, width=0.496\linewidth]{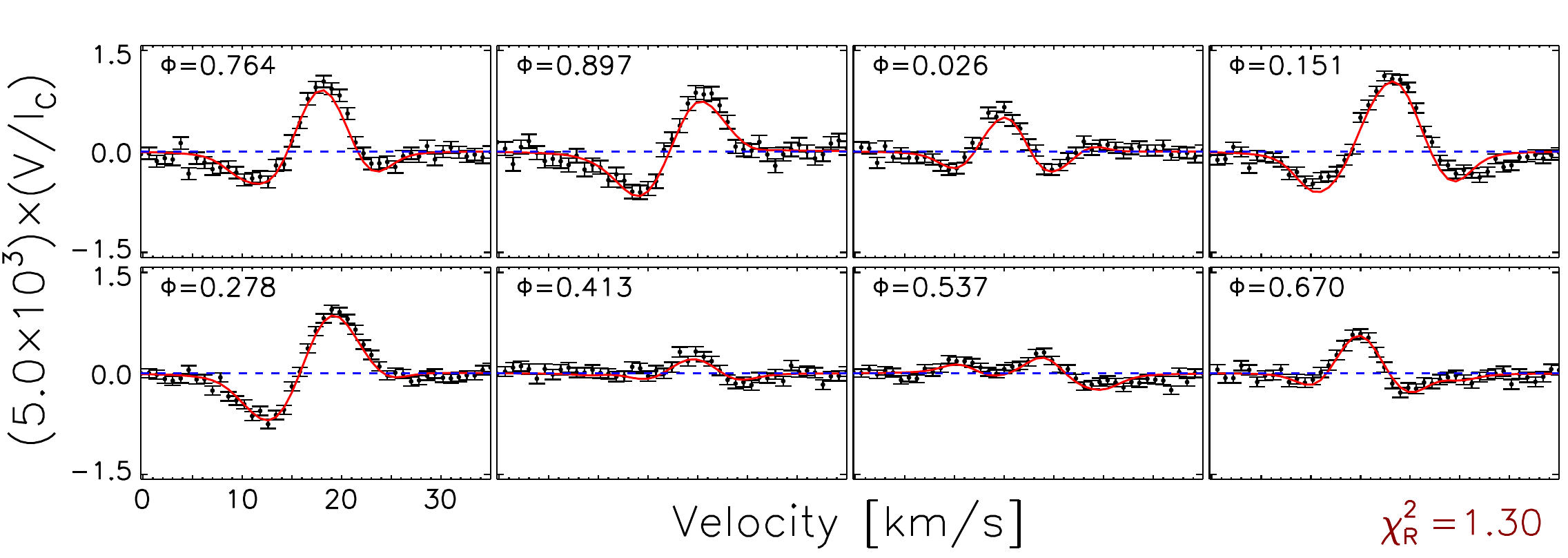}\hspace{2.5pt}\includegraphics[trim=0.5cm 6.1cm 1.0cm 0.2cm, clip=true, width=0.496\linewidth]{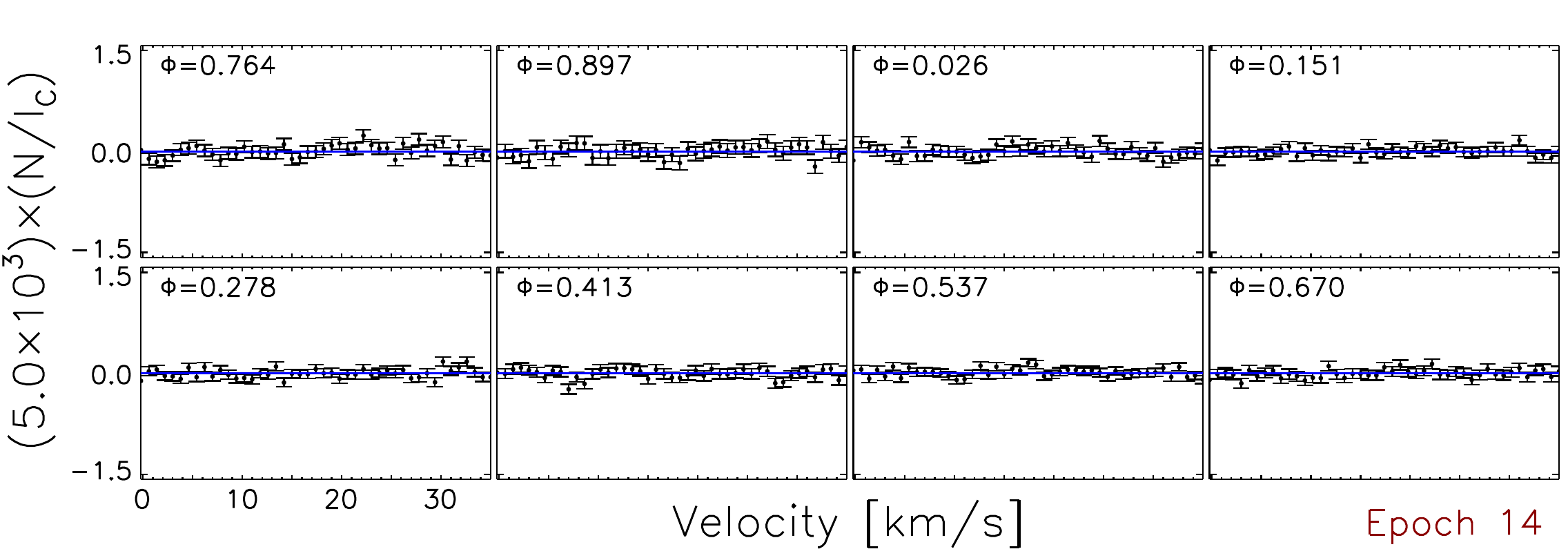}
\includegraphics[trim=0.5cm 6.1cm 1.0cm 0.2cm, clip=true, width=0.496\linewidth]{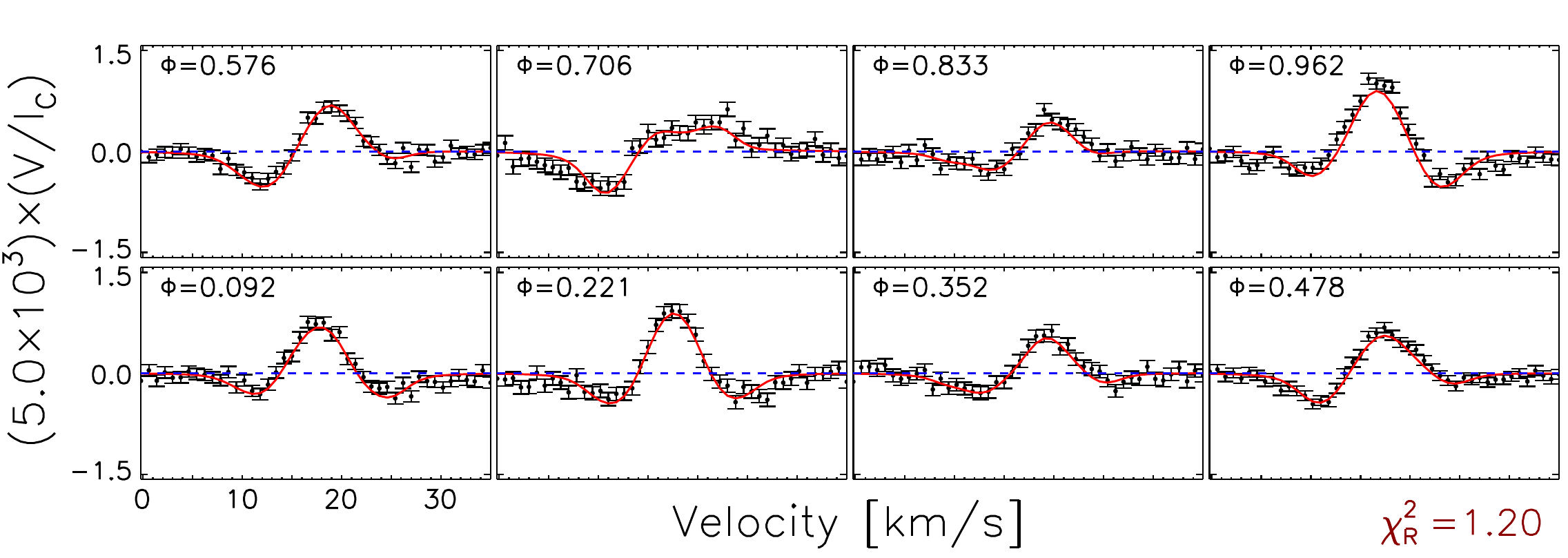}\hspace{2.5pt}\includegraphics[trim=0.5cm 6.1cm 1.0cm 0.2cm, clip=true, width=0.496\linewidth]{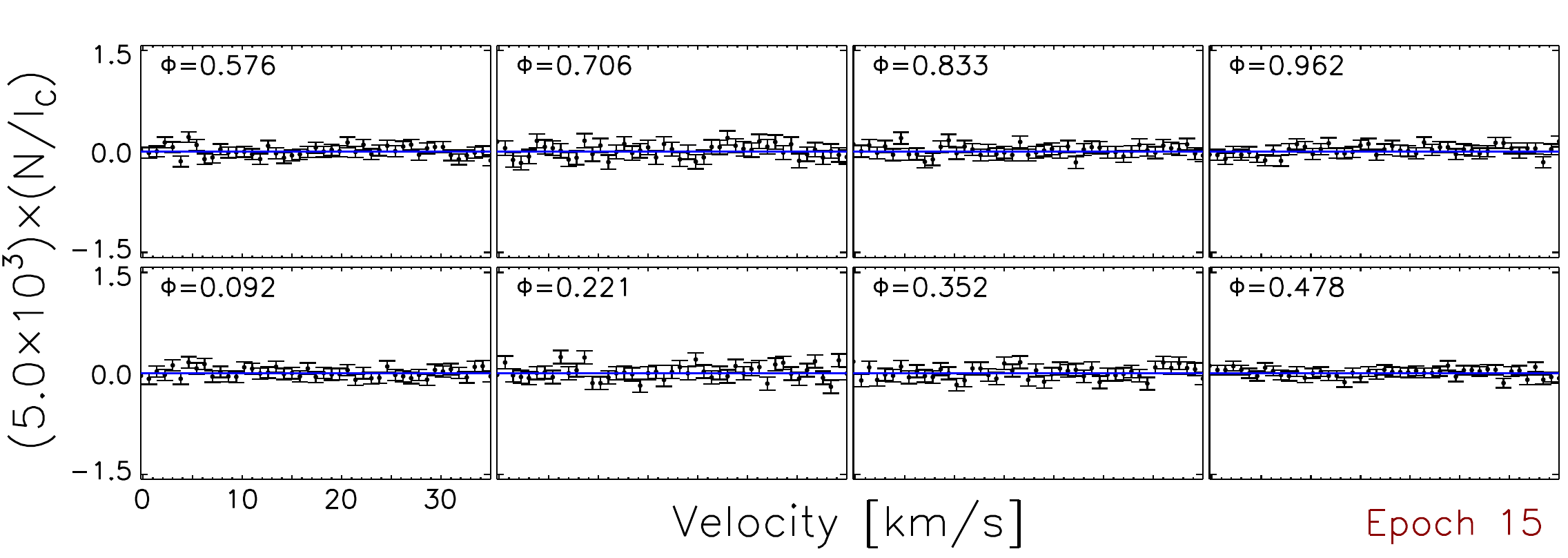}
\caption{Recovered LSD profiles from the spectropolarimetric observations of $\iota$~Hor (Epochs 11 to 15). See caption of Fig.~\ref{fig_4}.}\label{fig_6}
\end{figure*}

\section{Symmetric and Antisymmetric ZDI reconstructions}\label{app:ZDI-maps}

\begin{figure*}
\mbox{\hspace{0.172cm}}\includegraphics[trim=0.5cm 2.6cm 0.5cm 0.1cm, clip=true, width=\linewidth]{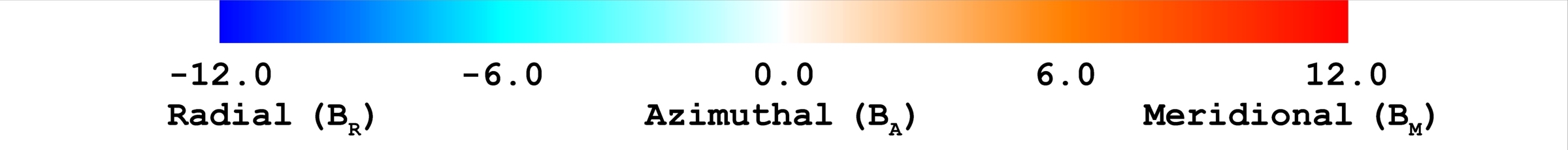}
\includegraphics[trim=0.0cm 1.0cm 1.0cm 9.5cm, clip=true, width=0.495\linewidth]{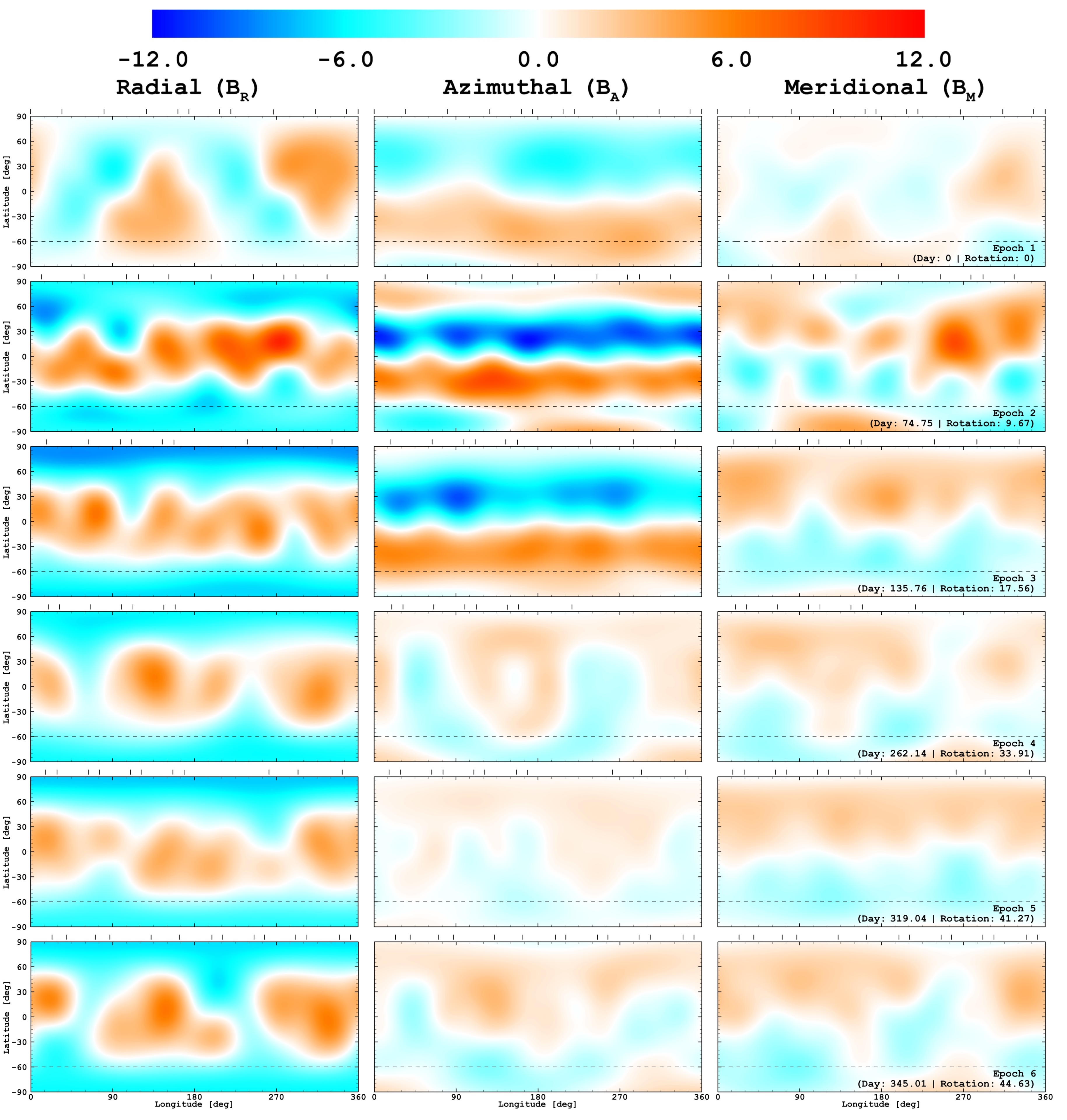}\hspace{2.5pt}\includegraphics[trim=0.0cm 1.0cm 1.0cm 9.5cm, clip=true, width=0.495\linewidth]{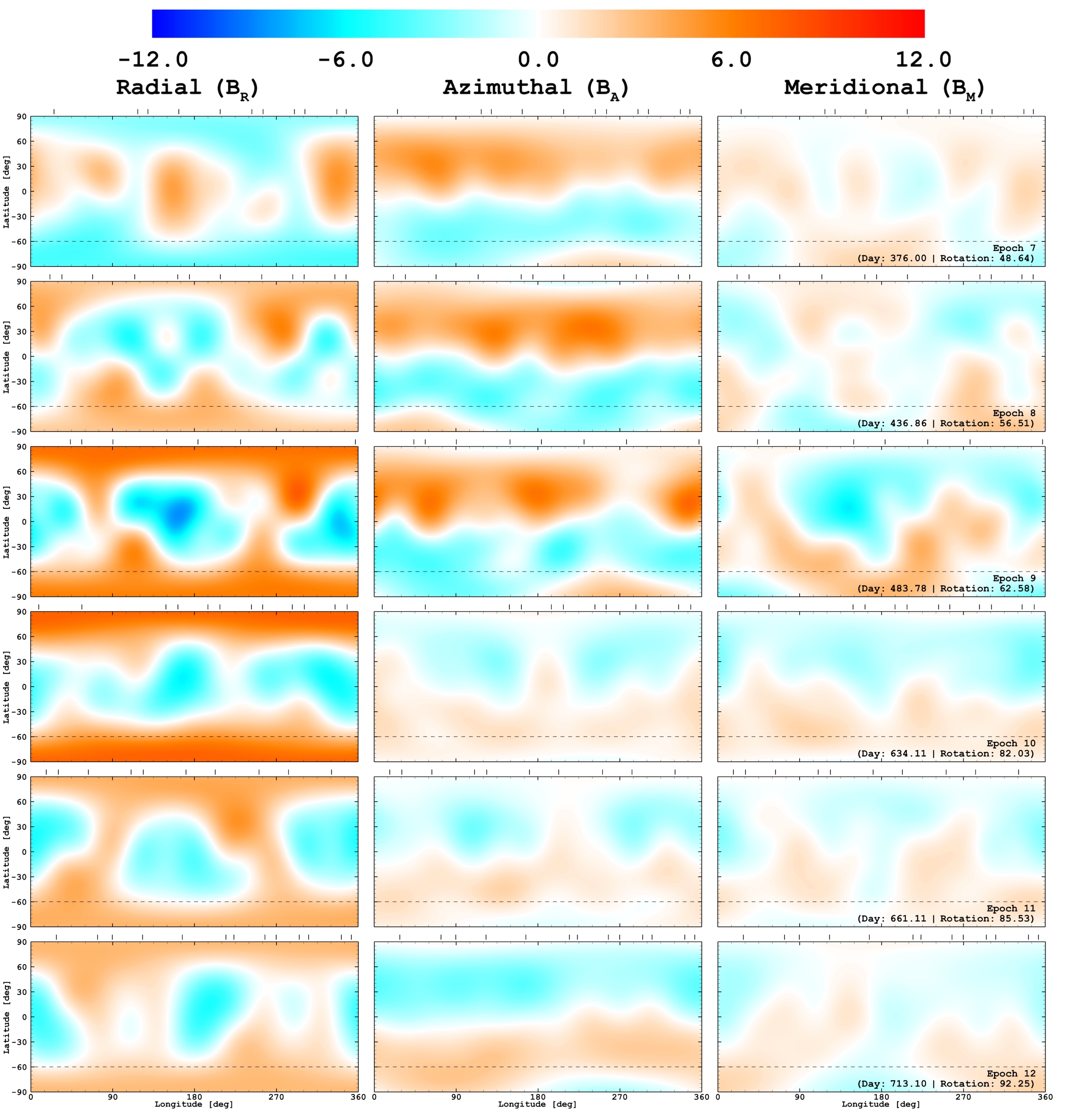}\vspace{-0.1cm}
\caption{Maps of the large-scale magnetic field of \ihor~(in latitude-longitude equirectangular projection) pushing for symmetry during the ZDI reconstruction. Columns contain the radial ($B_{\rm R}$, left), azimuthal ($B_{\rm A}$, middle), and meridional ($B_{\rm M}$, right) field components, with the color scale indicating the magnitude (in Gauss) and field polarity in each case. Each row corresponds to a different observing epoch as indicated (Epochs 1 to 6; left | Epochs 7 to 12; right), with the observed phases denoted by black tick marks in longitude (upper x-axes). The campaign day and number of stellar rotations (measured from ${\rm BJD} = 2457300.78580$ to the first night of observations of a given epoch) are listed in each map. The segmented line shows the visibility limit imposed by the inclination of the star ($i = 60$~deg).}\label{fig_A1}
\end{figure*}

\begin{figure*}
\mbox{\hspace{0.172cm}}\includegraphics[trim=0.5cm 2.6cm 0.5cm 0.1cm, clip=true, width=\linewidth]{ColorBar_RB.jpeg}
\includegraphics[trim=0.0cm 1.0cm 1.0cm 9.5cm, clip=true, width=0.495\linewidth]{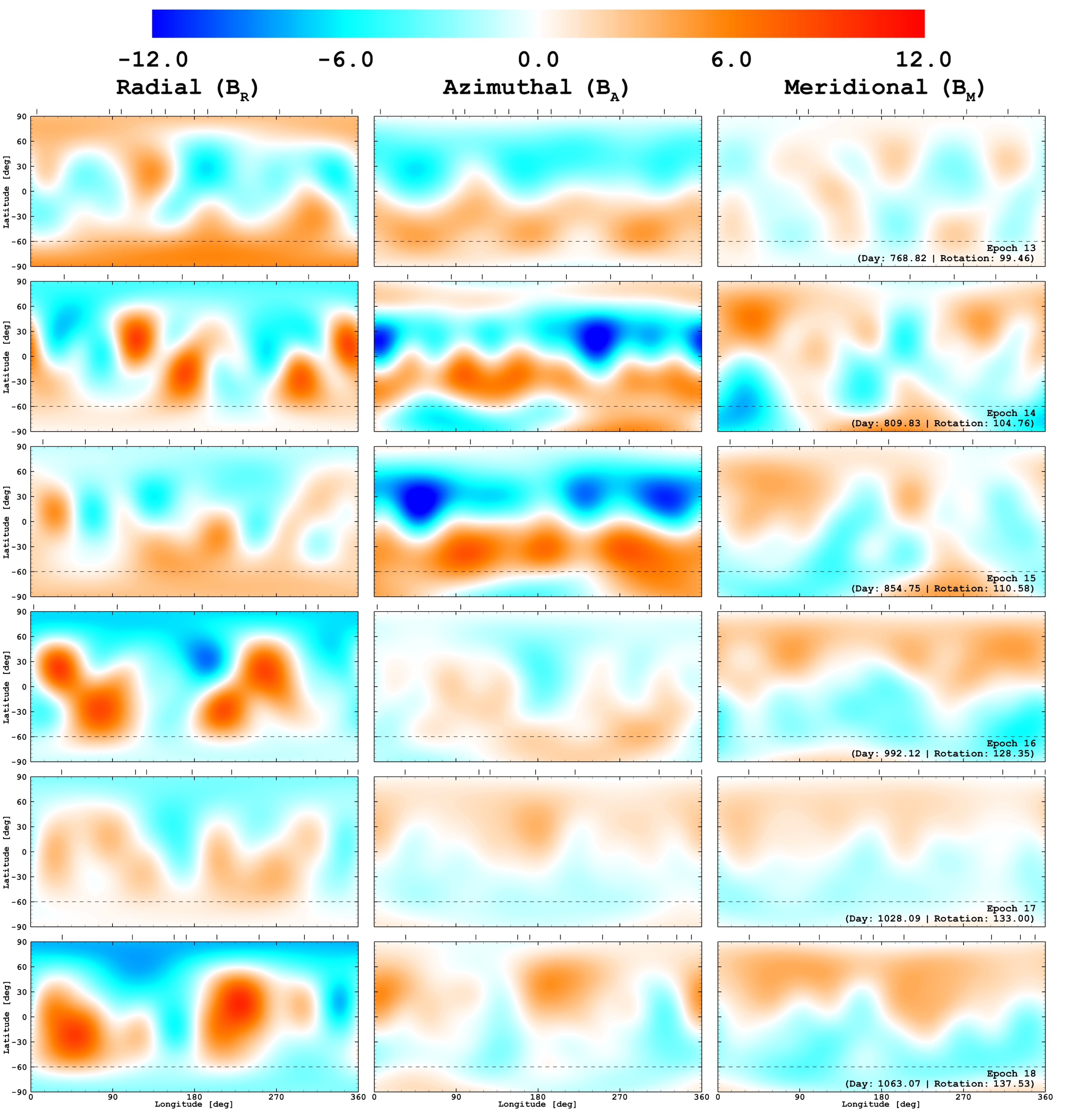}\hspace{2.5pt}\includegraphics[trim=0.0cm 1.0cm 1.0cm 9.5cm, clip=true, width=0.495\linewidth]{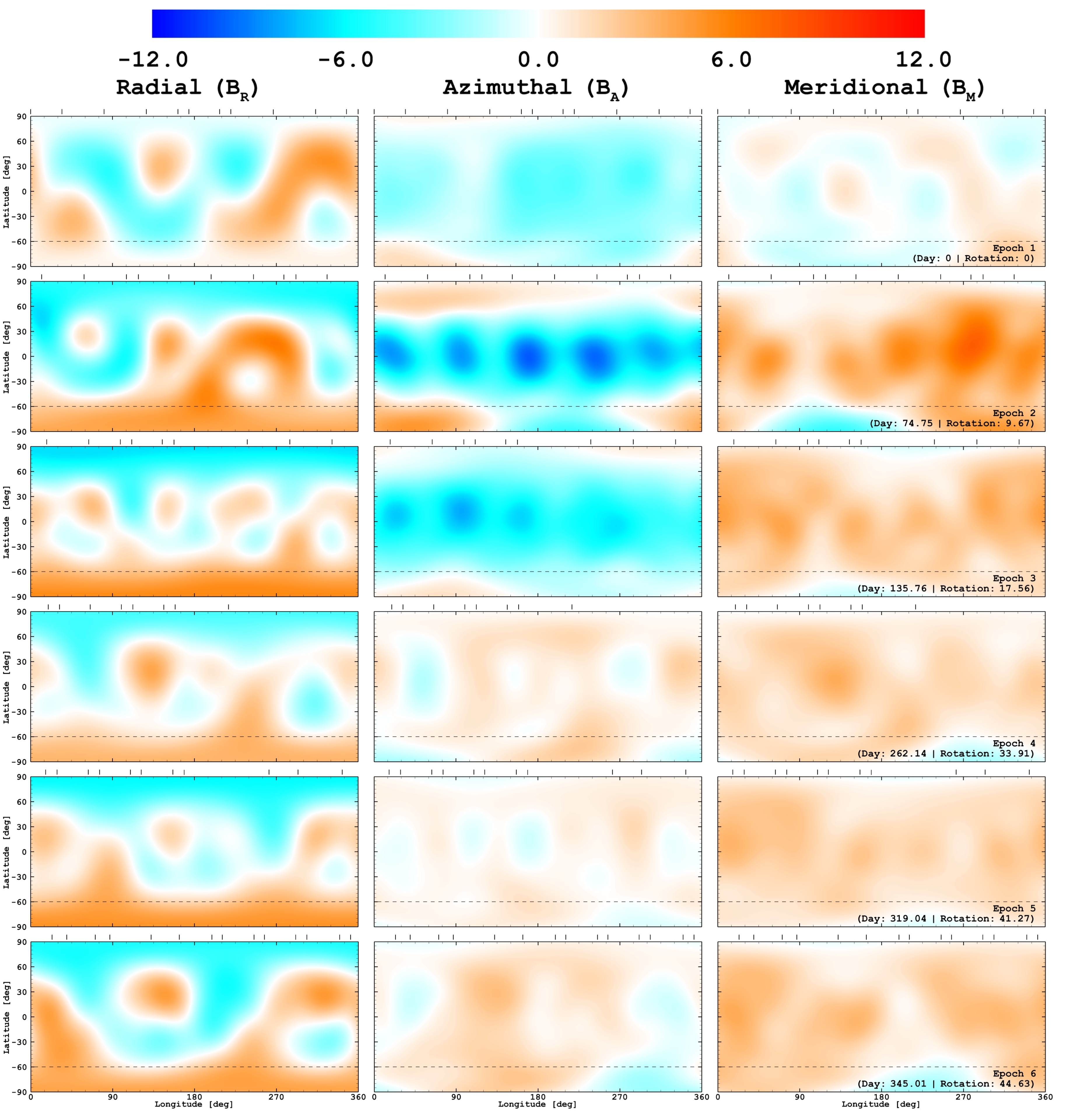}\vspace{-0.1cm}
\caption{Maps of the large-scale magnetic field of \ihor~(in latitude-longitude equirectangular projection) pushing for symmetry (Epochs 13 to 18; left) and antisymmetry (Epochs 1 to 6; right) during the ZDI reconstruction. See caption of Fig.~\ref{fig_A1}.}\label{fig_A3}
\end{figure*}

\begin{figure*}
\mbox{\hspace{0.172cm}}\includegraphics[trim=0.5cm 2.6cm 0.5cm 0.1cm, clip=true, width=\linewidth]{ColorBar_RB.jpeg}
\includegraphics[trim=0.0cm 1.0cm 1.0cm 9.5cm, clip=true, width=0.495\linewidth]{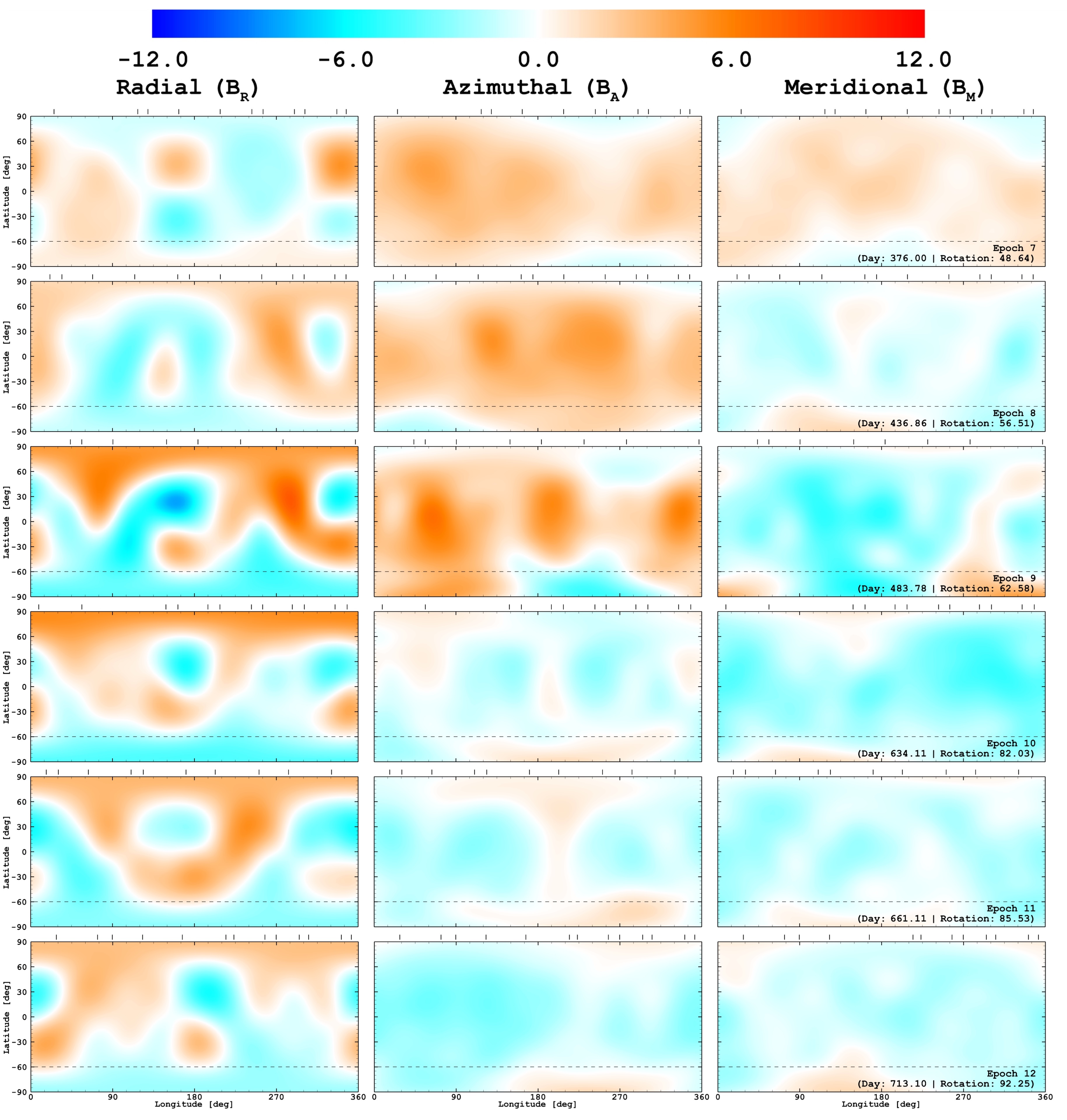}\hspace{2.5pt}\includegraphics[trim=0.0cm 1.0cm 1.0cm 9.5cm, clip=true, width=0.495\linewidth]{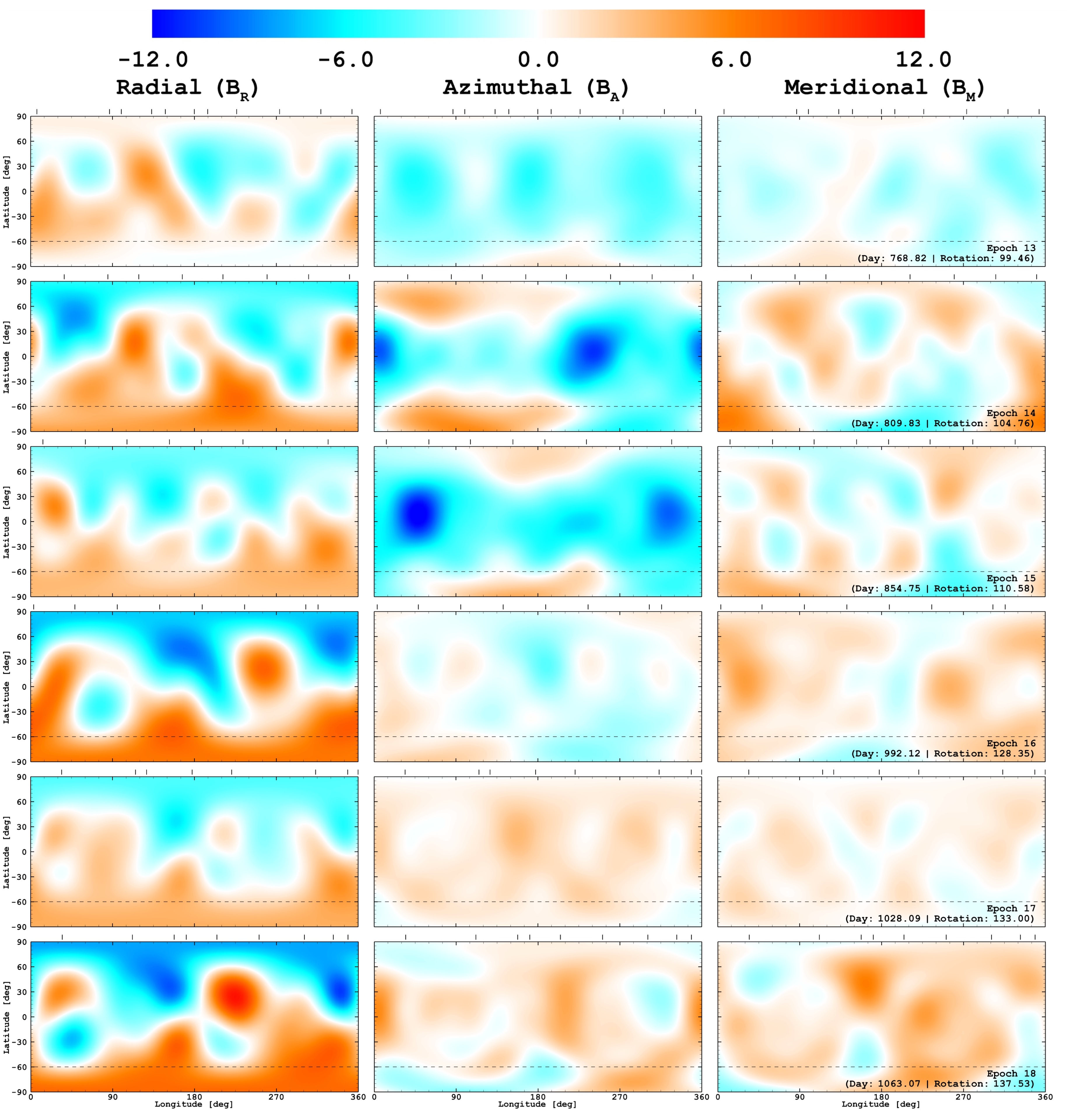}\vspace{-0.1cm}
\caption{Maps of the large-scale magnetic field of \ihor~(in latitude-longitude equirectangular projection) pushing for antisymmetry during the ZDI reconstruction (Epochs 7 to 12; left | Epochs 12 to 18; right). See caption of Fig.~\ref{fig_A1}.}\label{fig_A5}
\end{figure*}

\end{appendix}

\end{document}